\documentclass[manuscript]{aastex631}

\usepackage{newtxtext,newtxmath}
\newcommand{\kms}{km s$^{-1}$}
\newcommand{\dego}{$^\circ$}
\newcommand{\msun}{M$_\odot$}

\shorttitle{Recent mass ejection from W Hya}
\shortauthors{Hoai et al. 2022}

\begin{document}

\title{Recent mass ejection from AGB star W Hya}

\correspondingauthor{Do T. Hoai}

\author[0000-0002-3816-4735]{Do T. Hoai}
\email{dthoai@vnsc.org.vn}
\affiliation{Department of Astrophysics, Vietnam National Space Center, Vietnam Academy of Science and Technology, \\
18, Hoang Quoc Viet, Nghia Do, Cau Giay, Ha Noi, Vietnam}

\author[0000-0002-0311-0809]{Pham T. Nhung}
\email{pttnhungs@vnsc.org.vn}
\affiliation{Department of Astrophysics, Vietnam National Space Center, Vietnam Academy of Science and Technology, \\
18, Hoang Quoc Viet, Nghia Do, Cau Giay, Ha Noi, Vietnam}

\author[0000-0002-8979-6898]{Pierre Darriulat}
\affiliation{Department of Astrophysics, Vietnam National Space Center, Vietnam Academy of Science and Technology, \\
  18, Hoang Quoc Viet, Nghia Do, Cau Giay, Ha Noi, Vietnam}

\author[0000-0002-2808-0888]{Pham N. Diep}
\affiliation{Department of Astrophysics, Vietnam National Space Center, Vietnam Academy of Science and Technology, \\
  18, Hoang Quoc Viet, Nghia Do, Cau Giay, Ha Noi, Vietnam}

\author[0000-0002-5913-5554]{Nguyen B. Ngoc}
\affiliation{Department of Astrophysics, Vietnam National Space Center, Vietnam Academy of Science and Technology, \\
  18, Hoang Quoc Viet, Nghia Do, Cau Giay, Ha Noi, Vietnam}

\author[0000-0002-8408-4816]{Tran T. Thai}
\affiliation{Department of Astrophysics, Vietnam National Space Center, Vietnam Academy of Science and Technology, \\
18, Hoang Quoc Viet, Nghia Do, Cau Giay, Ha Noi, Vietnam}

\author[0000-0002-3773-1435]{Pham Tuan-Anh}
\affiliation{Department of Astrophysics, Vietnam National Space Center, Vietnam Academy of Science and Technology, \\
18, Hoang Quoc Viet, Nghia Do, Cau Giay, Ha Noi, Vietnam}

\begin{abstract}
 We analyse archival ALMA (Atacama Large Millimeter/submillimeter Array) observations of two molecular line emissions, $^{12}$CO(3-2) and $^{29}$SiO(8-7),  from oxygen-rich AGB (Asymptotic Giant Branch) star W Hya. Together with results of earlier VLT observations at visible and infrared wavelengths, our results suggest a two-component picture of the morpho-kinematics of the circumstellar envelope (CSE), one stable over time, at the scale of centuries, and the other variable, at the scale of years. The stable component consists of an approximately spherical shell of gas and dust expanding radially to a terminal velocity of $\sim$5 \kms\ at a distance of $\sim$30 au from the star. It is found to display comparable features as seen in the CSE of R Dor, a star similar to W Hya. The variable component projects on the plane of the sky over a region confined to the neighbourhood of the star and elongated toward the north. Its very high density and sudden acceleration suggest an interpretation in terms of mass ejection initiated a few years ago. We discuss its properties in relation with earlier observations of dust formation in the same region. Our results offer a picture of the wind of W Hya that differs significantly from the picture that could be suggested by earlier analyses, giving evidence for a mass ejection that had been previously overlooked and underscoring new relations between the dust and gas emissions. They have an impact on the evidence published earlier for the presence of CO masers. They favour an interpretation in terms of convective cell ejections playing the main role in the generation of the nascent wind, the stable component of the CSE being seen as the result of many successive such events occurring in different directions at short time intervals.
\end{abstract}

\keywords{stars: AGB and post-AGB, circumstellar matter, stars: individual: W Hya, radio lines: stars}

\section{Introduction}

W Hya is an oxygen-rich Asymptotic Giant Branch (AGB) star at a distance of only $104^{+14}_{-11}$ pc from the Sun \citep{vanLeeuwen2007}; a shorter distance of 78$\pm$6 pc \citep{Knapp2003} is often used in the published literature. It is a semi-regular variable with a period of $\sim$361 days \citep{Samus2017}, often quoted as a Mira \citep{Lebzelter2005} belonging to spectral class M7.5e-M9ep. Single dish observations suggest a mass-loss rate at the level of $\sim$10$^{-7}$ \msun\,yr$^{-1}$ \citep{Maercker2008, Khouri2014a} and a main sequence mass between 1 and 1.5 \msun\ \citep{Khouri2014b, Danilovich2017}.

The close neighbourhood of the star has been recently observed with the Very Large Telescope (VLT) at wavelengths covering from visible to near-infrared, using NACO \citep{Norris2012}, SPHERE /ZIMPOL \citep{Ohnaka2016, Ohnaka2017, Khouri2020}, AMBER/VLTI \citep{Ohnaka2016,Ohnaka2017, Hadjara2019} and MIDI/VLTI \citep{ZhaoGeisler2015}. These observations have given evidence for a clumpy and dusty layer, displaying important variability at short time scale. Dust grains, mostly aluminium composites, are found to have sizes ranging between 0.1 and 0.5 $\mu$m. \cite{Vlemmings2017,Vlemmings2019} using the Atacama Large Millimeter/submillimeter Array (ALMA) to observe continuum and CO($\nu$=1,$J$=3-2) emissions close to the star have given evidence for shocks induced by pulsations and convective cell ejections, producing an effective line broadening of the molecular emission observed on lines of sight close to the centre of the star \citep{Hoai2021}.

Some of the most relevant observations are displayed in the Appendix (Table \ref{taba1}). These observations have made it possible to refine and improve our earlier understanding of the morpho-kinematics of the CSE of W Hya, which \cite{Khouri2015} had summarized in a model based on a large number of earlier observations over a broad range of wavelengths, including from Herschel and other single-dish telescopes \citep{Khouri2014a,Khouri2014b}. In particular, the analysis of \cite{Ohnaka2016,Ohnaka2017} suggests the presence of a shell of large transparent grains, such as of Al$_2$O$_3$, with a 550 nm optical depth of 0.6$\pm$0.2 and inner and outer radii of 1.3 $R_*$ ($R_*$=25 mas) and 10$\pm$2 $R_*$, respectively; it gives evidence for time variability, with predominance of small (0.1 $\mu$m) grains at minimum light and larger (0.5 $\mu$m) grains at pre-maximum light; it gives also evidence for anisotropy with a dust density enhancement of a factor $\sim$4 in the north, over a cone of $\sim$45\dego\ half-opening angle. These results have been confirmed by ALMA millimetre observations of the emission of AlO lines \citep{Takigawa2017}, found to be confined within $\sim$3 $R_*$ from the centre of the star, in contrast with the emission of the $^{29}$SiO(8-7) line, observed to extend to much larger radii. Together with \cite{Takigawa2019}, these authors suggest that the dust shell near the star (2 to 3 $R_*$) is dominated by alumina-rich dust grains, and that the overall wind acceleration, up to some 5.5 \kms, occurs beyond this shell where silicate dust does not form efficiently, SiO molecules remaining mostly in the gas phase; such inefficient silicate formation is claimed to be consistent with the SiO/AlO ratio estimated from mid-infrared dust emission, where only $\sim$6\% of the Si atoms should condense as silicate dust if all Al atoms are bound in aluminium oxide.

Finally, \cite{Vlemmings2021} have recently observed CO masers at a Doppler velocity of 5.5 \kms\ using ALMA millimetre observations of the $^{12}$CO(3-2) line emission.

The present article revisits earlier analyses of archival ALMA observations of the $^{29}$SiO(8-7) and $^{12}$CO(3-2) molecular emissions in order to draw a significantly improved picture of the morpho-kinematics of the CSE of the star. Section 2 gives a brief summary of observations and data reduction. Section 3 describes the observed morpho-kinematics of the CSE. Section 4 uses simple radiative transfer considerations to shed light on these observations. Section 5 comments on continuum emission. Section 6 is an attempt at drawing a global picture that accounts for as many as possible observations reported both here and in the earlier literature. Section 7 discusses the implications in terms of the dynamics at stake in the W Hya CSE and concludes.   

\section{Observations and data reduction}
The present work uses archival observations of W Hya from project ADS/JAO.ALMA \#2015.1.01446.S (PI: A. Takigawa), which were carried out for a total of $\sim$2 hours on source between 30 November and 5 December 2015 (stellar pulsation phase $\varphi$=0.3) with ALMA in Cycle 3. The antennas, respectively 33 and 41 in number, were configured in such a way that the baseline lengths were distributed in two groups, one covering between $\sim$15 m and $\sim$200 m, the other between $\sim$400 m and $\sim$8 km; the former group included 10 antennas in a circle of 100 m radius providing insufficient $uv$ coverage for addressing properly the short spacing problem. As a result, reliable imaging is limited to projected angular distances smaller than 0.15-0.20 arcsec from the star \citep{Hoai2021}. The emissions of the $^{29}$SiO(8-7) line, with a frequency of 342.9808 GHz, and of the $^{12}$CO(3-2) line, with a frequency of 345.7960 GHz, were observed in different spectral windows, with channel spacing of 0.854 \kms\ and 0.4236 \kms, respectively. The data have been calibrated with CASA\footnote{https://casa.nrao.edu/} using standard scripts provided by ALMA without continuum subtraction. Imaging, using GILDAS\footnote{https://www.iram.fr/IRAMFR/GILDAS/}/IMAGER\footnote{https://imager.oasu.u-bordeaux.fr} with robust weighting parameter of 1, produces beams of 52$\times$38 mas$^2$ (PA=99\dego) and 50$\times$42 mas$^2$ (PA=101\dego) for the SiO and CO lines respectively. Noise levels ($\sigma$) are of $\sim$2 and $\sim$3 mJy\,beam$^{-1}$, respectively.

The observations of the CO line have been analysed earlier by  \cite{Vlemmings2017,Vlemmings2019,Vlemmings2021} who used the data in conjunction with observations of a second epoch (project 2016.A.00029.S) on 25 Nov 2017 with an on-source observing time of $\sim$35 min. The observations of the SiO line have been analysed earlier by  \cite{Takigawa2017},  \cite{Danilovich2019} and \cite{Hoai2021}. Details related to the observations, calibration and reduction of the data have been presented in these publications.

We use coordinates centred on the peak of continuum emission, $x$ pointing east, $y$ pointing north and $z$ pointing away from Earth. The projected distance to the star is calculated as $R$=$\sqrt{x^2+y^2}$. Position angles, $\omega$, are measured counter-clockwise from north. The Doppler velocity, $V_z$, spectra are referred to a systemic velocity of 40.4 \kms.

\section{Morpho-kinematics of the CSE}

\subsection{A radially expanding shell}

\begin{figure*}
  \centering
  \includegraphics[height=4.2cm,trim=.7cm 1.cm 2.4cm 2cm,clip]{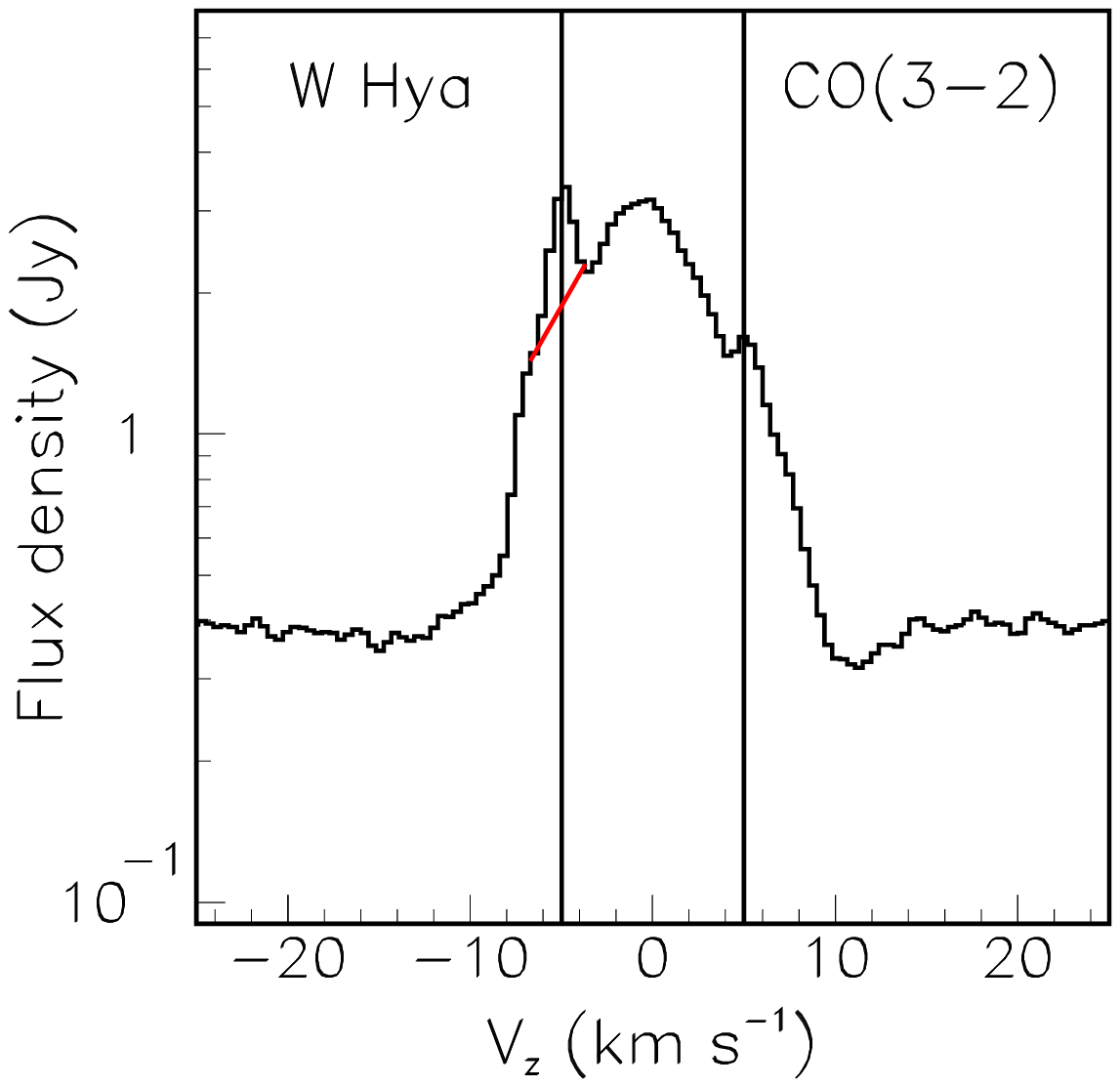}
  \includegraphics[height=4.2cm,trim=.7cm 1.cm 2.4cm 2cm,clip]{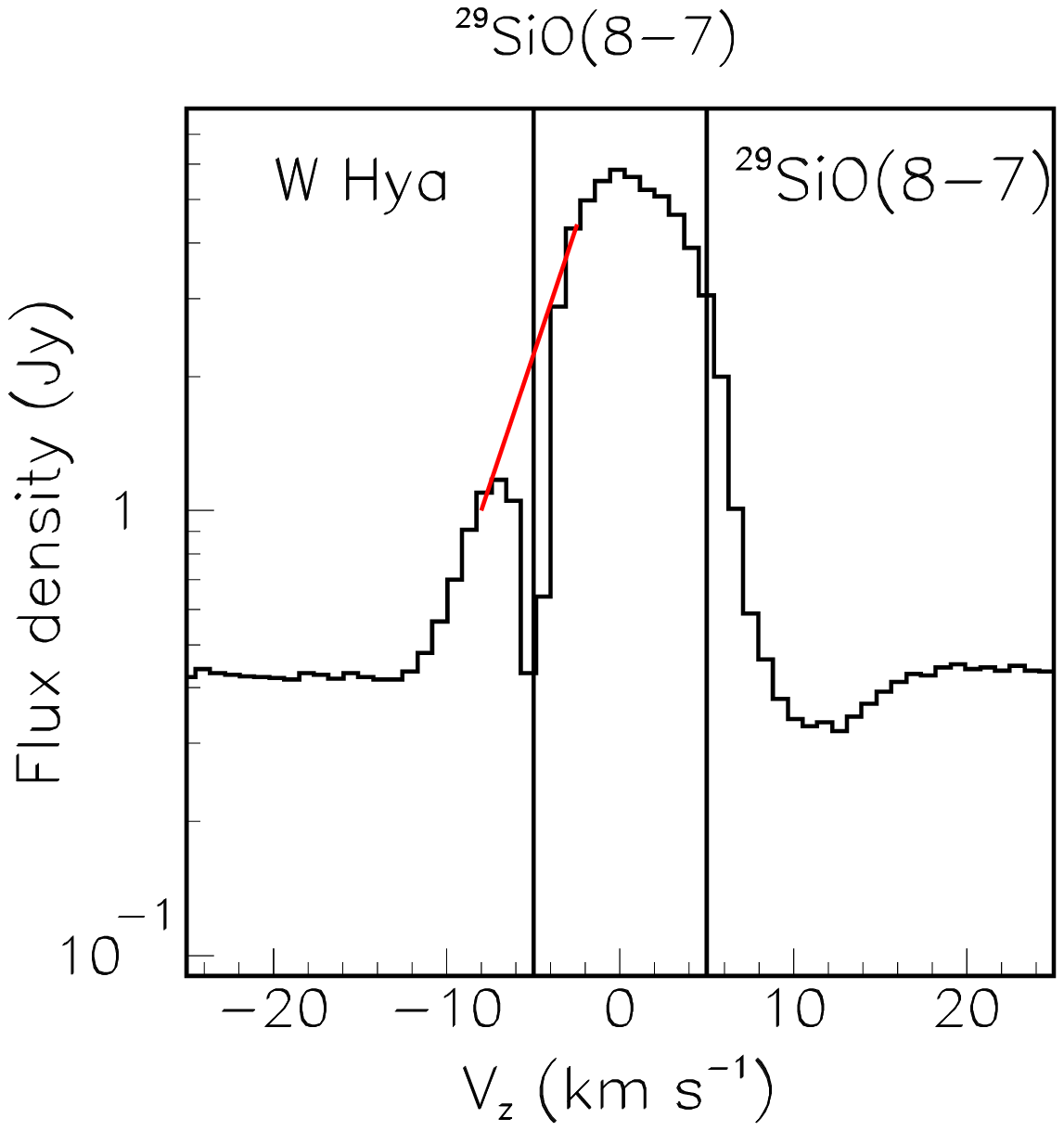}
  \includegraphics[height=4.2cm,trim=.7cm 1.cm 2.4cm 2cm,clip]{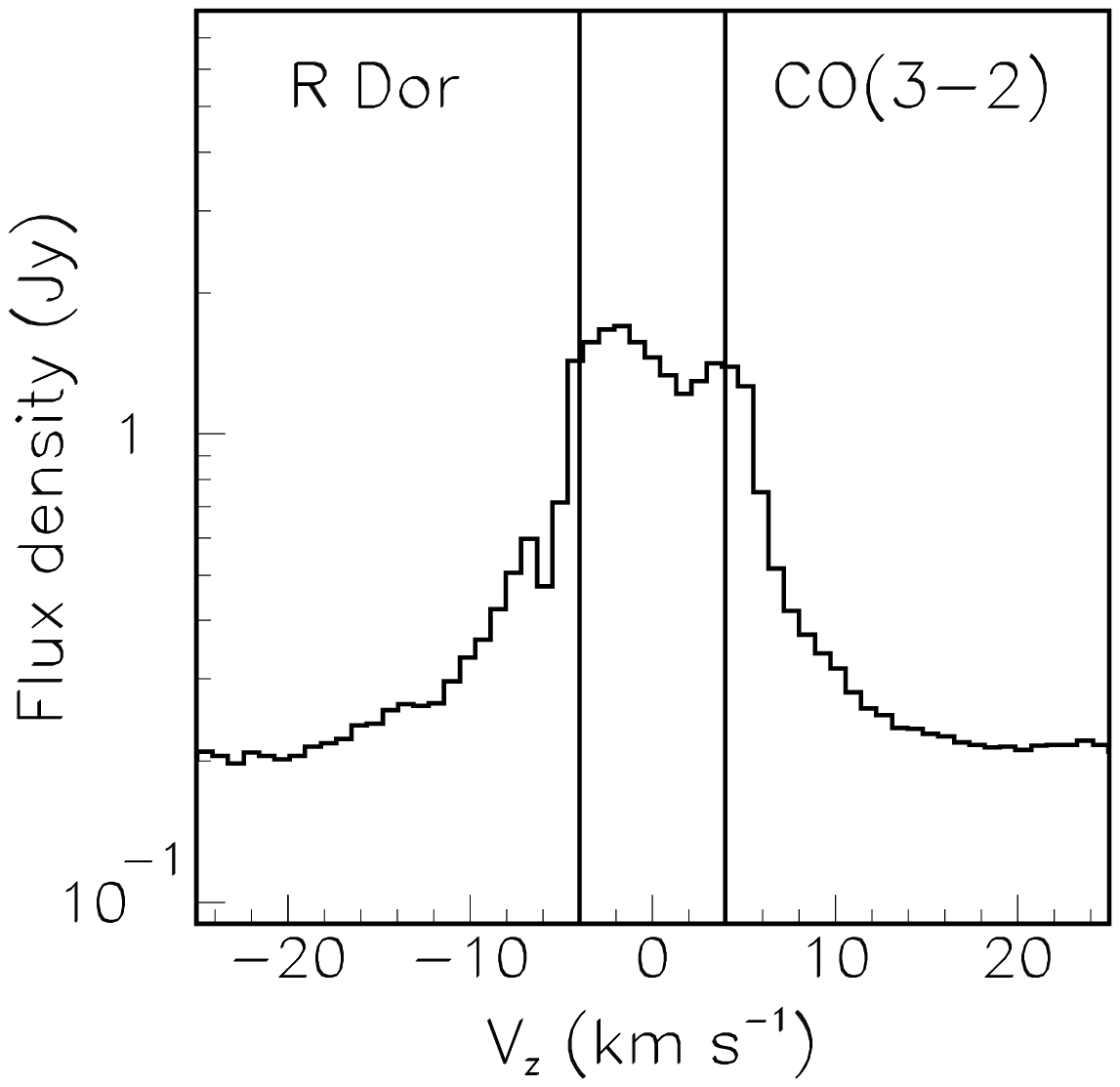}
  \includegraphics[height=4.2cm,trim=.7cm 1.cm 2.4cm 2cm,clip]{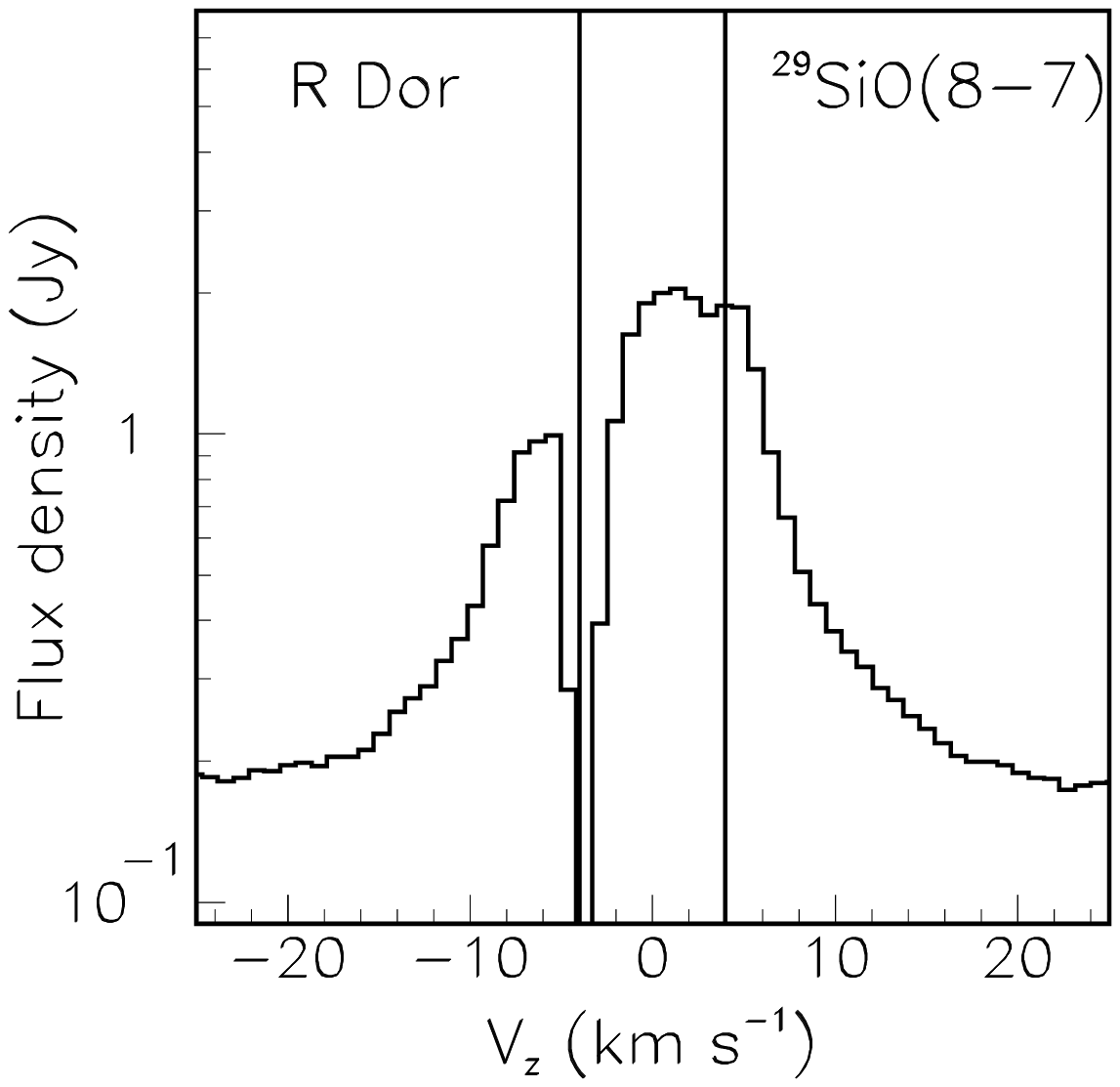}
  \caption{Doppler velocity spectra integrated over the circle $R$$<$0.2 arcsec for the CO(3-2) (left) and $^{29}$SiO (8-7) (centre-left) line emissions of W Hya. The red lines show the interpolation over the peak range of the emission observed on the sides of the peak (see Section 3.3). The right panels show the same line emissions for R Dor, integrated over a circle $R$$<$0.35 arcsec and divided by a factor 3.1, accounting for the different distances (104 pc for W Hya vs 59 pc for R Dor). Vertical lines are shown at $V_z$=$\pm$5 \kms\ for W Hya and $\pm$4 \kms\ for R Dor. }
  \label{fig1}
\end{figure*}

Doppler velocity spectra integrated over the circle $R$$<$0.2 arcsec are displayed in the left panels of Figure \ref{fig1}. Contrasting with the small angular range being covered, these spectra probe the CSE along the line of sight over large distances where temperature and density conditions allow for emission of the observed lines. Both spectra show a clear peak at $V_z$$\sim$$-5$ \kms, SiO in absorption and CO in emission. This suggests the presence of a gas layer between star and observer, optically thin for CO and thick for SiO emissions, over which the wind has reached a terminal velocity of $\sim$5 \kms: a similar situation is observed in several other AGB stars as illustrated in the right panels of Figure \ref{fig1} by the case of R Dor \citep{Nhung2021} where the terminal velocity is $\sim$4 \kms. In both stars, the CO spectrum shows a small emission peak at terminal velocity on the red-shifted side, suggesting that the expanding layer is in fact part of a spherical shell surrounding the star. 

However, in the case of W Hya, such interpretation raises a few questions: a) the Doppler velocity of $\sim$5 \kms\ is the same as the velocity where \cite{Vlemmings2021} observe a CO maser; b) the wind terminal velocity evaluated by  \cite{Khouri2014a,Khouri2014b} is 7.5 \kms\ rather than 5 \kms; c) while similar, the spectra of the two stars display small but significant differences; in particular the CO emission peak of W Hya is narrower than that of R Dor and the SiO absorption peak is shallower. We address these questions in the remaining of the present section.

\begin{figure*}
  \centering
  \includegraphics[height=4.5cm,trim=.2cm 1.cm 1.5cm 1.5cm,clip]{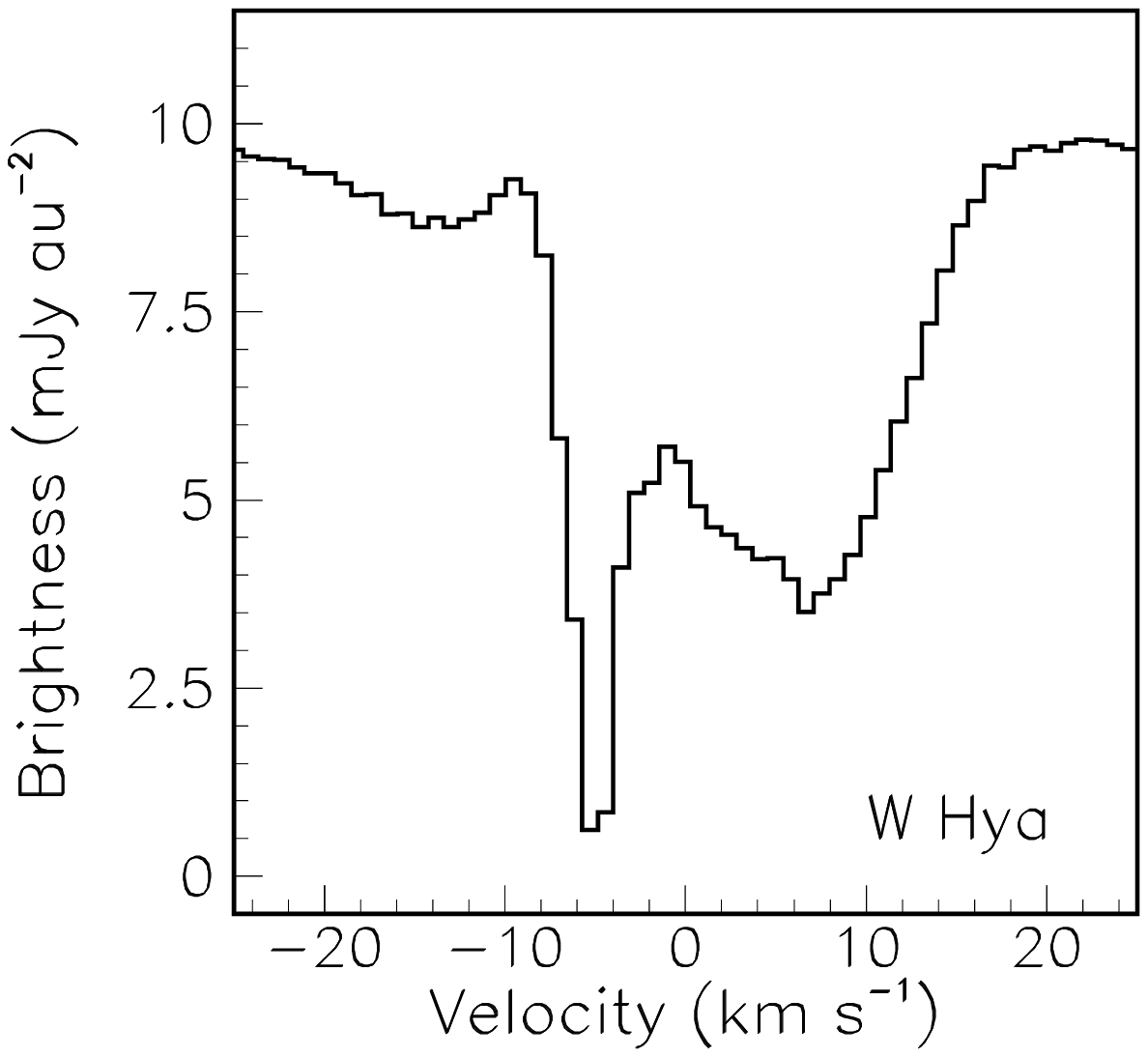}
  \includegraphics[height=4.5cm,trim=.2cm 1.cm 1.5cm 1.5cm,clip]{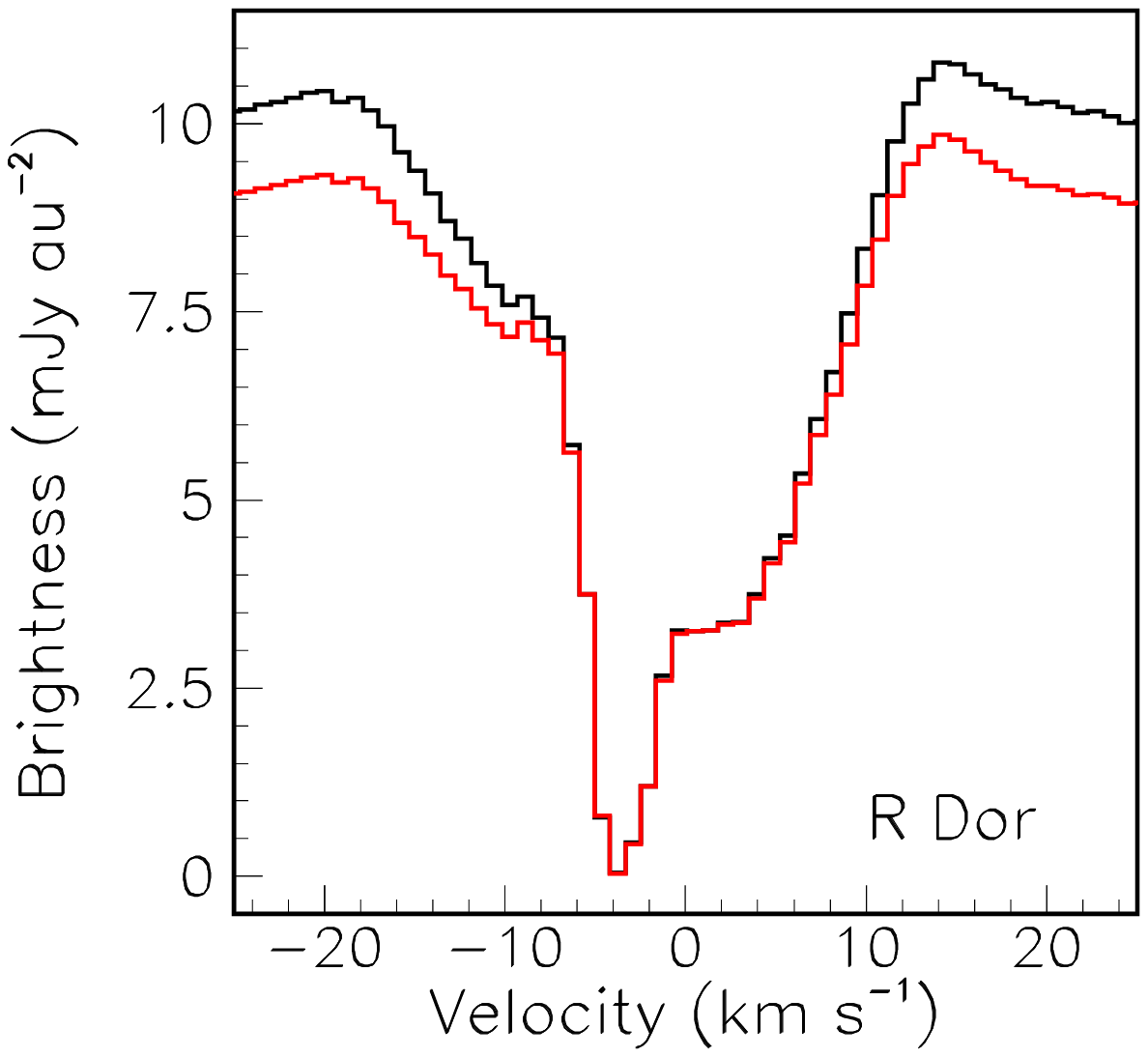}
  \caption{   Comparing the spectral distributions of the central brightness in W Hya (left) and R Dor (right). The difference in distances is taken into account by proper scaling and the difference in beam sizes (53$\times$40 mas$^2$ for W Hya and 41$\times$35 mas$^2$ for R Dor) is taken into account by averaging over 30$\times$30 mas$^2$ for W Hya and over both 45$\times$45 mas$^2$ (black) and 55$\times$55 mas$^2$ (red) for R Dor, providing an estimate of the attached uncertainty. }
  \label{fig2}
\end{figure*}

Spectral maps of both lines in a square $|x,y|$$<$75 mas centred on the star are displayed in Figure \ref{figa1} of the Appendix. Significant continuum contribution is seen to extend to $R$$\sim$50 mas, in agreement with earlier analyses \citep{Vlemmings2017,Vlemmings2019}. In all panels, the distributions show a peak between $\sim$$-$4 and $\sim$$-$6 \kms, in emission for the CO line and in absorption for the SiO line. Under the peak, the CO spectra display a broad distribution, covering between $V_z$$\sim$$-$10 and $+$10 \kms; red-shifted absorption is seen over the stellar disc, particularly strong for the SiO line, suggesting the presence of in-falling gas. Also shown in this figure is a same set of spectral maps for the $^{29}$SiO(8-7) emission of R Dor \citep{Nhung2021}, scaled to account for the difference in distances. It displays remarkable similarity with the W Hya set as illustrated in Figure \ref{fig2} for the spectra covering the stellar discs. Yet, significant differences are visible: the absorption peak associated with the terminal velocity is broader for R Dor than for W Hya and the broad absorption spectrum underneath the peak is more red-shifted for W Hya than for R Dor.

\subsection{Terminal velocity}

The absence of evidence for a terminal velocity exceeding $\sim$5 \kms\ may cast doubt on the evaluation of 7.5 \kms\ proposed by \citet{Khouri2014a}. These authors fit single dish infrared to sub-millimetre observations of CO line profiles (Herschel, APEX, SMT and SEST) using simple non-LTE radiative transfer modelling with an initial CO/H$_2$ abundance of 4$\times$10$^{-4}$. The best fit gives a terminal velocity of 7.5$\pm$0.5 \kms, a turbulent velocity of 1.4$\pm$0.1 \kms\ and a radial dependence of the temperature of the form $T$[K]=2500$\times$(23/$r$[mas])$^{0.65\pm0.05}$. It requires a photo-dissociation radius of the CO molecules 2.5 times smaller than expected from the \citet{Mamon1988} modelling and a slow start of the wind acceleration for expansion velocities below 5.5-6.0 \kms, followed by a fast injection of momentum to reach the terminal velocity. These results contrast with those obtained by \citet{Takigawa2017}, who suggest that acceleration takes place  beyond an AlO-rich dust shell that covers between 50 and 70 mas from the centre of the star and reaches quickly a terminal velocity of 5.5 \kms. Also, \citet{Vlemmings2021} find it difficult to adopt the radial dependence of the expansion velocity proposed by \citet{Khouri2014a} when modelling the CO masers: their model requires a larger velocity at the inner envelope, $\sim$4.2 \kms, a steeper acceleration and a larger turbulent velocity component. In summary, large uncertainties are attached to the results obtained by \citet{Khouri2014a} using single dish line profiles: the Doppler velocity spectra of the $^{12}$CO(3-2) and $^{29}$SiO(8-7) ALMA observations presented here can reasonably be interpreted as evidence for a terminal velocity of 5-6 \kms\ rather than 7.5 \kms. In this context, we remark that the escape velocity from a solar mass star is $\sim$20 \kms\ beyond the dust shell and reaches 5 \kms\ at only $\sim$700 mas: the precise form of the radial dependence of the velocity in this interval depends on the details of the forces acting on the dust grains and of their collisions with the gas molecules, in a way which is far from being well understood. We also note that the observation of narrow peaks at terminal velocity implies that this velocity has been reached early enough to dominate much of the explored regions of the line of sight.    We  remark that an OH maser observed in 1986 at Nan\c{c}ay \citep{Etoka2001} is at this same velocity of 5-6 \kms.

Of course, we cannot exclude that further acceleration, beyond 5 \kms, takes place at distances from the star where the emission of the CO line has become too weak to be detected, because of the density or/and the temperature have become too low. This however would be irrelevant to the arguments made in the present article, the expansion velocity of $\sim$5 \kms\ being effectively terminal over the explored radial range.

\subsection{Peak emission of the CO line}

\begin{figure*}
  \centering
  \includegraphics[height=5cm,trim=.0cm 1.cm 1.5cm .5cm,clip]{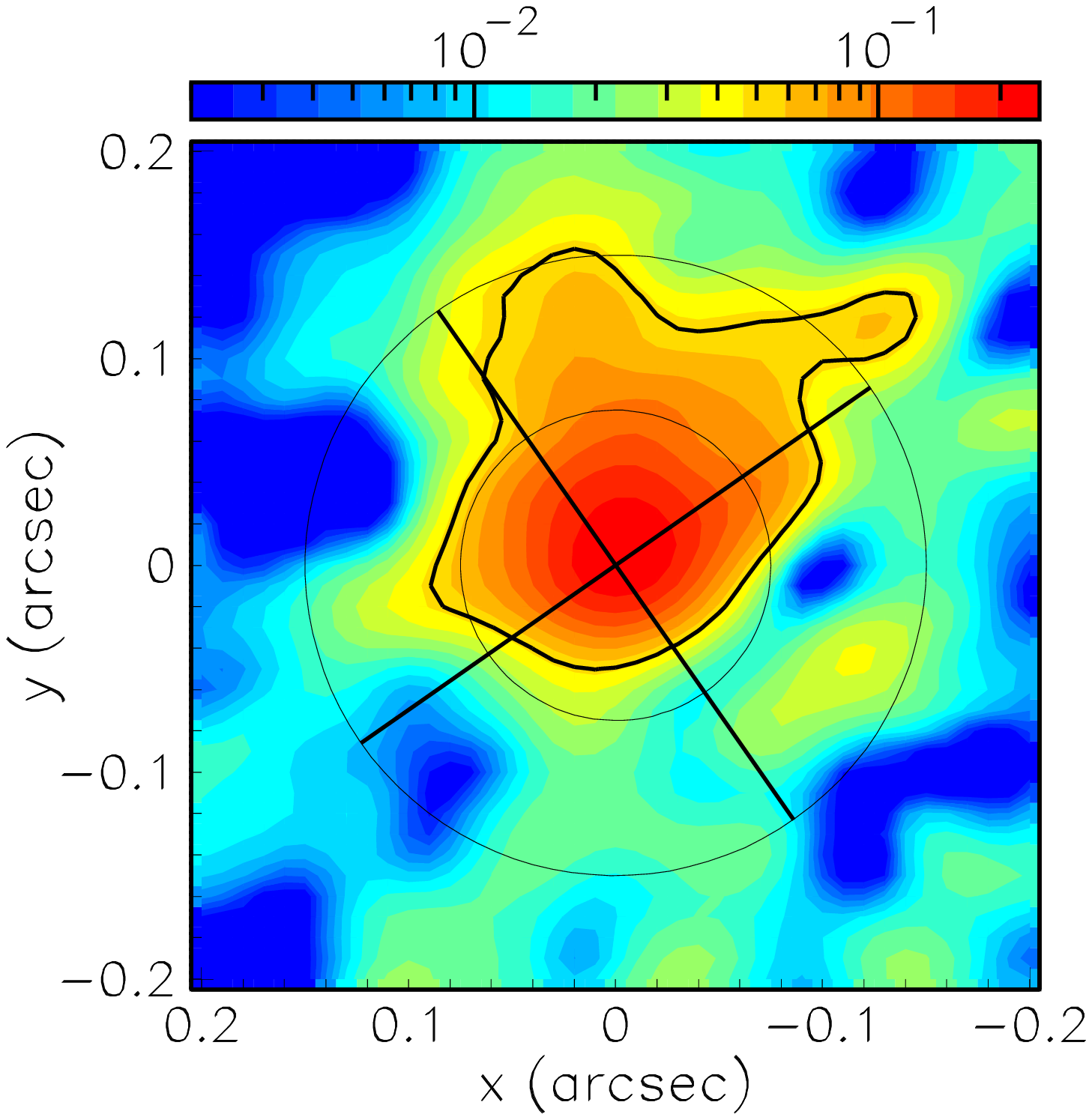}
  \includegraphics[height=5cm,trim=.0cm 1.cm 1.5cm .5cm,clip]{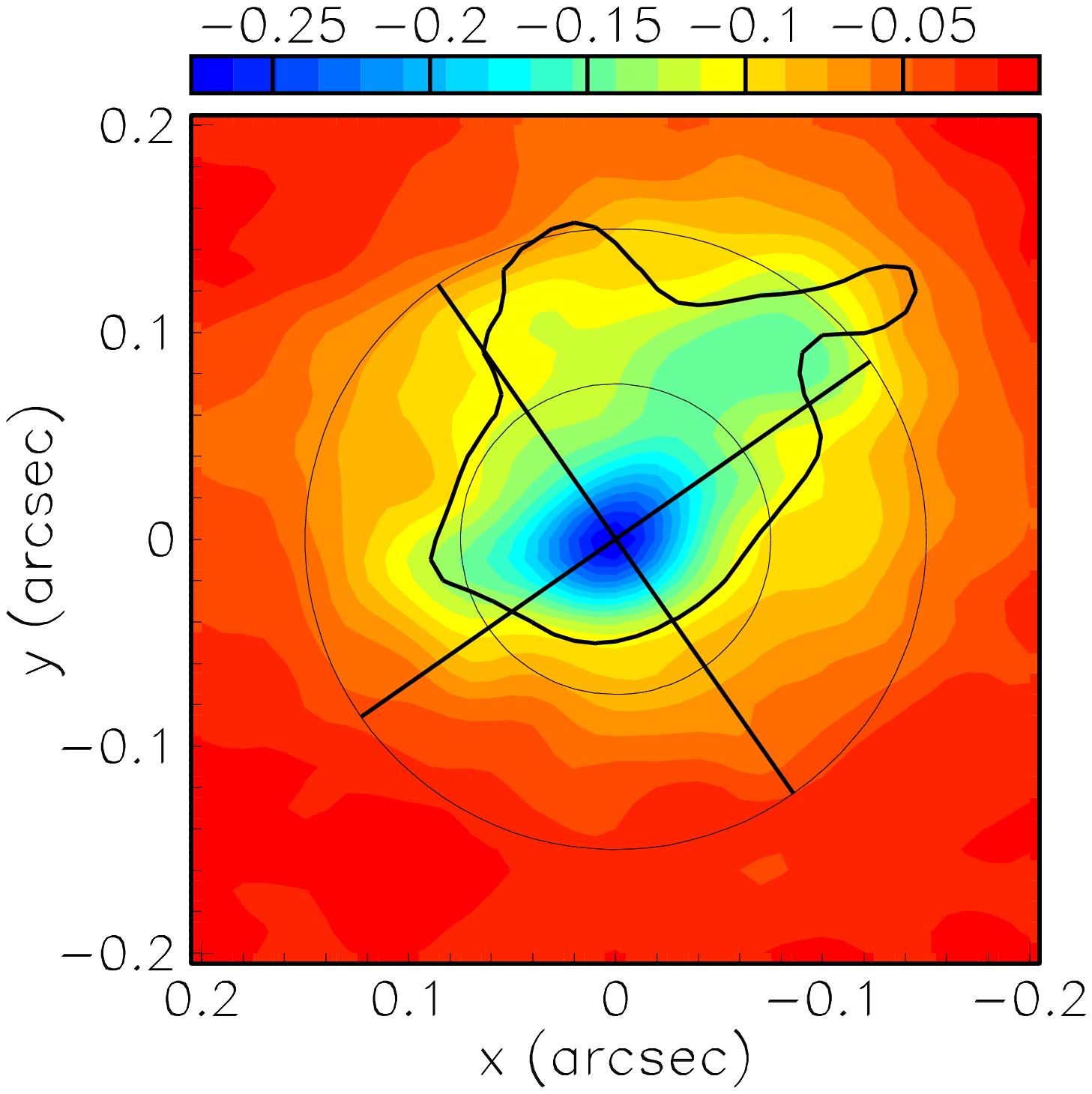}
  \includegraphics[height=5cm,trim=.0cm 1.cm 1.5cm .5cm,clip]{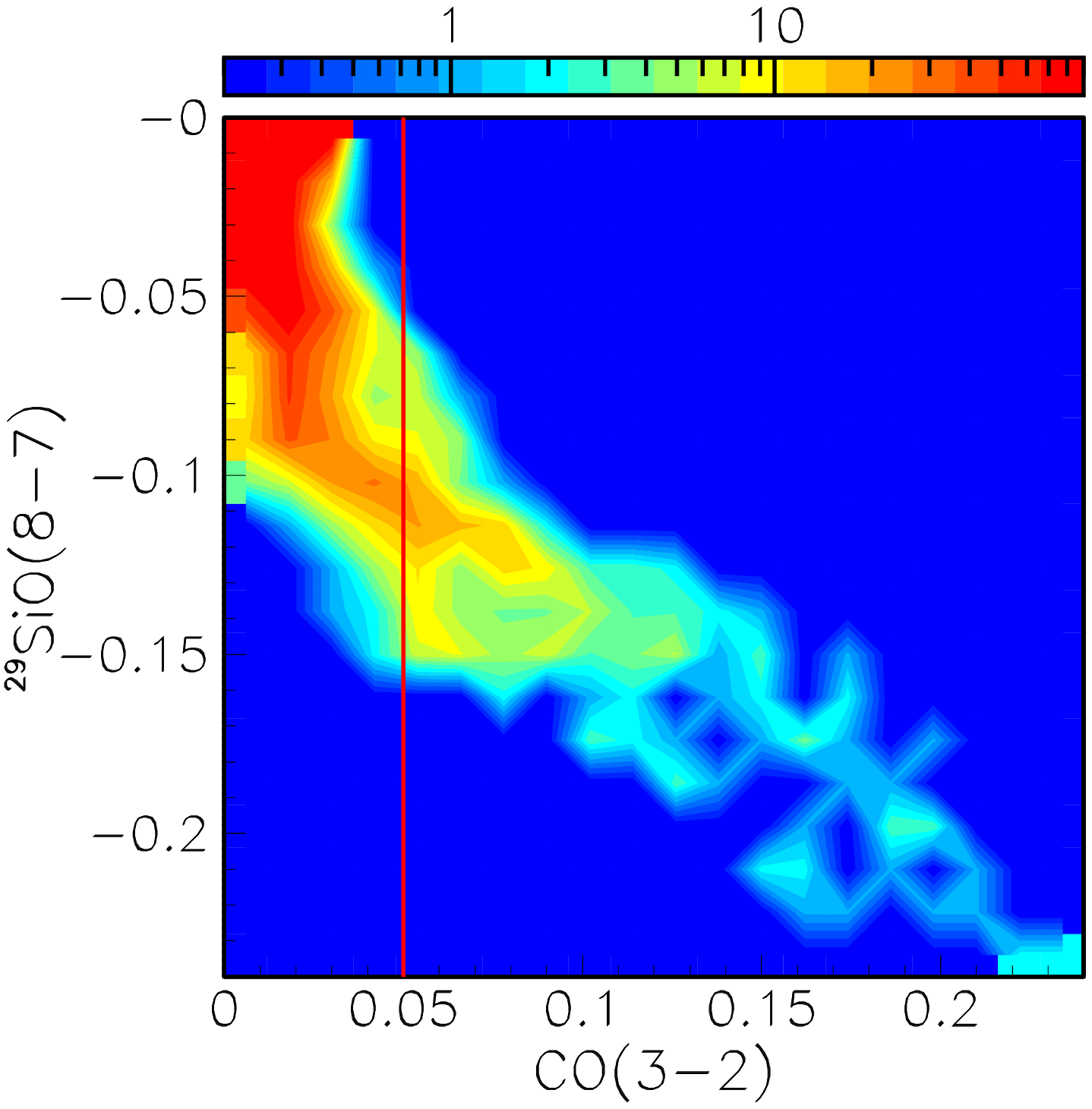}
  \caption{Left: W Hya CO intensity map of the peak integrated between $-$4 and $-$6 \kms\ with linearly interpolated background subtracted. Colour scale is Jy beam$^{-1}$ \kms. Centre: same as left but for SiO; the contour of the CO line at 50 mJy beam$^{-1}$ \kms\ is shown on both the left and central panels and defines region C. Right: correlation between the intensities mapped in the left and central panels. For each pixel, we plot the CO intensity in abscissa and the SiO intensity in ordinate. The colour scale is in number of pixels per bin. The red line shows the intensity associated with the CO contour displayed in the left and central panels.}
  \label{fig3}
\end{figure*}

       Between $\sim$$-$6 and $\sim$$-$4 \kms, CO emission and SiO emission (Figure \ref{fig1}) show an emission and an absorption peak, respectively. In order to obtain an approximate evaluation of what they correspond to, we evaluate their contribution by subtracting the linear interpolation over the peak range of the emission observed on the sides of the peak (Figure \ref{fig1}). The result is mapped in Figure \ref{fig3}, giving evidence for very similar morphologies, the CO intensity being positive and the SiO intensity negative. The right panel of Figure \ref{fig3} gives indeed evidence for a clear correlation between the peak emissions of the two lines. This is a surprising result in relation with the maser interpretation proposed by \citet{Vlemmings2021} for the emission of the CO line: it is not clear why the morphology observed for the CO maser emission should match the region of SiO absorption.

\begin{figure*}
  \centering
  \includegraphics[height=3.6cm,trim=.5cm 1.cm 0cm 1cm,clip]{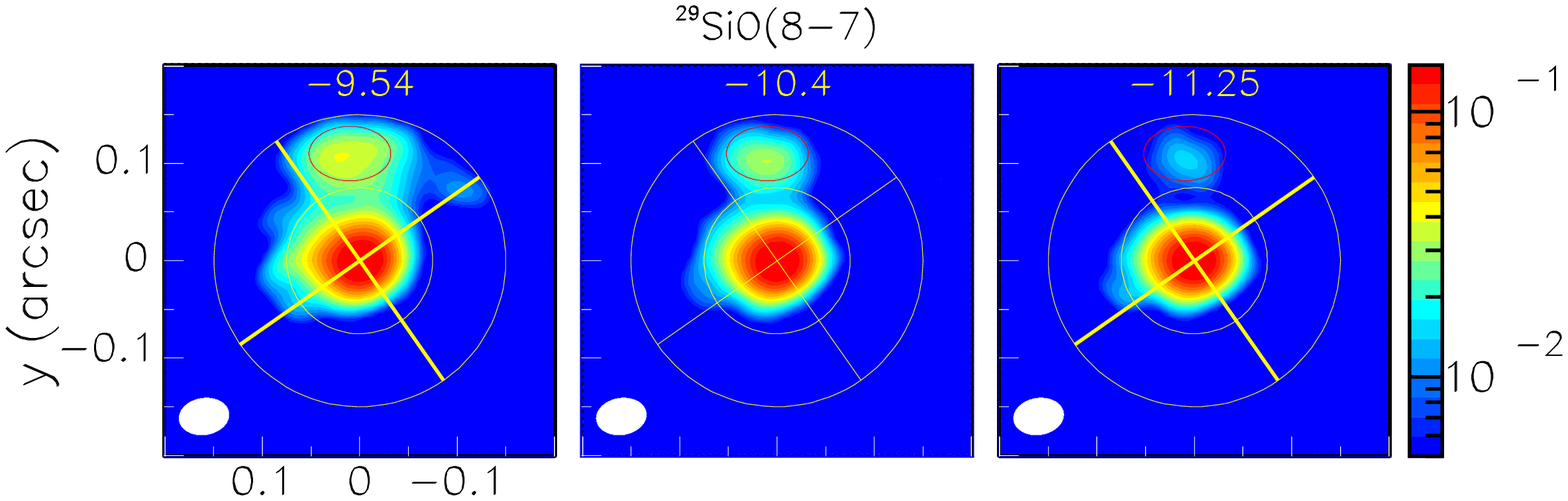}
  \includegraphics[height=4cm,trim=.5cm .0cm 0cm 1cm,clip]{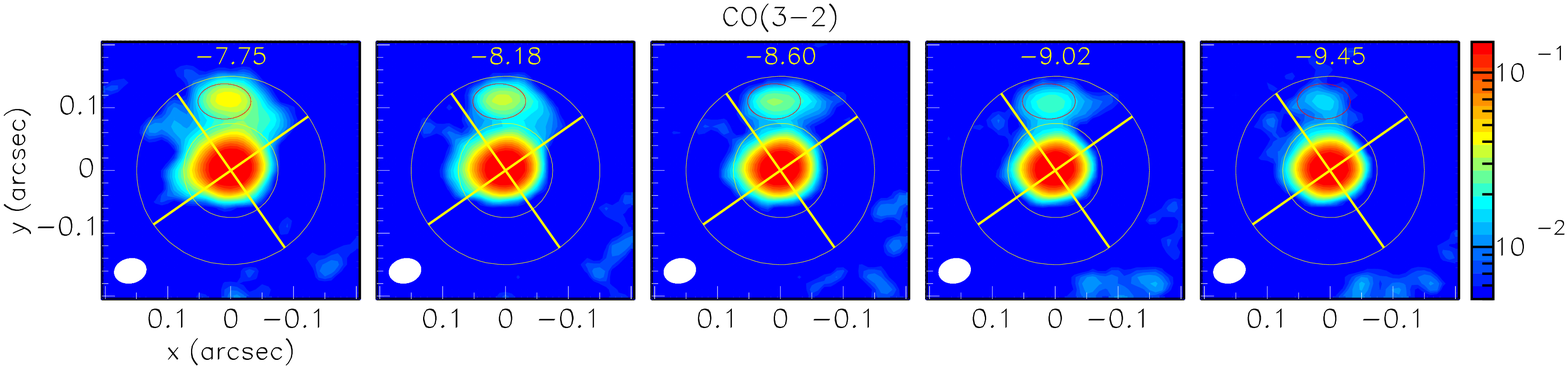}
  \caption{Blue blob emission: channel maps of the $^{29}$SiO(8-7) emission around $-$10.4 \kms\ and of the CO emission around $-$8.6 \kms. The colour scales are in Jy beam$^{-1}$.}
  \label{fig4}
\end{figure*}

The peak emission displayed in Figure \ref{fig3} shows clear anisotropy, with enhancement in a wedge covering approximately from 305\dego\ to 35\dego\ counter-clockwise in position angle. We refer to this region of enhanced emission as ``region C''; concretely, for convenience, we define it as the region where the intensity of the emission of the CO peak exceeds 50 mJy beam$^{-1}$ \kms.  It qualitatively matches the anisotropy of the dust distribution observed by \citet{Ohnaka2016, Ohnaka2017} at two epochs bracketing the observation of the molecular lines (July 8th 2015 and March 23rd 2016 compared with end of November 2015). Moreover, it also matches the observation of a blue-shifted blob of SiO emission by \citet{Hoai2021} at Doppler velocities around 10 \kms, for which these authors failed to find a sensible interpretation. As illustrated in Figure \ref{fig4}, which displays channel maps of both the CO and SiO lines in the range of relevant Doppler velocities, a similar blob is observed in the CO emission, however at less negative Doppler velocities.

\subsection{Overall view}
\begin{figure*}
  \centering
  \includegraphics[width=8.7cm,trim=.0cm .5cm 1.5cm 1.cm,clip]{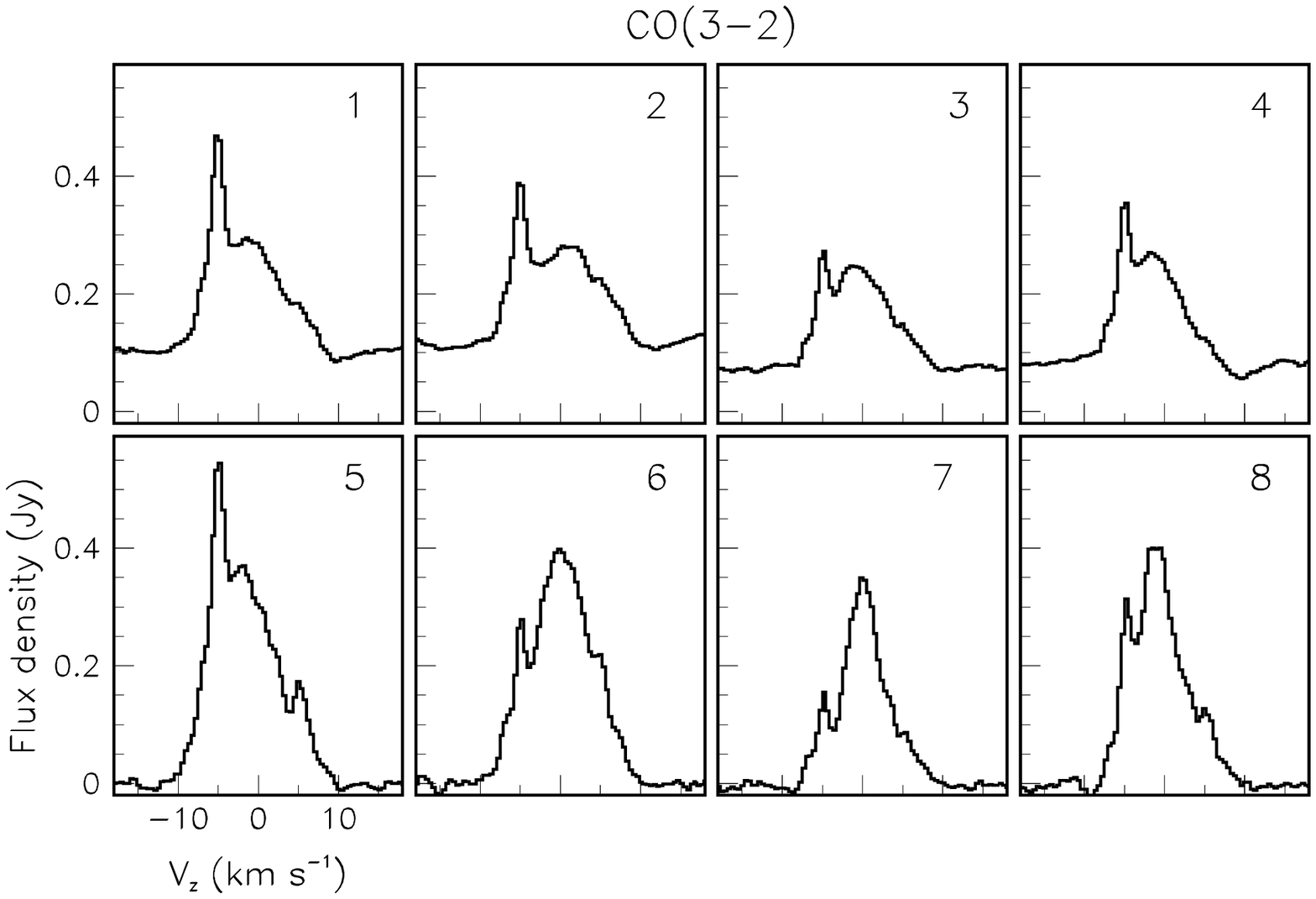}
  \includegraphics[width=8.7cm,trim=.0cm .5cm 1.5cm 1.cm,clip]{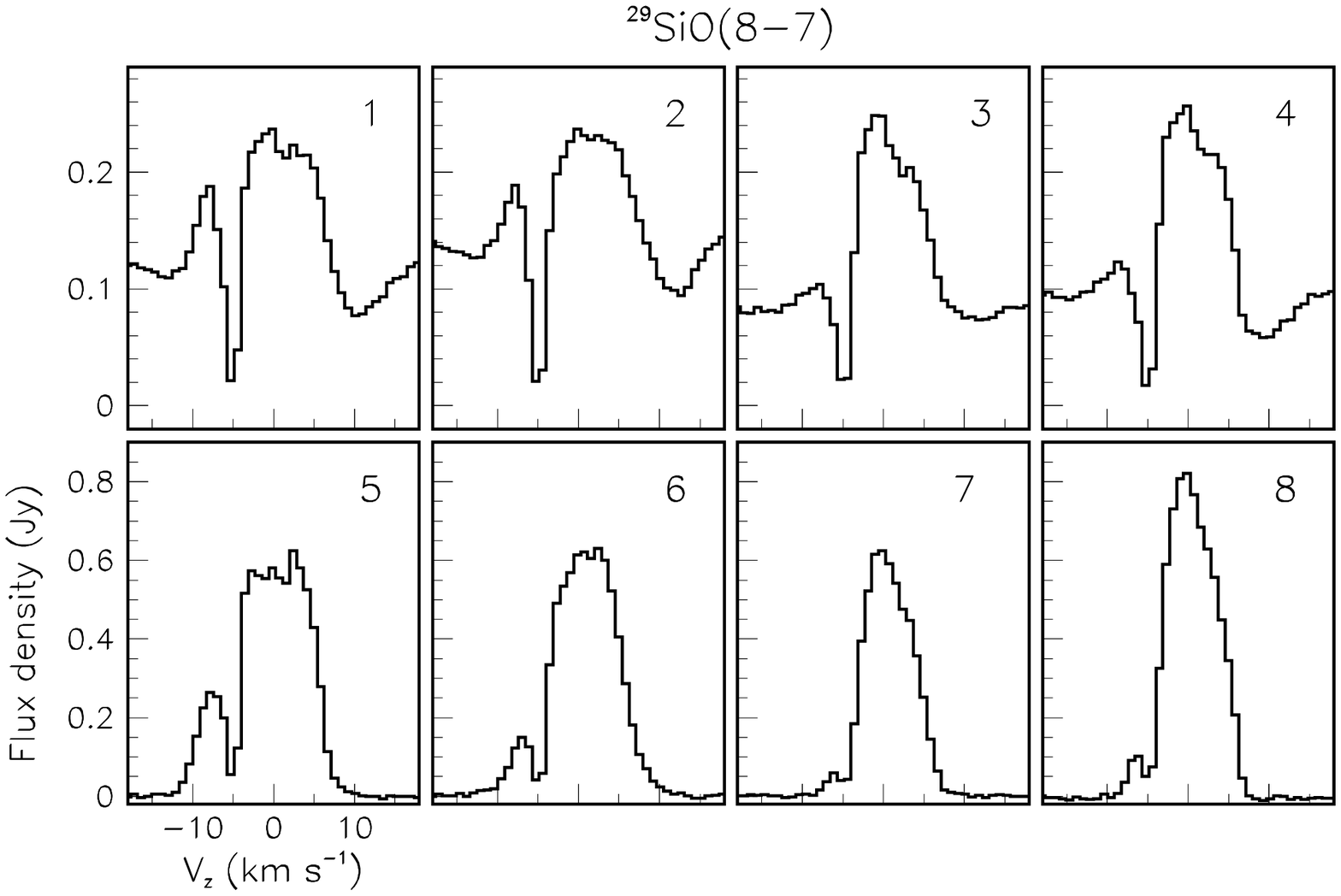}
  \caption{Doppler velocity spectra integrated in each of 8 quadrants for the $^{12}$CO(3-2) emission (left) and $^{29}$SiO(8-7) emission (right) of W Hya. The quadrants cover the projected radial ranges 0$<$$R$$<$75 mas (upper row of each pair) and 75$<$$R$$<$150 mas (lower row of each pair), respectively. In each row the quadrants are numbered in increasing order starting from north counter-clockwise, from 1 to 4 for the inner quadrants and from 5 to 8 for the outer quadrants. Quadrants 1 and 5 cover position angles $|\omega-350$\dego|$<$45\dego.}
  \label{fig5}
\end{figure*}

In order to shed light on these observations, we consider separately 8 regions covering the environment of the star within 150 mas (6 $R_*$) projected distance from its centre: four quadrants covering 0$<$$R$$<$75 mas and four others covering 75$<$$R$$<$150 mas. The quadrants, shown in Figure \ref{fig3}, are numbered in increasing order from north counter-clockwise, from 1 to 4 for the inner quadrants and from 5 to 8 for the outer quadrants. Quadrants 1 and 5 cover the wedge $|\omega-350$\dego$|<$45\dego. Figure \ref{fig5} displays, for each quadrant and for each of the CO and SiO lines, the observed Doppler velocity spectrum. Together with the inner quadrants, outer quadrant 5 covers region C where the CO peak emission is enhanced (Figure \ref{fig3}) and where the blue-shifted blobs are observed (Figure \ref{fig4}). The inner quadrants cover the range where dust is being formed and starts accelerating the gas \citep{Ohnaka2017, Takigawa2017} and where shocks anticorrelated with the dust clouds are seen to heat the extended atmosphere of the star \citep{Vlemmings2017}. Tables \ref{tab1} and \ref{tab2} list a number of parameters that describe the spectra displayed in Figure \ref{fig5}.    
They are described as the sum of three components: a flat continuum emission averaged over 15<|$V_z$|<26 \kms, a broad component and a peak, positive for CO and negative for SiO. To separate the contribution of the peak from that of the broad component we define a peak interval as 4<|$V_z$|<6 \kms\ over which we evaluate the contribution of the broad component by linear interpolation from the sides of the peak interval. The peak is then redefined as the difference between the observed distribution and this linearly interpolated distribution, the broad component being defined accordingly. Tables \ref{tab1} and \ref{tab2} list the continuum level and, for the broad component, its value at maximum, its FWHM and its central velocity; for the peak, Table \ref{tab1} lists the peak emission at maximum and Table \ref{tab2} the emission at maximal absorption. Uncertainties are typically $\pm$1 \kms\ on Doppler velocities and $\pm$10 mJy on flux densities.

\begin{deluxetable*}{cccccc}
\tablenum{1}
\tablecaption{Parameters describing the $^{12}$CO(3-2) spectra of quadrants 1 to 8 (see text).\label{tab1}}
\tablehead{\colhead{Quadrant} & \colhead{Continuum} & \colhead{Peak height} & \colhead{} & \colhead{Broad emission} & \colhead{}\\
 \colhead{} & \colhead{} & \colhead{} & \colhead{Maximum} & \colhead{FWHM} & \colhead{Central velocity}\\
\colhead{} & \colhead{(Jy)} & \colhead{(Jy)} & \colhead{(Jy)} & \colhead{(\kms)}&\colhead{(\kms)}}
\startdata
1&
0.11&
0.25&
0.18&
10&
$-$2\\
2&0.12&
0.19&
0.16&
11&
+1\\
3&
0.07&
0.11&
0.18&
10&
$-$1\\
4&
0.08&
0.18&
0.19&
11&
$-$2\\
5&
0&
0.28&
0.37&
10&
$-$2\\
6&
0&
0.12&
0.40&
9&
0\\
7&
0&
0.09&
0.35&
6&
0\\
8&
0&
0.15&
0.40&
6&
$-$2\\
\enddata
\end{deluxetable*}

\begin{deluxetable*}{cccccc}
\tablenum{2}
\tablecaption{Parameters describing the  $^{29}$SiO(8-7) spectra of quadrants 1 to 8 (see text).\label{tab2}}
\tablehead{\colhead{Quadrant} & \colhead{Continuum} & \colhead{Flux at maximal} & \colhead{} & \colhead{Broad emission} & \colhead{}\\
 \colhead{} & \colhead{} & \colhead{ absorption} & \colhead{Maximum} & \colhead{FWHM} & \colhead{Central velocity}\\
 \colhead{} & \colhead{(Jy)} & \colhead{(mJy)} & \colhead{(Jy)} & \colhead{(\kms)}&\colhead{(\kms)}}
\startdata
1&
0.12&
25&
0.24&
14&
$-$1\\
2&
0.14&
25&
0.23&
13&
0\\
3&
0.08&
25&
0.24&
12&
0\\
4&
0.10&
25&
0.25&
12&
$-$1\\
5&
0&
50&
0.58&
13&
0\\
6&
0&
50&
0.62&
11&
1\\
7&
0&
50&
0.62&
8&
0\\
8&
0&
50&
0.82&
9&
0\\
\enddata
\end{deluxetable*}

\begin{figure*}
  \centering
  \includegraphics[height=3.8cm,trim=.5cm 2.3cm 2.3cm 1.9cm,clip]{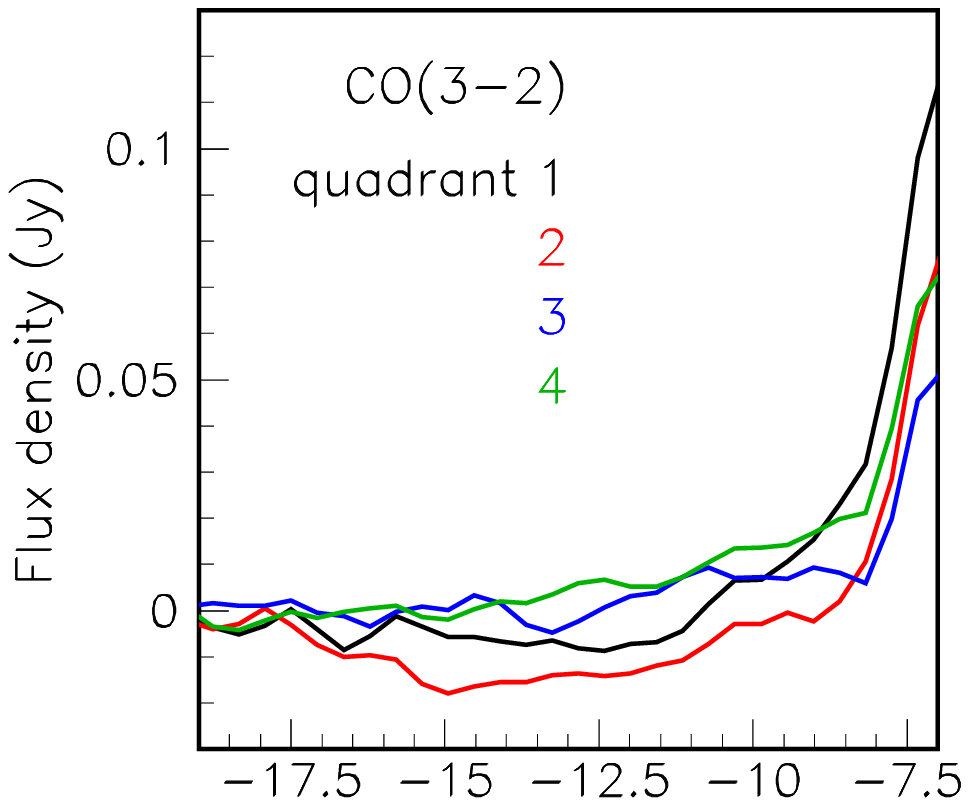}
  \includegraphics[height=3.8cm,trim=1.5cm 2.3cm 2.3cm 1.9cm,clip]{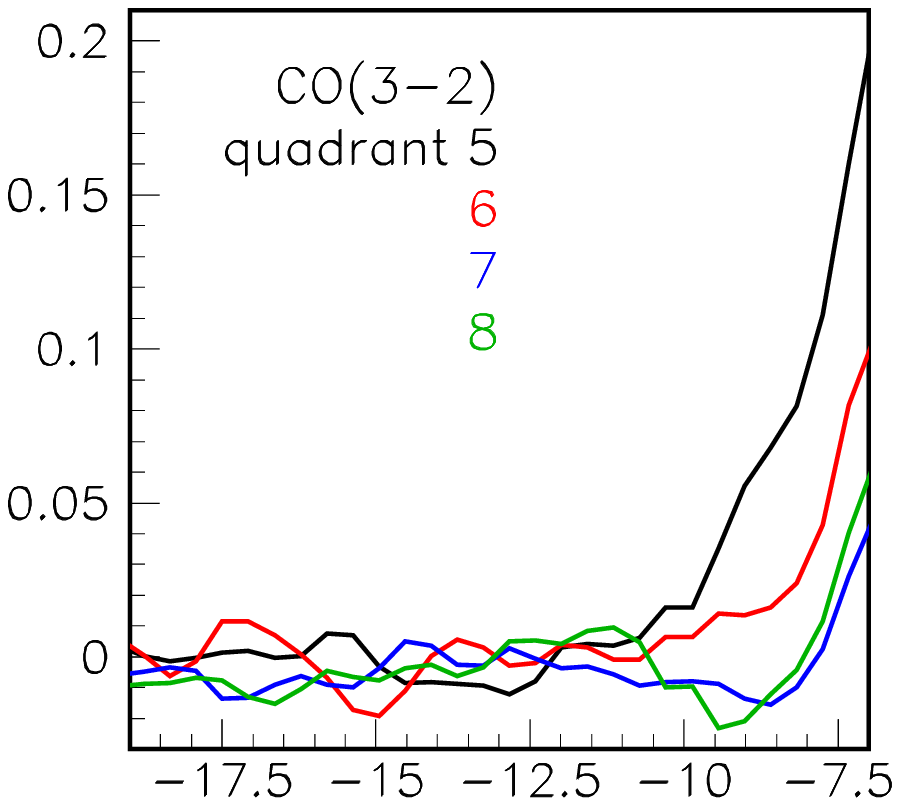}
  \includegraphics[height=3.8cm,trim=.5cm 2.3cm 2.3cm 1.9cm,clip]{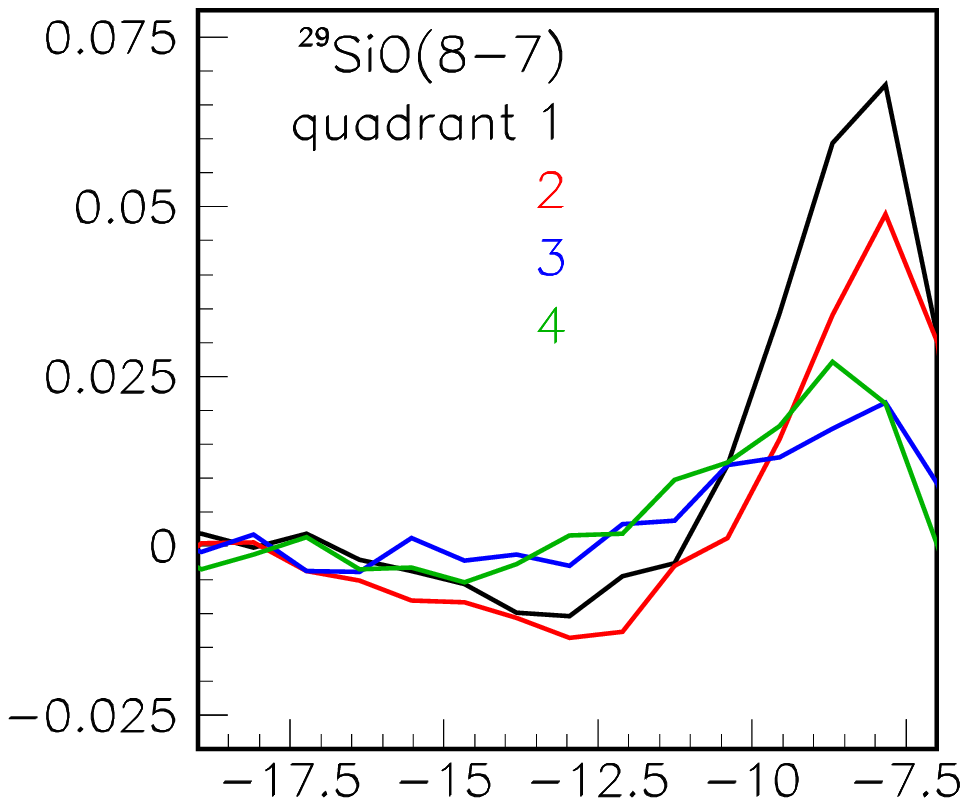}
  \includegraphics[height=3.8cm,trim=1.5cm 2.3cm 2.3cm 1.9cm,clip]{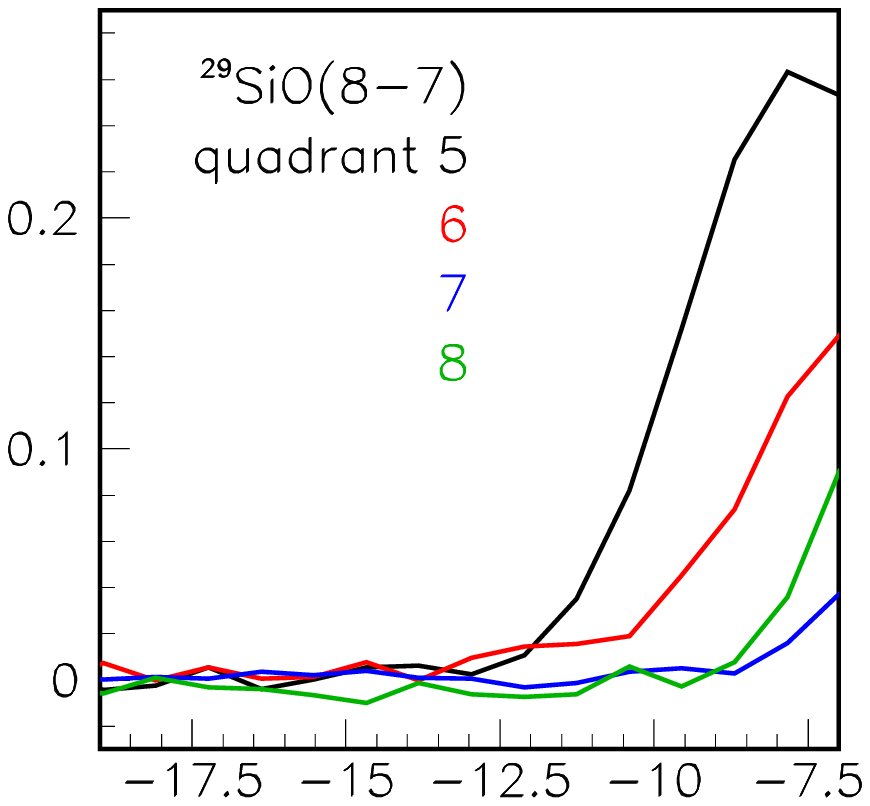}\\
  \includegraphics[height=4.05cm,trim=.5cm 1.7cm 2.3cm 1.9cm,clip]{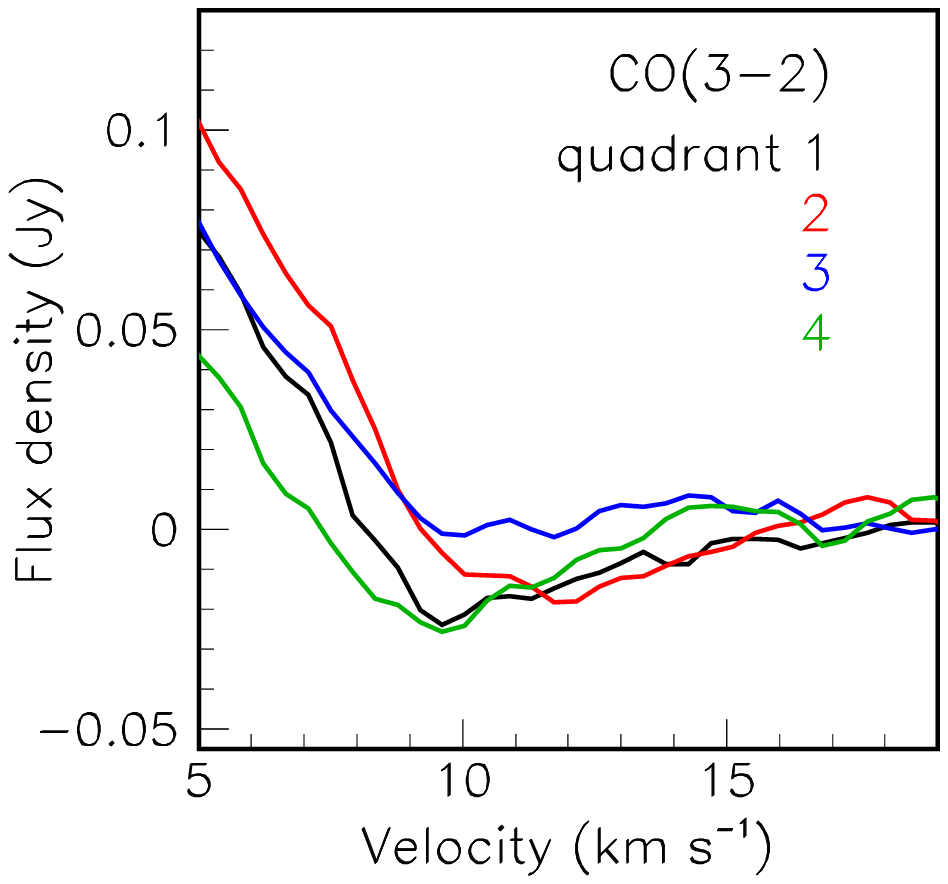}
  \includegraphics[height=4.05cm,trim=1.5cm 1.7cm 2.3cm 1.9cm,clip]{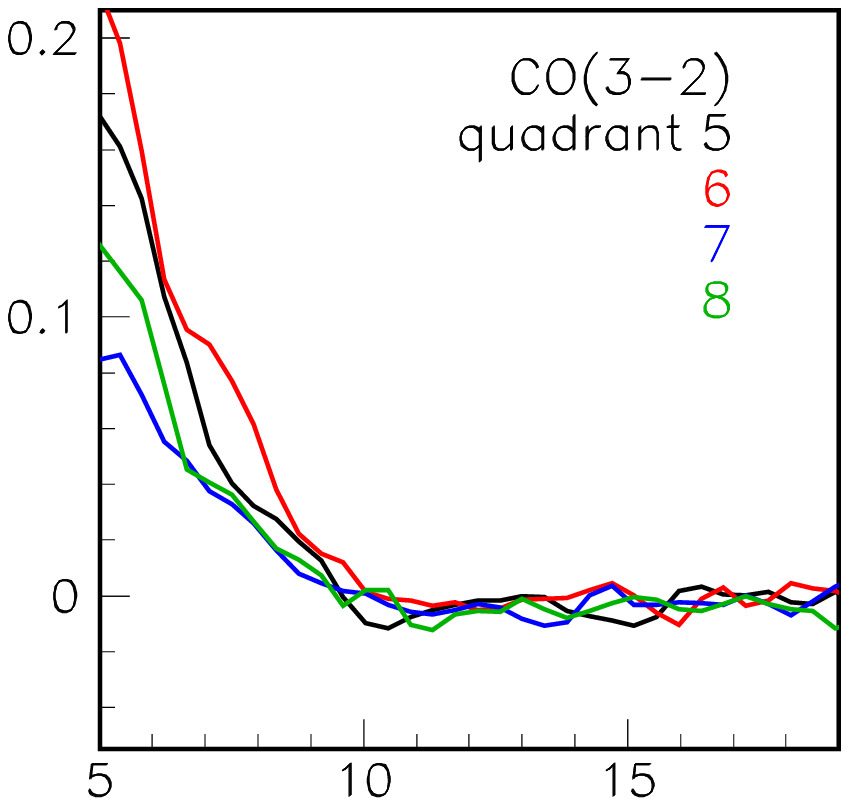}
  \includegraphics[height=4.05cm,trim=.5cm 1.7cm 2.3cm 1.9cm,clip]{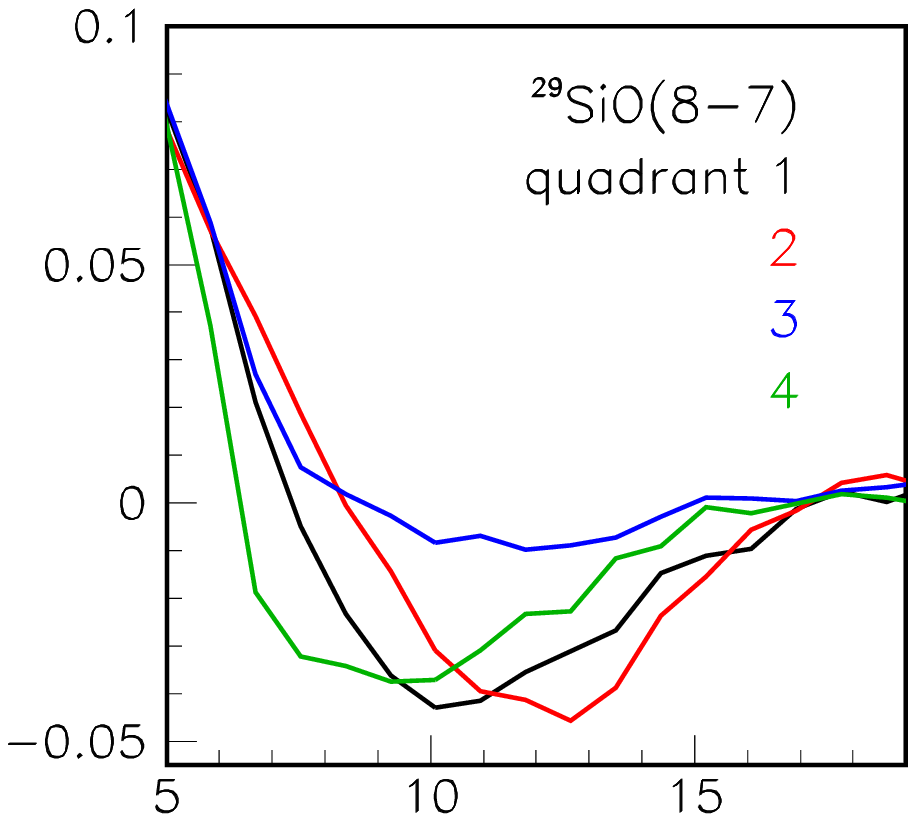}
  \includegraphics[height=4.05cm,trim=1.5cm 1.7cm 2.3cm 1.9cm,clip]{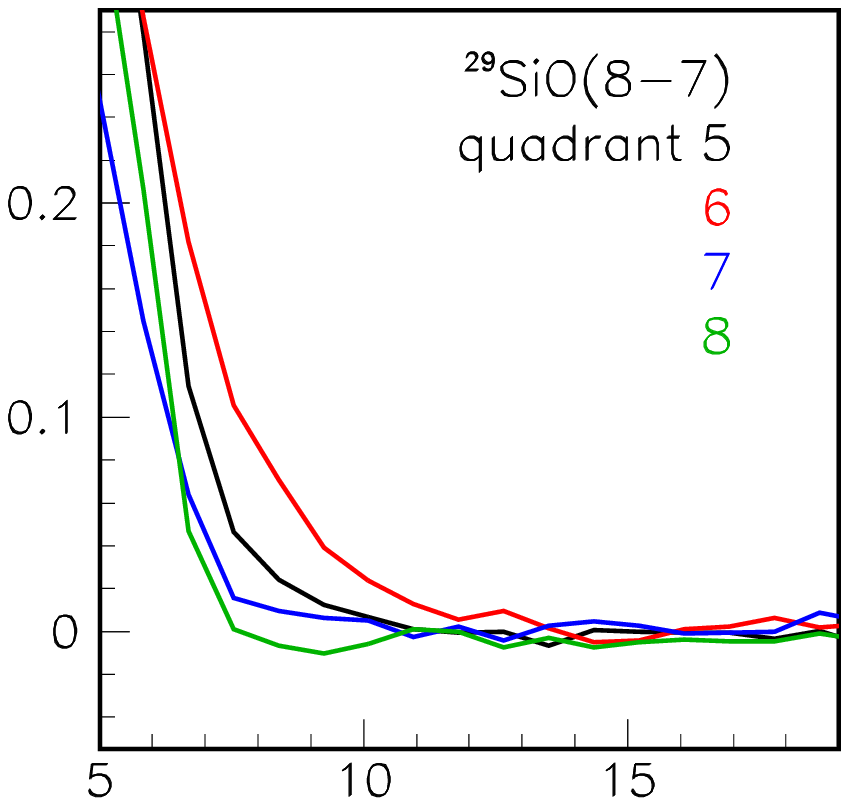}
  
  \caption{Continuum-subtracted blue-shifted parts (upper panels) and red-shifted parts (lower panels) of the spectra (Jy) displayed in Figure \ref{fig5}. }
  \label{fig6}
\end{figure*}

\begin{figure*}
  \centering
  \includegraphics[height=5cm,trim=.2cm 1.cm 1.5cm 1.9cm,clip]{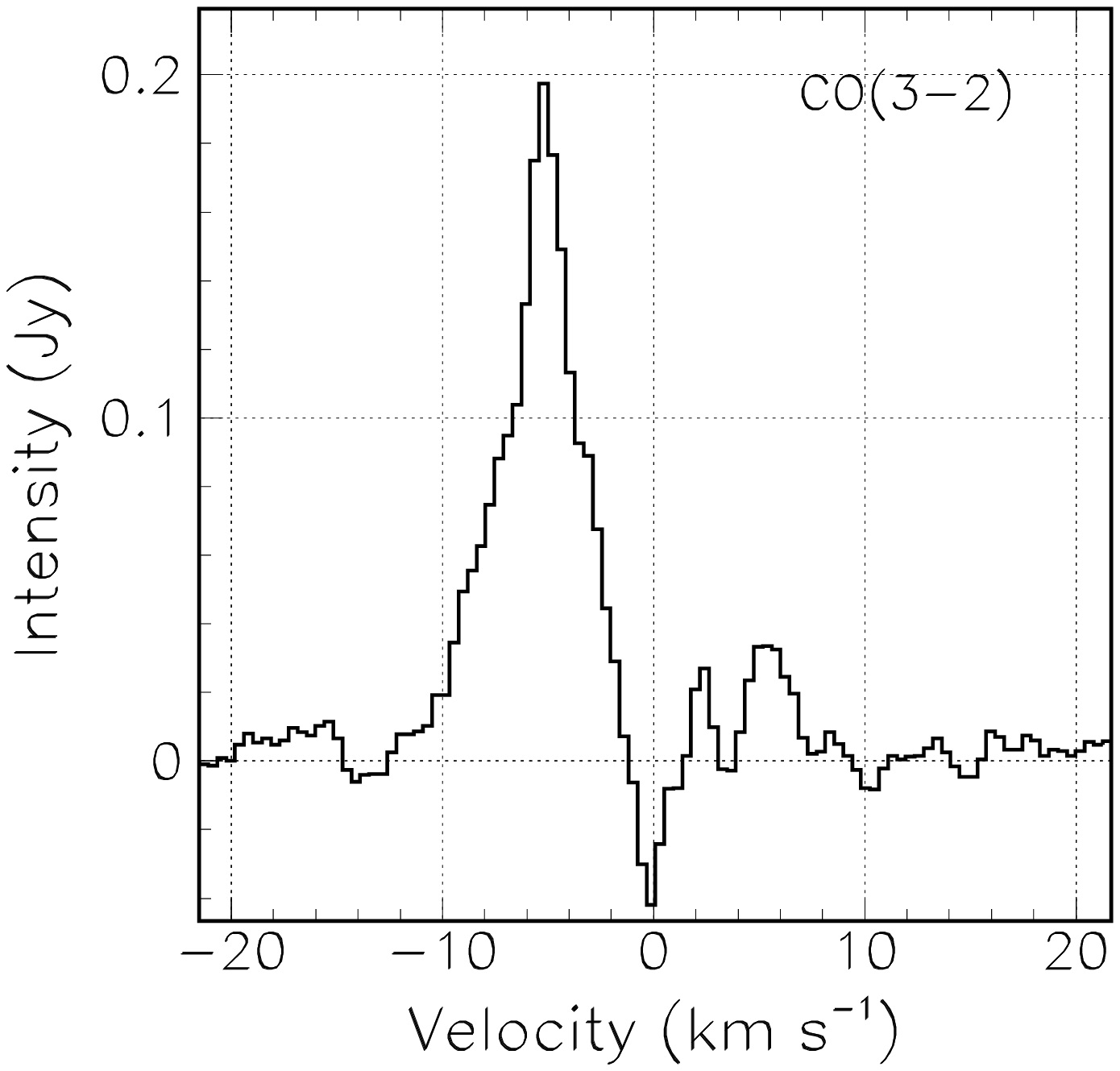}
  \includegraphics[height=5cm,trim=.2cm 1.cm 1.5cm 1.9cm,clip]{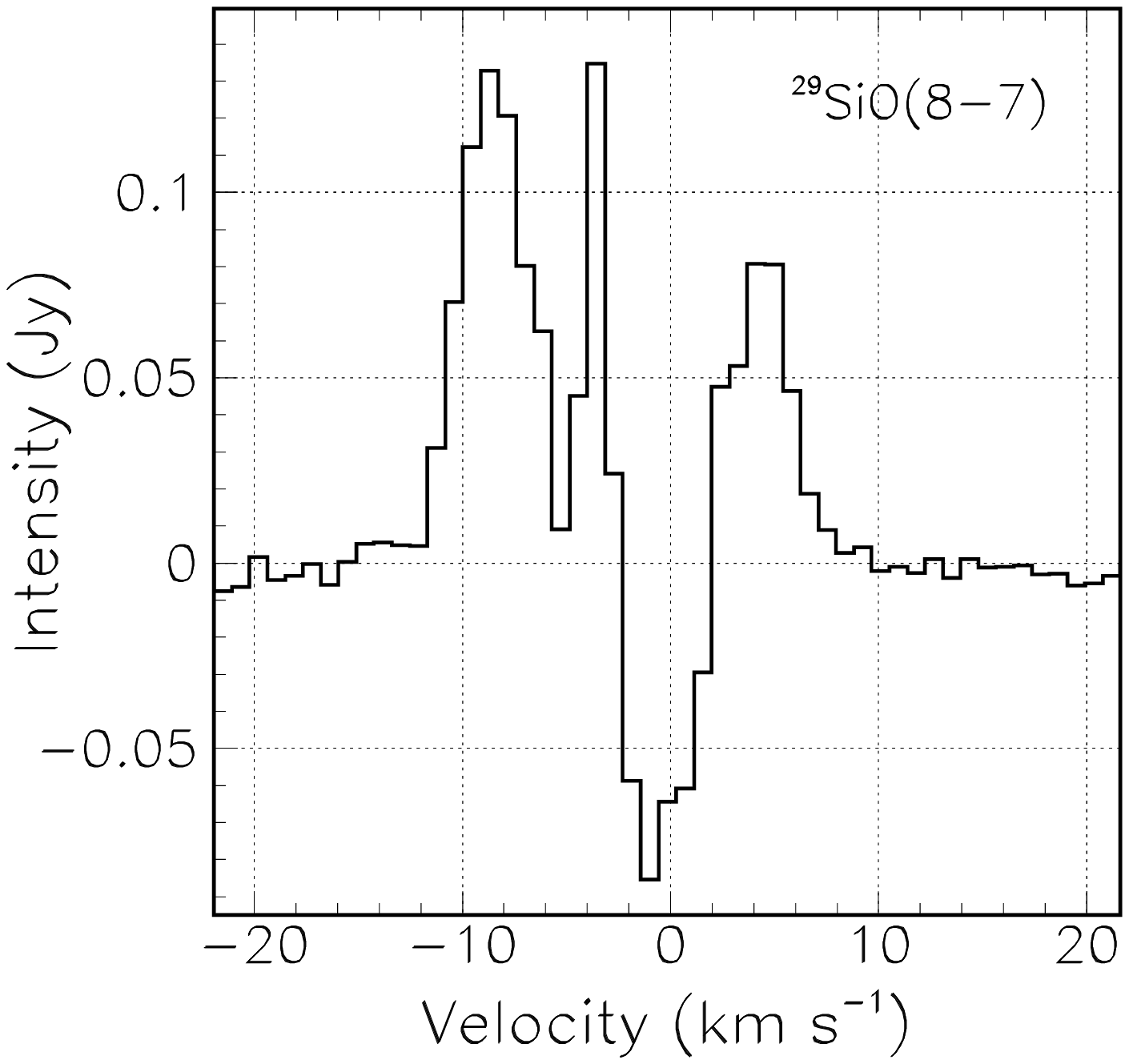}
  \caption{Difference between the Doppler velocity spectrum integrated over the blob region (75$<$$R$$<$150 mas, $|\omega|$$<$30\dego) and that integrated over the rest of the 75$<$$R$$<$150 mas ring and divided by 5 to account for the broader $\omega$ interval. Left panel is for the CO line and right panel for the SiO line.}
  \label{fig7}
\end{figure*}

The same exercise using sextants rather than quadrants is reported in the Appendix (Figure \ref{figa2}, Table \ref{taba2} and \ref{taba3}) where a parameterized description of the spectra, in the form of a flat continuum and a Gaussian for each of the peak and broad component, is presented. We used it to cross-check the results obtained from the quadrant analysis. These results call for a number of comments.

a) Continuum is seen to contribute to the inner quadrants exclusively and the levels observed in the SiO and CO bands agree to a precision of $\pm$7 mJy \kms\ arcsec$^{-2}$. They display anisotropy comparable with the dust distribution observed by \citet{Ohnaka2017}, namely a depression in the south-western direction: the ratio between the average emission in sextants 4 and 5 and that in the other four sextants (see Appendix) is 58\% and the ratio between the average emission in quadrants 3 and 4 and that in quadrants 1 and 2 is 67\%. This seems to contradict the detailed analysis of \citet{Vlemmings2017} which gives evidence for a south-western hot spot dominating the asymmetry of the continuum emission. We discuss it further in Section 5.

b) The emission peaks of the CO spectra are at $-$5.1$\pm$0.2 \kms\ and the absorption peaks of the SiO spectra at $-$4.7$\pm$0.2 \kms. The difference is 0.4$\pm$0.2 \kms, barely significant and possibly associated with an extra acceleration in the region probed by the CO line emission compared with that probed by the SiO line emission. The  rms deviation from the peak mean velocity is 1.0$\pm$0.4 \kms\ for CO and 0.8$\pm$0.2 \kms\ for SiO. The red-shifted peak of CO emission is precisely at $+$5 \kms, supporting the shell picture.

c) The amplitude of the CO emission peak is 3 times larger in quadrants 1 and 5 than in the opposite quadrants, 3 and 7. The quadrants in-between have intermediate peak heights: at the level of 3/4 of quadrant 1 in the inner circle and at the level of 1/2 of quadrant 5 in the outer ring. Accounting for the factor 3 difference in area between the inner circle and the outer ring, the mean brightness of the peak is 4.5 times larger in the inner quadrants than in outer quadrants 6 to 8. These asymmetries quantify the anisotropy illustrated in the left panel of Figure \ref{fig3}. They suggest an enhancement of density and/or an anticipated reach of the terminal velocity in region C. We discuss this further in Section 4.

d) The minima of the absorption peaks of the SiO spectra are at the approximate level of $\sim$25 mJy in the inner circle and $\sim$50 mJy in the outer ring, independently from the position angle. They reveal the important optical thickness and provide information on temperature that is further discussed in Section 4.

e) The emission above continuum and outside of the terminal velocity interval, between $\sim$$-$6 and $\sim$$-$4 \kms, is centred at $\sim$$-$1 \kms\ for the CO line and at $\sim$0 for the SiO line, with a precision (rms) of $\pm$1 \kms. In quadrants 1 to 6, it has a FWHM of 10$\pm$1 \kms\ for the CO line and of 12.5$\pm$1.0 \kms\ for the SiO line. In quadrants 7 and 8, the FWHM is only 2/3 of these values. The flux density at maximum displays little dependence on position angle. In the inner circle, it is $\sim$0.18 and $\sim$0.24 Jy for the CO and SiO lines, respectively and in the outer ring, $\sim$0.38 and $\sim$0.66 Jy. In terms of brightness, on average, the outer ring brightness is $\sim$83\% of that of the inner circle. Altogether, this emission displays less anisotropy and less inhomogeneity than in the terminal velocity interval, the main deviations from uniformity being a broader spectrum for SiO emission than for CO emission and for the C region than opposite to it. Broader SiO than CO spectra are commonly observed in other oxygen-rich AGB stars, the difference being caused by shorter distances to the star (where the line-width is larger) being probed by SiO emission than by CO emission.

f) The high $V_z$ wings of the CO and SiO emissions are illustrated in Figure \ref{fig6}. They show an important excess of flux density on the blue-shifted side of quadrant 5. The contribution of the blue blob in this region is shown in Figure \ref{fig7}, which displays the difference between the Doppler velocity spectrum integrated over the blob region (75$<$$R$$<$150 mas, $|\omega|$$<$30\dego) and that integrated over the rest of the 75$<$$R$$<$150 arcsec ring and divided by 5 to account for the broader $\omega$ interval. The blob is seen as a blue-shifted source of intense emission centred on the terminal velocity and covering some $\sim$10 \kms. The CO emission shows a narrow peak at terminal velocity over a broader distribution, each reaching $\sim$0.1 Jy at maximum. The SiO emission is more difficult to interpret. The persistence of an absorption peak at terminal velocity implies that the SiO line is much more opaque over the blob than on the rest of the ring. The structure seen at Doppler velocities exceeding $\sim$$-$3 \kms\ results from the different shapes displayed by the spectra (Figure \ref{fig5}).  .

\section{Temperature dependence and radiative transfer modelling of the observed line emissions}

\subsection{Formalism}

In the LTE approximation, for a given velocity $V_z$ and a brightness $I_0$ [Jy arcsec$^{-2}$] on the entrance side of a thin slab of column density $N$ [molecules cm$^{-3}$ arcsec] and temperature $T$ [K], the brightness $I$ [Jy arcsec$^{-2}$] at the exit side, reads
\begin{equation}
  I = I_0e^{-\tau}+S(1-e^{-\tau})=S+(I_0-S)e^{-\tau}
\end{equation}
where $S$ is the source function defined as $S$=$\varepsilon/\tau$ with $\varepsilon$ [Jy arcsec$^{-2}$] describing the unabsorbed emission of the slab and  $\tau$ being its optical thickness. In the LTE approximation, $S$ is the Planck function: $S$=$\varepsilon/\tau$=$(2hf^3)/(c^2)$(e$^{\Delta{E}/T}-1)^{-1}$=$f$[GHz]$^3$/$(2.9\times10^7)$(e$^{\Delta{E}/T}-1)^{-1}$ Jy arcsec$^{-2}$. The frequency $f$ and the energy $\Delta{E}$ of the transition take similar values for the CO and SiO lines: 345.80 and 342.98 GHz, and 16.7 and 16.6 K, respectively. When the column density increases, $I$ varies from $I_0$ to $\varepsilon/\tau$. In case of absorption, $I_0$$>$$\varepsilon/\tau$, $I$ is always larger than $\varepsilon/\tau$, a lower limit to the emission of an optically thick shell: however large absorption may be, the outer part of the layer cannot be prevented from shining. When the energy of the transition is much smaller than the temperature $T$, one can approximate
  
  $\varepsilon/\tau$ $\sim$ $f^3T/(2.9 10^7\Delta{E})$ $\sim$ $T$[K]/11.8.
  
  Figure \ref{fig2} shows that the minimal brightness observed in a 30$\times$30 mas$^2$ region is $\sim$0.5 mJy au$^{-2}$=5.5 Jy arcsec$^{-2}$, which must therefore exceed $T$[K]/11.8, implying $T$$<$65 K (75 K using the exact relation). The same result is obtained with the minimal flux density of $\sim$25 mJy observed in each of the 4.4$\times$10$^{-3}$ arcsec$^2$ inner quadrants. This result is independent from the value of $\varepsilon$, namely of the column density. At $r$=1 arcsec, this temperature is only one third of that obtained from the law used by \citet{Khouri2014a} and \citet{Takigawa2017} close to the star: $T$[K]=2500$\times$(23/$r$[mas])$^{0.65}$, suggesting a steeper radial decrease of the temperature than commonly assumed in the  published literature.
  
\begin{figure*}
  \centering
  \includegraphics[height=4.2cm,trim=.5cm 1.cm 1.9cm 1.9cm,clip]{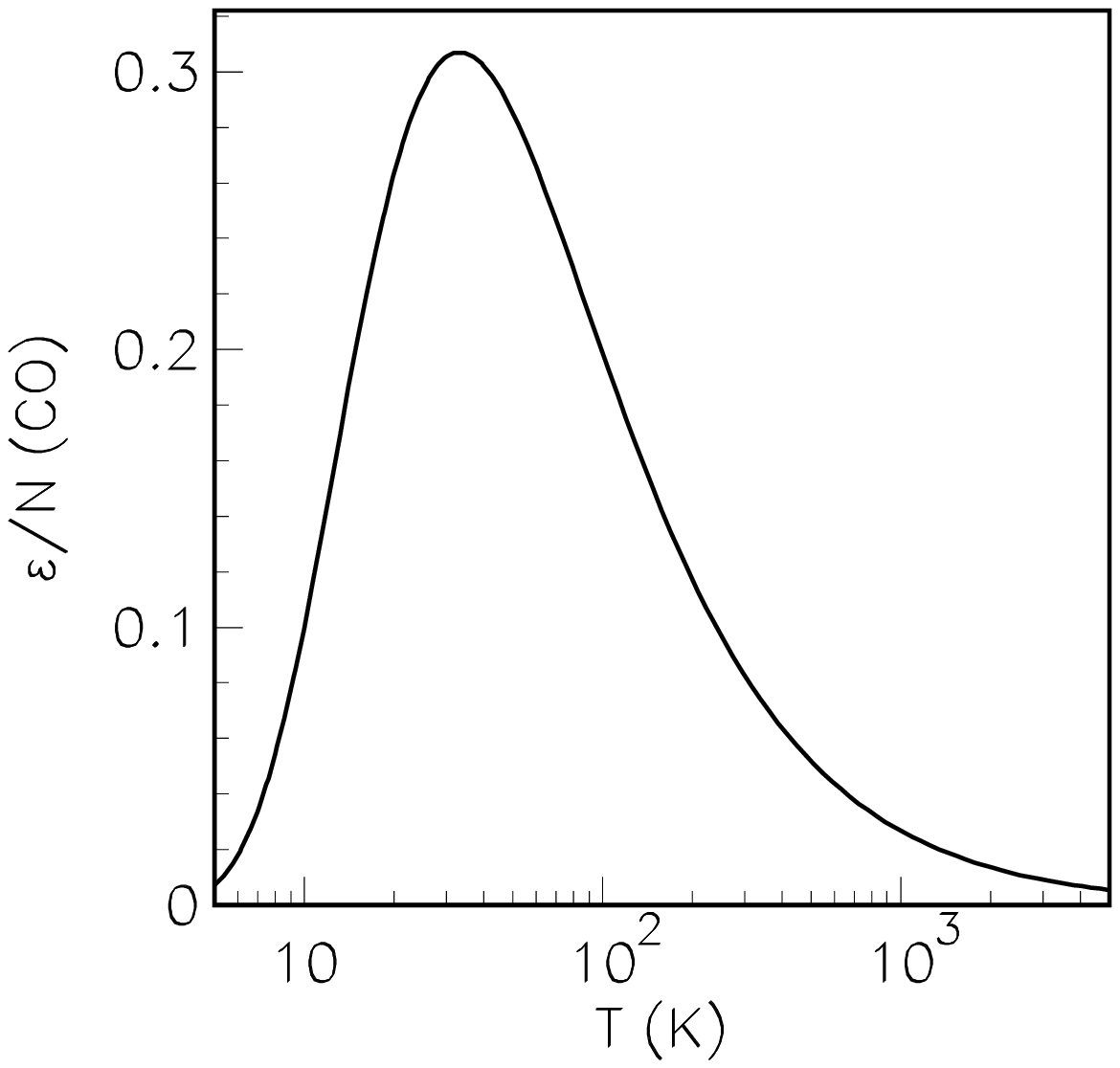}
  \includegraphics[height=4.2cm,trim=.5cm 1.cm 1.9cm 1.9cm,clip]{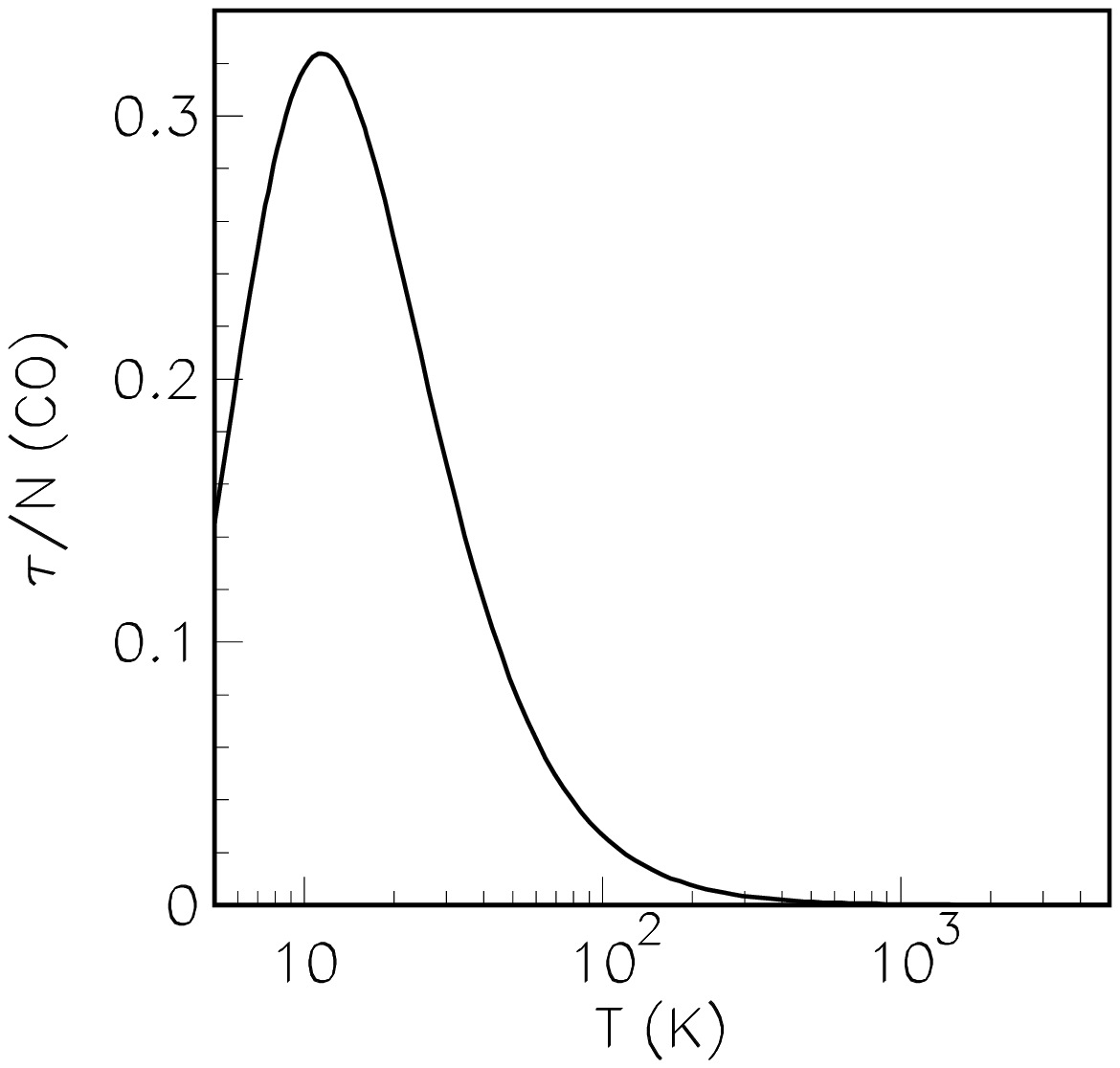}
  \includegraphics[height=4.2cm,trim=.5cm 1.cm 1.9cm 1.9cm,clip]{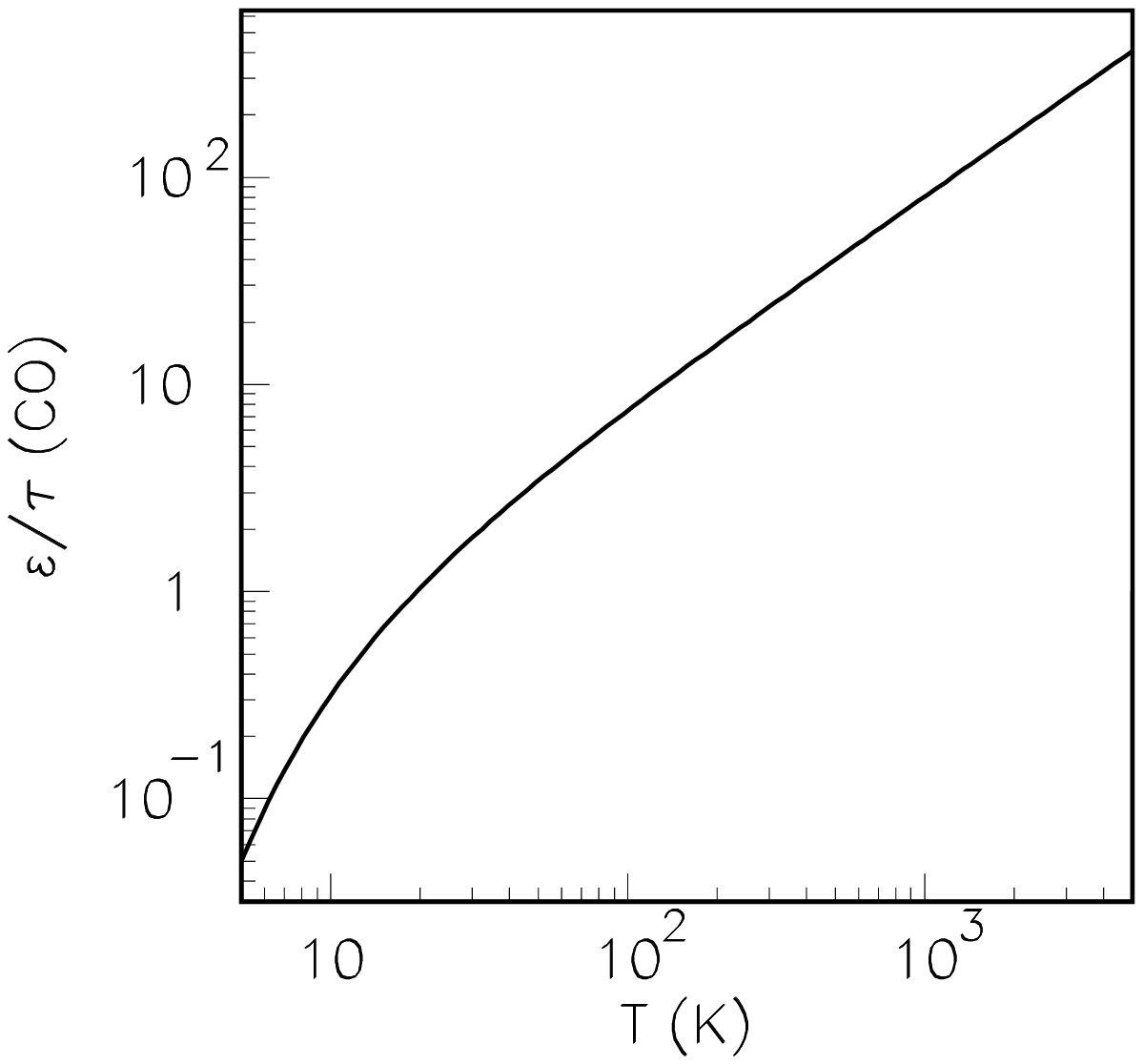}\\
  \includegraphics[height=4.2cm,trim=.5cm 1.cm 1.9cm 1.9cm,clip]{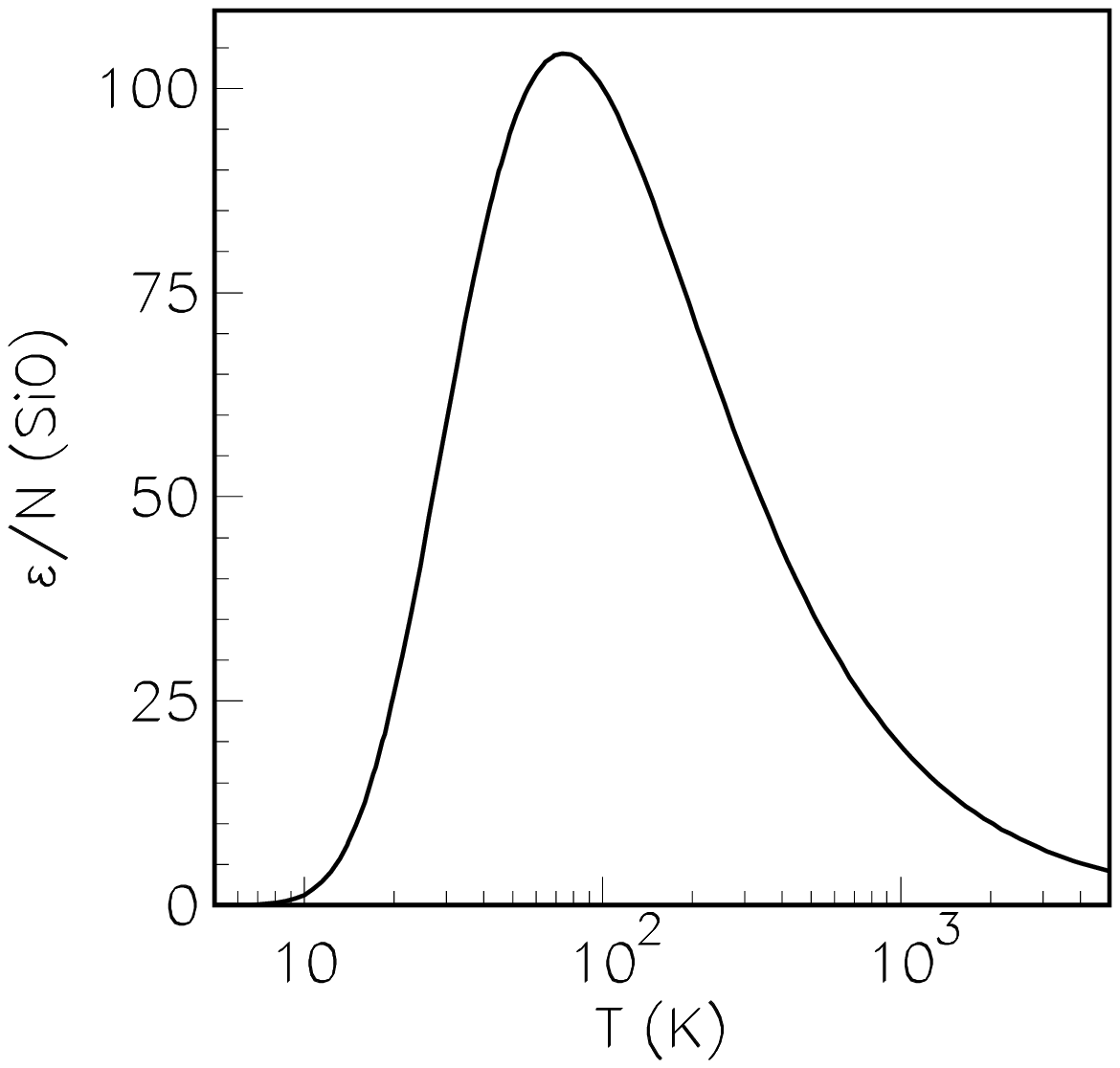}
  \includegraphics[height=4.2cm,trim=.5cm 1.cm 1.9cm 1.9cm,clip]{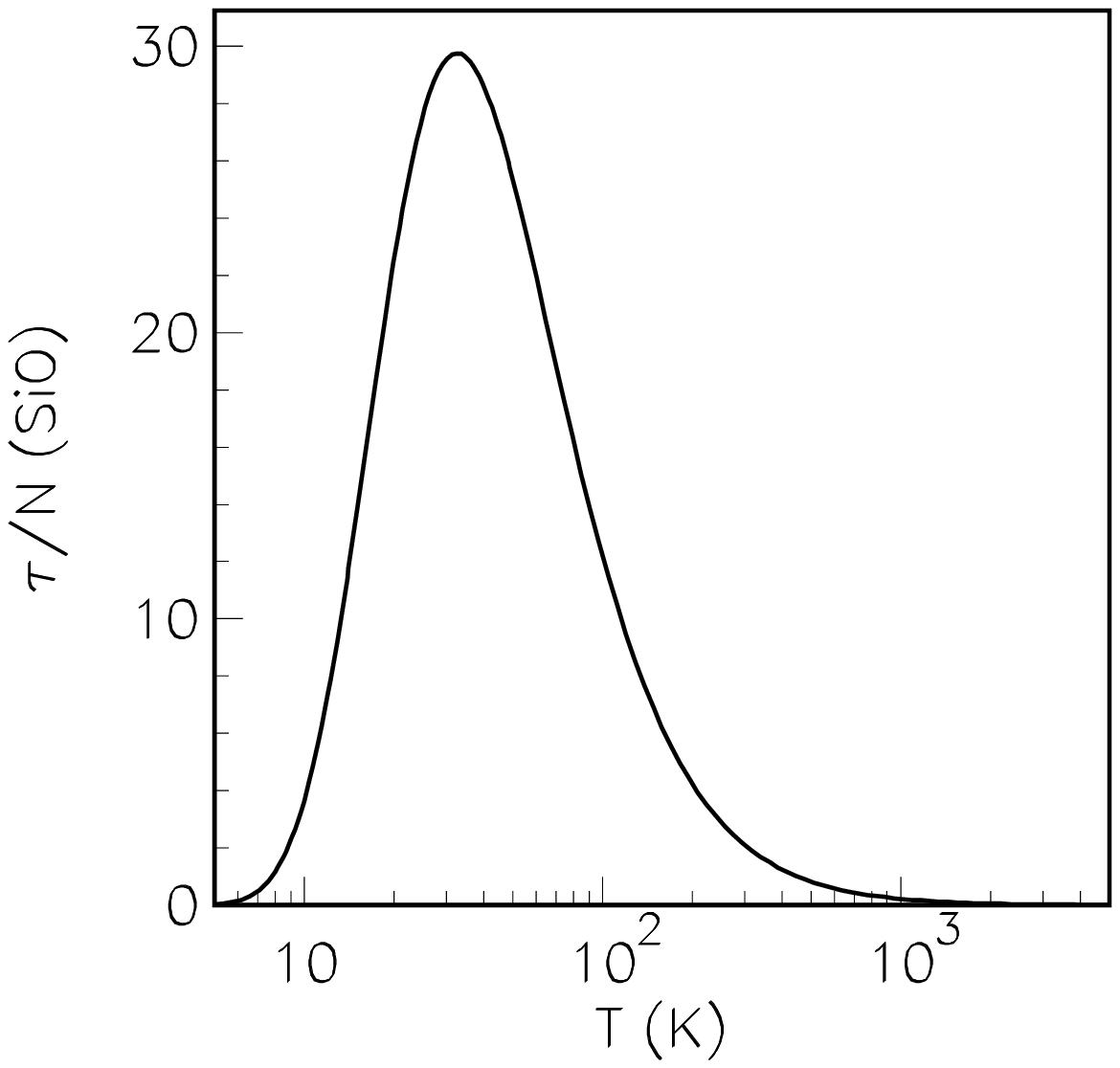}
  \includegraphics[height=4.2cm,trim=.5cm 1.cm 1.9cm 1.9cm,clip]{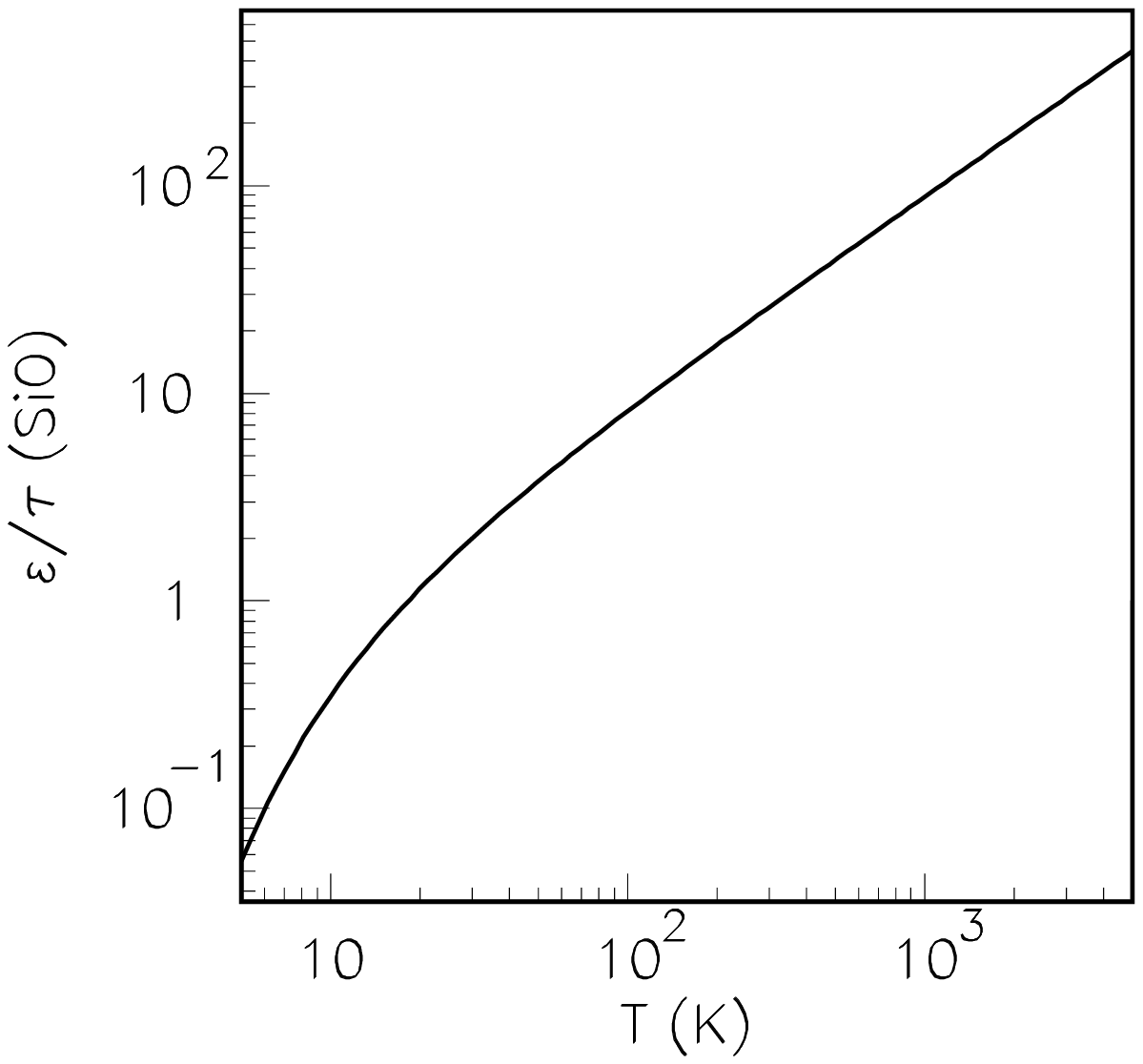}
  \caption{Dependence on temperature of $\varepsilon/N$ (left), $\tau/N$ (centre) and $\varepsilon/\tau$ (right) for the $^{12}$CO(3-2) (upper panels) and $^{29}$SiO(8-7) (lower panels), respectively. Units are Jy arcsec$^{-2}$ for $\varepsilon$ and mol cm$^{-3}$ arcsec for $N$. }
  \label{fig8}
\end{figure*}

  The emission parameter $\varepsilon$ [Jy arcsec$^{-2}$] is obtained from Relation 2 below:
  \begin{eqnarray}
    \varepsilon= & hf/(4\pi\Delta{f})(ndz) A_{\text{ji}} f_{\text{pop}} = hc/(4\pi\Delta{V})(ndz) A_{\text{ji}} f_{\text{pop}}  \\
               = & 5.5\,10^3\,\Delta{V}^{-1}\, d\,N\,A_{\text{ji}}\,f_{\text{pop}},    f_{\text{pop}}=k(2J+1)e^{-E_{\text{u}}/T}/T \nonumber \\ \nonumber
  \end{eqnarray}
  where $d$[pc] is the distance from the Sun to the star, $J$ and $E_u$[K] the spin and energy of the upper level, $A_{\text{ji}}$[s$^{-1}$] the Einstein coefficient, $k$[K] the ratio between the temperature and the partition function, $N$[molecules cm$^{-3}$ arcsec] the column density, $\Delta{f}$ and $\Delta{V}$ the line widths in frequency and velocity respectively and $T$[K] the temperature. The parameters for the CO and SiO lines are given in Table \ref{tab3}. The values of $\varepsilon_0$ and $\tau_0$ for $\Delta{V}$=1 \kms\ listed in the last columns are defined from the expressions $\varepsilon$=$\varepsilon_0Ne^{E_u/T}/T$ and $\tau$=$\tau_0Ne^{-E_u/T}/T(e^{\Delta{E}/T}-1)$. The dependence on temperature of $\varepsilon/N$ and $\tau/N$ is illustrated in Figure \ref{fig8}. 

  \subsection{Comparing W Hya and R Dor outside region C}
  
\begin{figure*}
  \centering
  \includegraphics[height=7cm,trim=.0cm .0cm 1.7cm .9cm,clip]{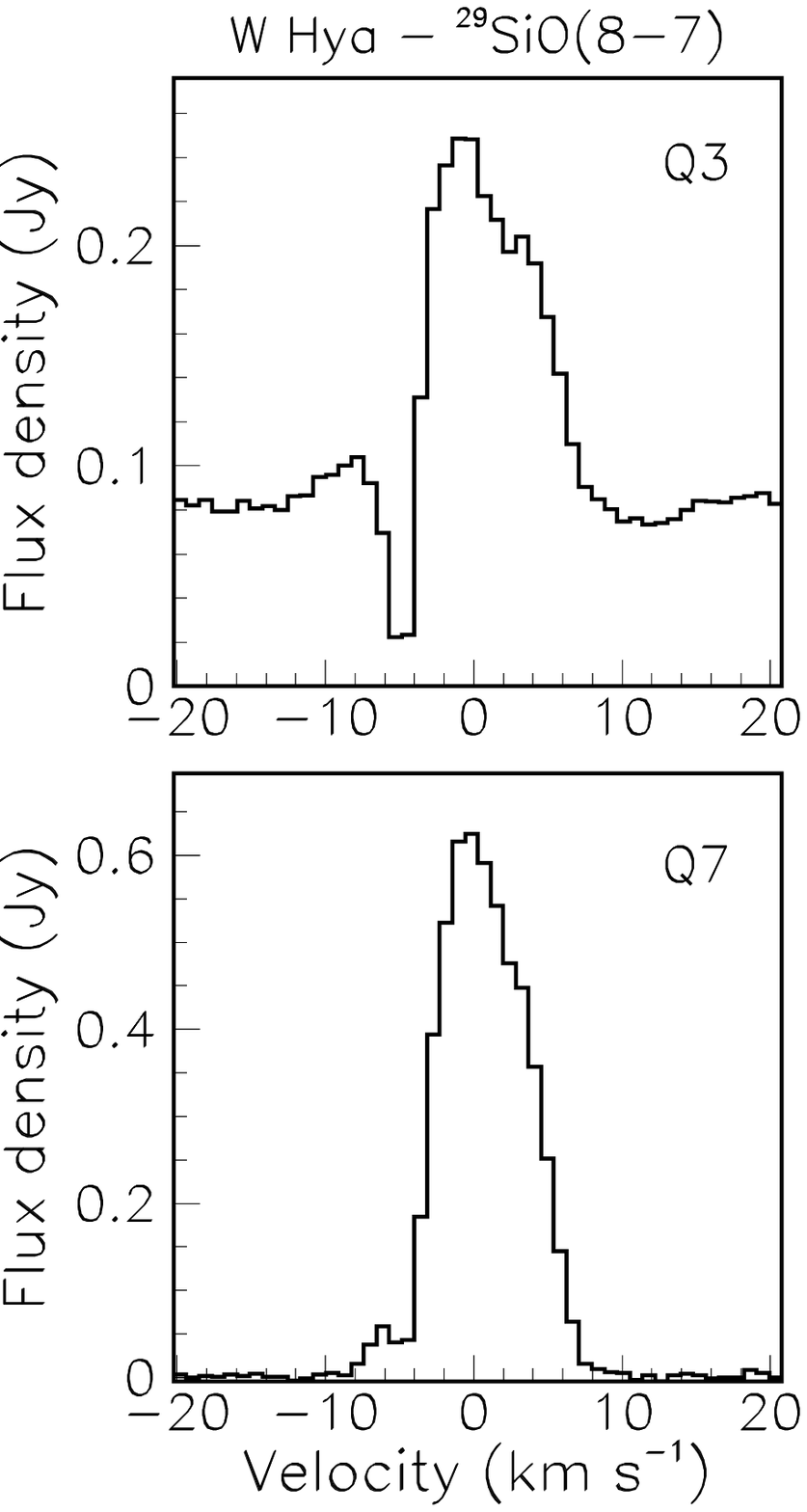}
  \includegraphics[height=7cm,trim=.0cm .0cm 1.7cm .9cm,clip]{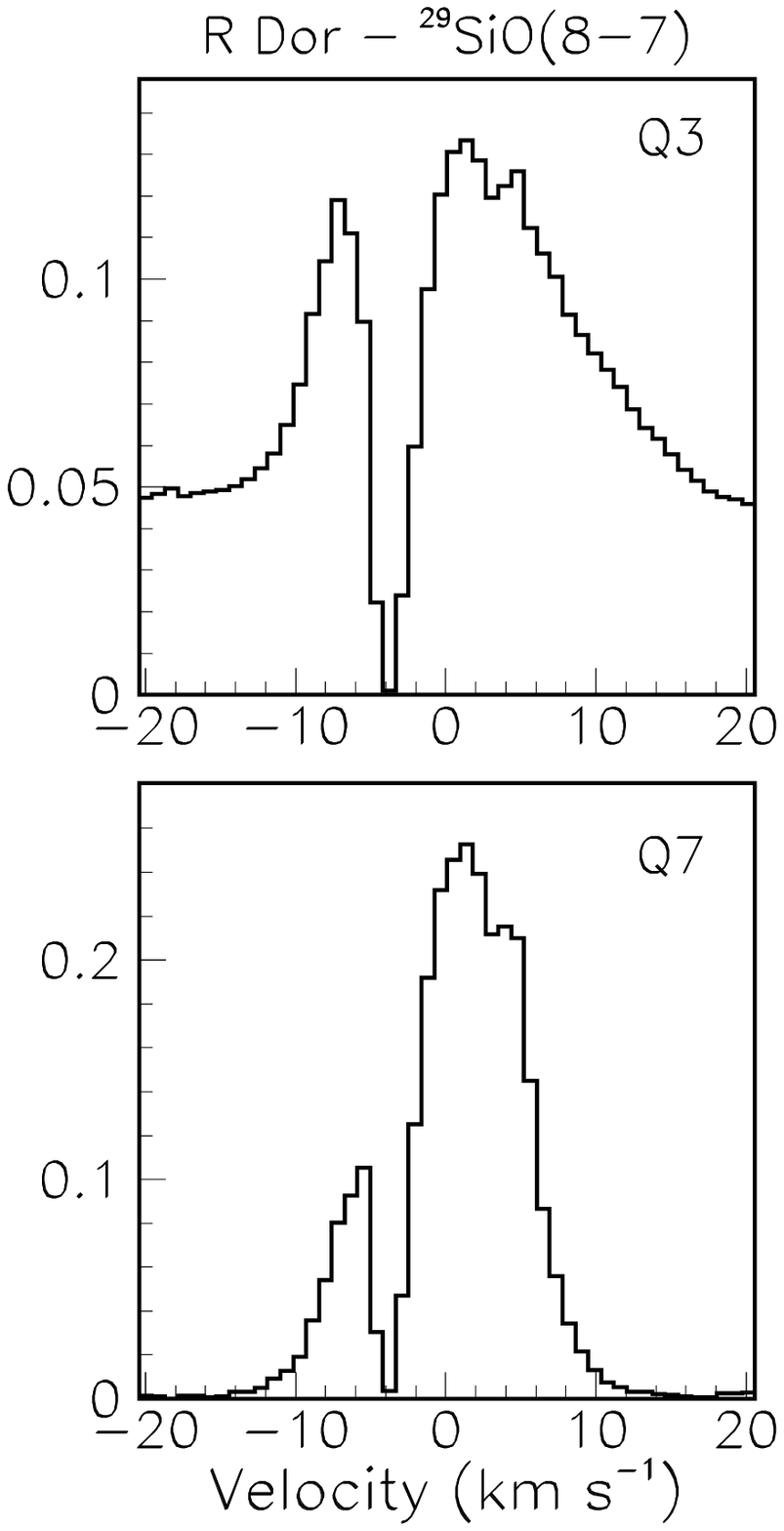}
  \includegraphics[height=7cm,trim=.0cm .0cm 1.7cm .9cm,clip]{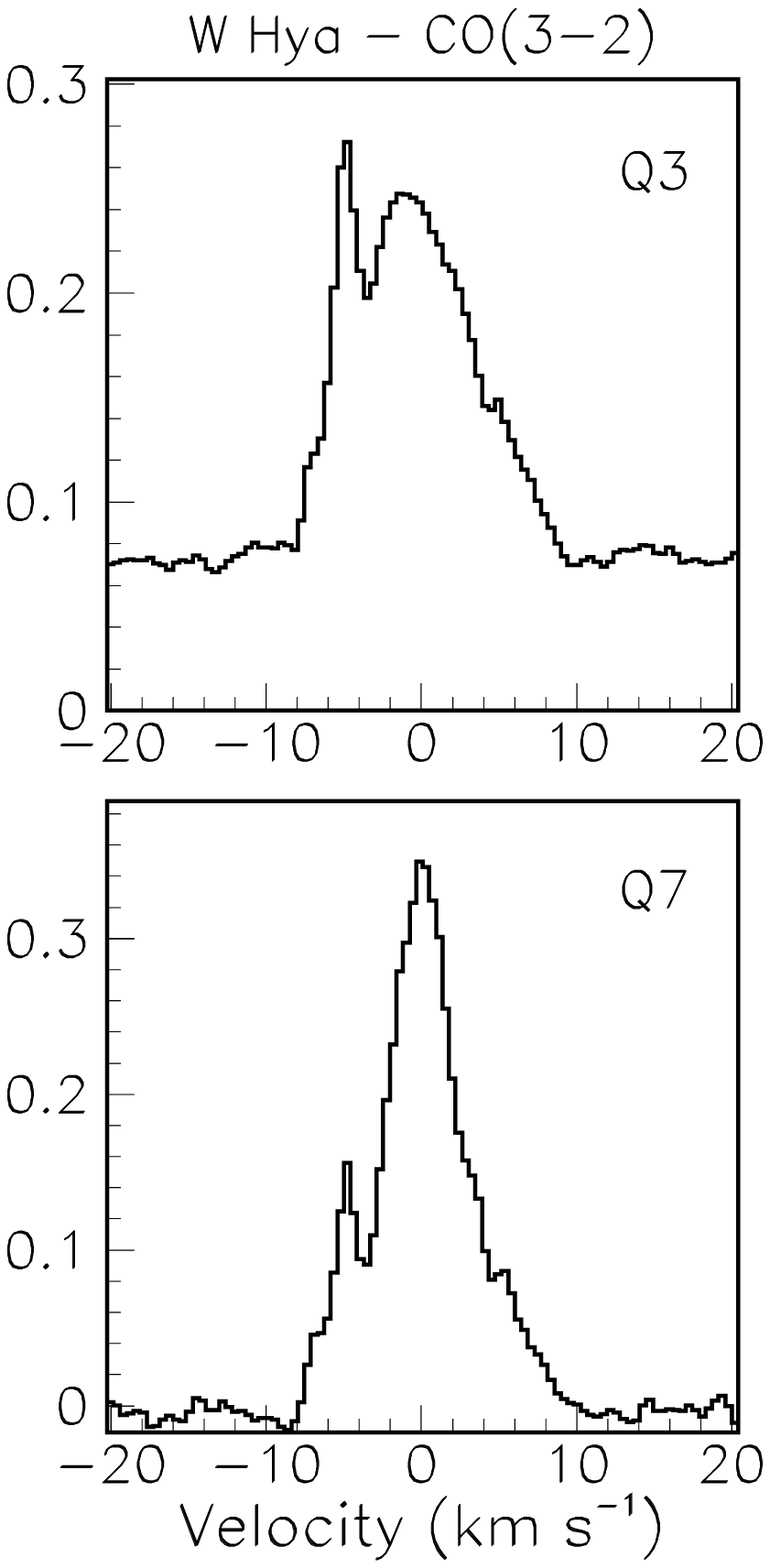}
  \includegraphics[height=7cm,trim=.0cm .0cm 1.7cm .9cm,clip]{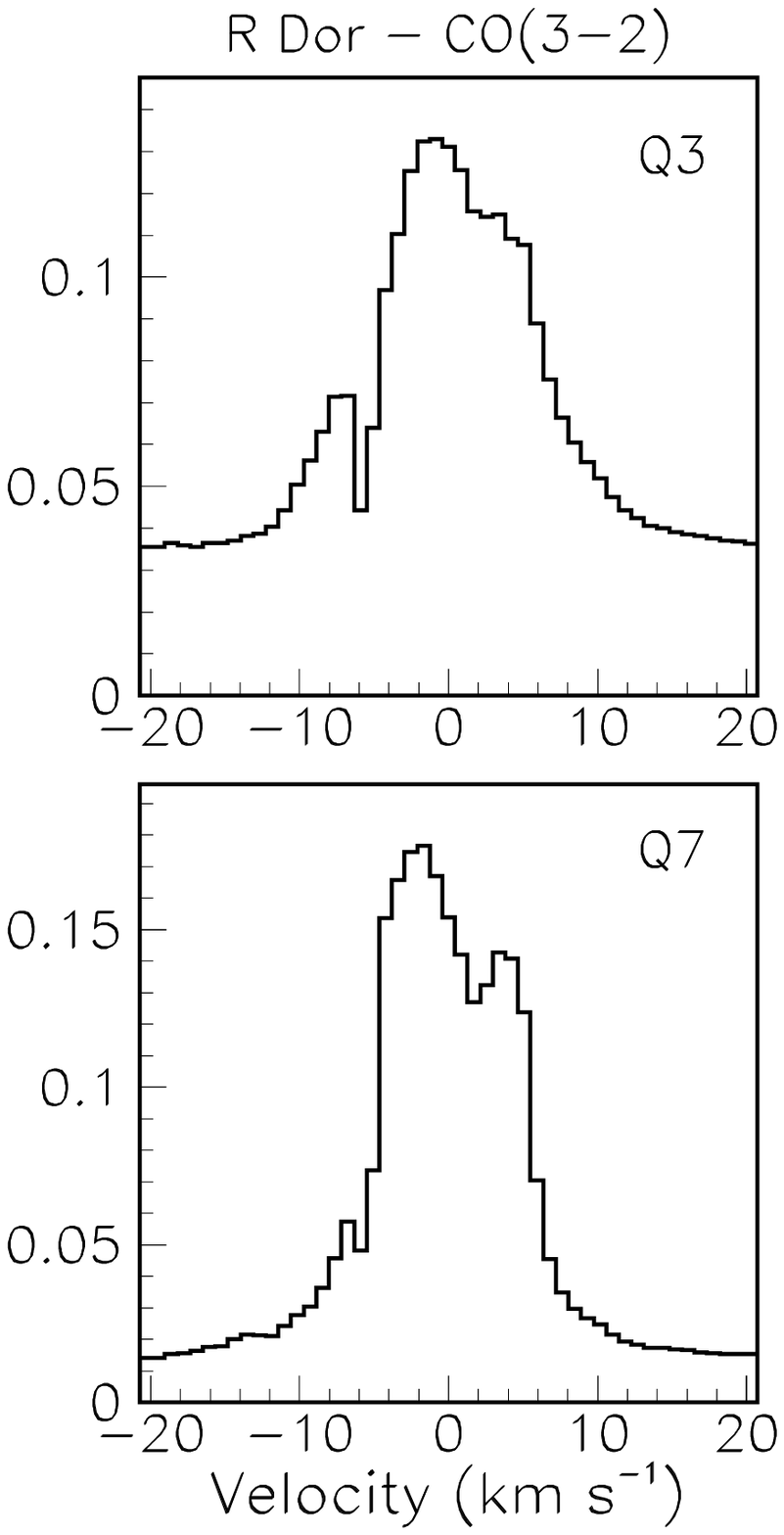}
  \caption{Comparison between W Hya and R Dor, $^{29}$SiO(8-7) and CO(3-2) lines, in quadrants 3 and 7, opposite to region C, as specified on top of the upper panels.   The R Dor data have been  scaled down for the different distance from the Earth (see text). The upper row is for the inner circle (quadrant 3), the lower row for the outer ring (quadrant 7).}
  \label{fig9}
\end{figure*}

  Figure \ref{fig9} compares the CO and SiO spectra in quadrants 3 and 7, opposite to region C, of W Hya to the equivalent for R Dor \citep{Nhung2021}, divided by 3.1 and integrated over a circle $R$$<$0.13 arcsec and a ring 0.13$<$$R$$<$0.26 arcsec, to account for the different distances from the Earth and divided by 4 to compare with a quadrant. In the outer ring, which is free of region C, the SiO R Dor spectra are seen to be lower (0.25 compared to 0.62 Jy) than the W Hya spectra and broader (width at 1/5 maximum $\sim$15 \kms\ compared to $\sim$10 \kms). Qualitatively, however, the similarity between the two spectra suggests that a same dynamics is at stake in both stars. The broader line-width of the R Dor spectra, as noted earlier by \citet{Hoai2021}, suggests a stronger impact of shocks induced by pulsations and convective cell ejections in R Dor than in W Hya away from region C. The difference in intensity ($\sim$3.7 vs $\sim$6.2 Jy \kms) suggests a higher density for W Hya than for R Dor. In the inner circle, the difference between W Hya and R Dor spectra is qualitatively the same as in the outer ring.
  
  The CO emission spectra are smeared by the broader beam size in the R Dor case (180$\times$140 mas$^2$ instead of 52$\times$40 mas$^2$), making the comparison with W Hya difficult.
  
\begin{deluxetable*}{cccccccccc}
\tablenum{3}
\tablecaption{Parameters related to the emissions of the CO and SiO lines. The values of frequency $f$ (GHz), transition energy $\Delta{E}$ (K), upper level energy $E_u$ (K), emission coefficient $A_{\text{ji}}$ (s$^{-1}$) and partition function are taken from CDMS \citep{Muller2005}. \label{tab3}}
\tablehead{ \colhead{Line} & \colhead{$f$(GHz)} & \colhead{$\Delta{E}$ (K)} & \colhead{$E_u$ (K)} & \colhead{2$J$+1} & \colhead{$A_{\text{ji}}$ (s$^{-1}$)} & \colhead{$k$ (K)} & \colhead{$d$ (pc)} & \colhead{$\varepsilon_0$} & \colhead{$\tau_0$}}
\startdata
$^{12}$CO(3-2)&
345.80&
16.7&
33.19&
7&
2.50$\times$10$^{-6}$&
2.77&
104&
27.7&
19.4\\
$^{29}$SiO(8-7)&
342.98&
16.6&
74.08&
17&
21.0$\times$10$^{-4}$&
1.03&104&
21.0$\times$10$^3$&
16.0$\times$10$^3$\\
\enddata
\end{deluxetable*}

In the remaining of this section, we use a very simple LTE model to compare the CSE of R Dor with that of W Hya in quadrant 7, away from region C. We restrict the analysis to the SiO data for R Dor, the beam size being too large for a reliable comparison with the CO data. The R Dor data have been scaled to correct for the different distances from the Earth and to account for the solid angle coverage of quadrant 7. The model assumes that the wind velocity $V$ is radial and increases linearly from $r$=50 mas ($\sim$5 au or $\sim$2 $R_*$) to a terminal velocity $V_{term}$ at $r$=$r_{term}$. The radial decrease of the temperature is taken of the form $T$=$T_0$/($r$/0.05)$^p$ where $T_0$ is the temperature at $r$=50 mas. The CO density takes a value $d_0$ at $r$=1 arcsec and scales in inverse proportion to $Vr^2$. The SiO density is taken as $\rho$ times the CO density. Its depletion is approximately accounted for by limiting the layer radius to 1.5 arcsec instead of 3 arcsec for CO. The line width is described in an ad hoc and very approximate way with Gaussian forms of different $\sigma$ values, as one would do for turbulence, although it is expected to result from shocks induced by pulsations and convective cell ejections in the innermost layers. In particular, the presence of in-falling gas is simply ignored. Beyond a distance $r$=$r_{turb}$, we assume a constant $\sigma$=$V_{turb2}$. Below this distance, we assume a radial dependence of $\sigma$ of the form $\sigma$=$V_{turb1}\exp[-0.5(3r/r_{turb})^2]$. Moreover, even closer to the star, at distances smaller than $r_{turb}/q_r$, we assume a broader value of $\sigma$=$q_VV_{turb1}$.

The main results are illustrated in Figure \ref{fig10} and summarised in Table \ref{tab4} together with the rms differences $\Delta$ between the model flux densities and those observed. We stress that a reliable evaluation of the model parameters would require data cubes covering a larger solid angle and the observation of at least another line of each of the two molecular species; in particular the numbers quoted after the $\pm$ sign should not be understood as uncertainties but simply as measures of the sensitivity of the best fit to the values taken by each parameter when the others are kept fixed. The quality of the fits is satisfactory. The main differences between the SiO emissions of the two stars are a steeper radial dependence of the temperature ($p$=1.44$\pm$0.04 instead of 0.74$\pm$0.05) and a larger region of line broadening ($r_{turb}$=0.84$\pm$0.03 instead of 0.31$\pm$0.07) for R Dor than for W Hya. The value of $\rho$ obtained for W Hya, (10$\pm$3)$\times$10$^{-3}$, corresponds to an abundance ratio $^{28}$SiO/$^{12}$CO$\sim$13\%, consistent with values commonly adopted in the published literature (e.g. \citet{VandeSande2018} quote $\sim$30\% for R Dor). In both cases, the terminal velocity is reached close to the star.

The fit to the R Dor SiO spectrum, although of apparently better quality than that to the W Hya spectrum, is actually less reliable, being unconstrained by the CO data. In particular, the value obtained for $V_{\text{turb2}}$ is unreasonably low. Moreover, the values of $\rho$ and $d_0$ are correlated, only their product matters (0.79 mol cm$^{-3}$ for W Hya and 0.40 mol cm$^{-3}$ for R Dor). The CO spectrum predicted for R Dor using the parameters of the best fit to the SiO spectrum is compared with observation in the rightmost panel of Figure \ref{fig10}. In order to account approximately for the larger beam size, we have subtracted from the observed quadrant 7 spectrum a fraction of the observed quadrant 3 spectrum such that the resulting continuum level cancels.  The high $V_z$ wings are well described but the bulk emission is slightly red-shifted in the model instead of blue-shifted in the data. Moreover the model predicts a sharp absorption near $V_z$=4 \kms, which is absent from the data. This is caused by the low $V_{turb2}$ value obtained from the best fit to the SiO spectrum and by the known difference of $V_{term}$ for the SiO and CO emissions, interpreted by \citet{Nhung2021} as evidence for persistence of some acceleration at large distances. Yet, these are small differences considering the crudeness of the exercise, the overall properties of the spectrum (mean and rms values of $V_z$, total flux density) being well predicted.

In summary, while very crude, this exercise has reliably shown that the morpho-kinematics of the CSE of W Hya, when observed outside region C, is qualitatively similar to that of R Dor and can be described by a model accounting for the main features usually displayed by oxygen-rich AGB stars.
	
\subsection{W Hya inside the C region}
        
\begin{figure*}
  \centering
  \includegraphics[height=4.2cm,trim=.5cm 1.3cm 2.2cm 1.8cm,clip]{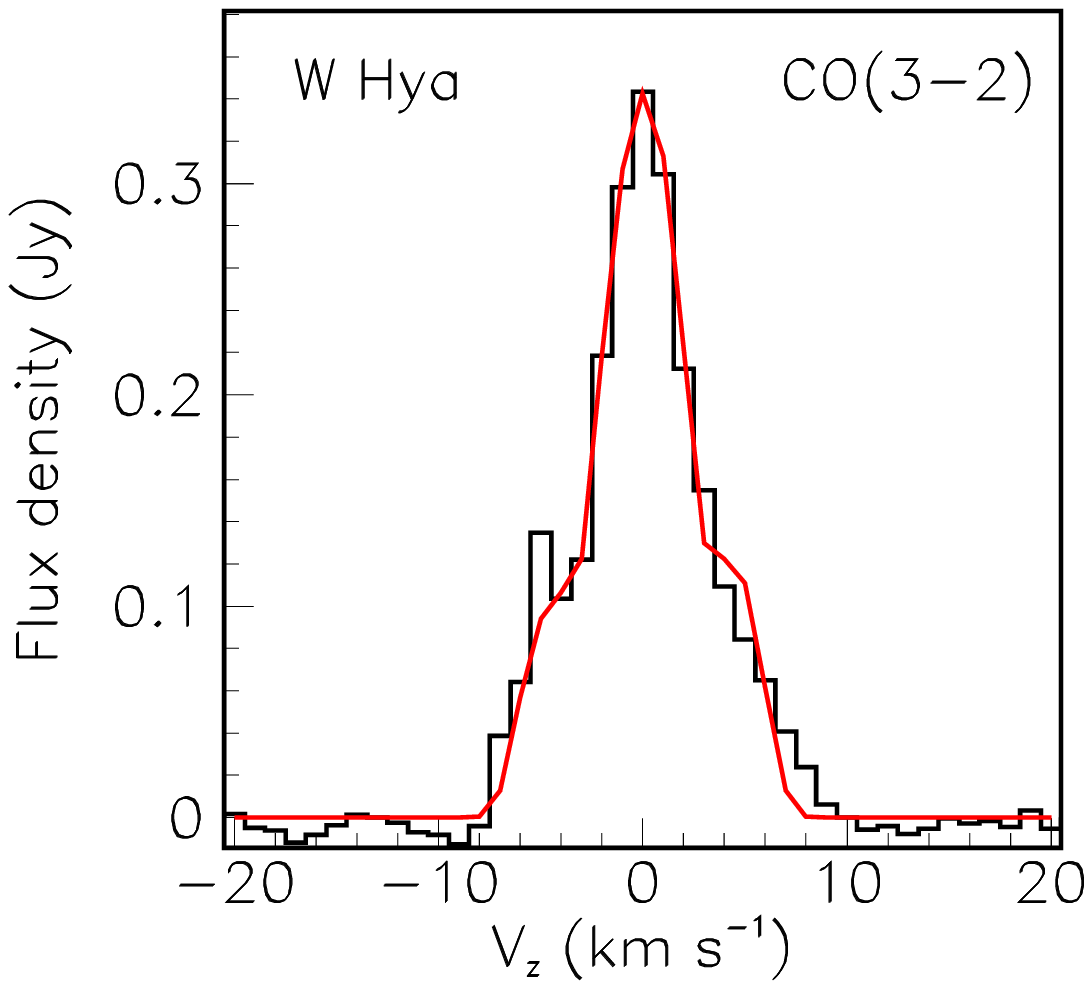}
  \includegraphics[height=4.2cm,trim=.5cm 1.3cm 2.2cm 1.8cm,clip]{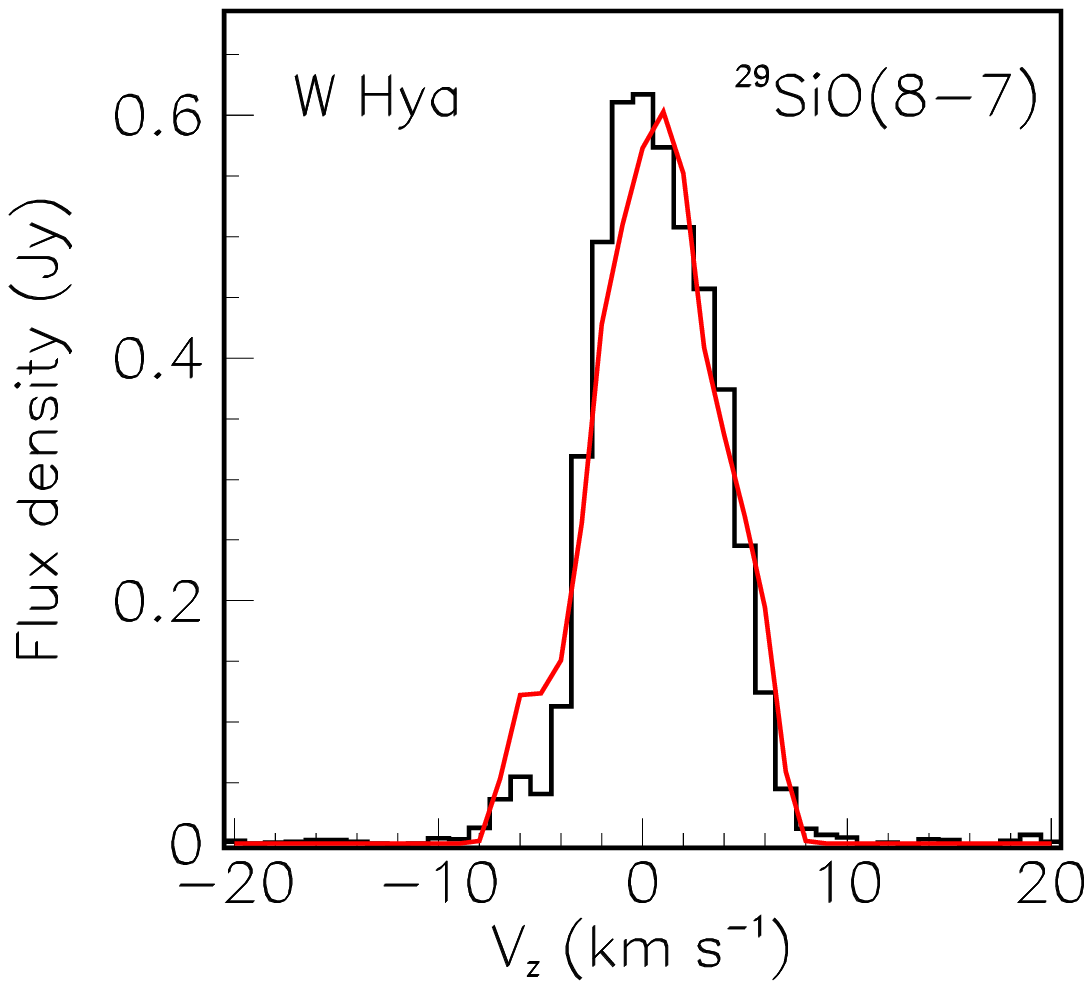}
  \includegraphics[height=4.2cm,trim=.5cm 1.3cm 2.2cm 1.8cm,clip]{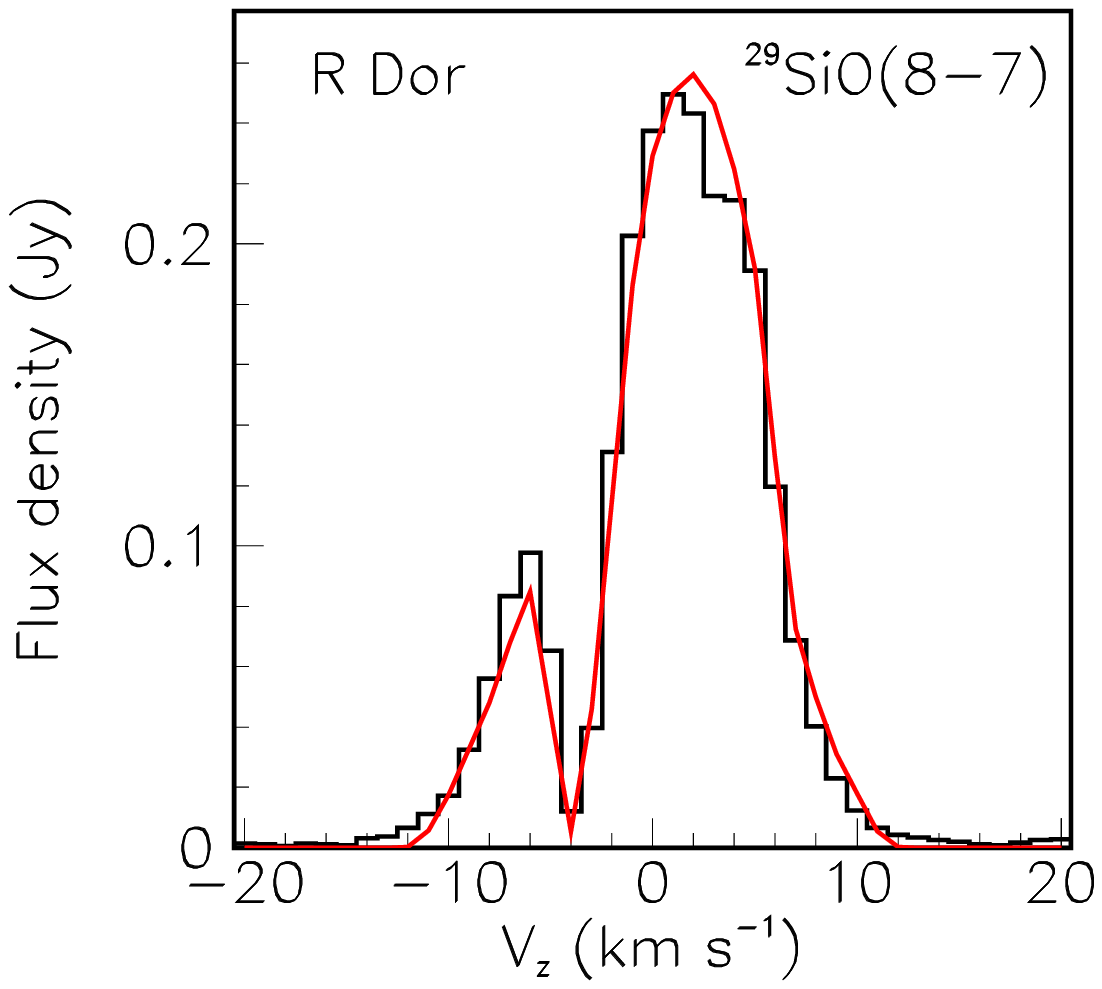}
  \includegraphics[height=4.2cm,trim=.5cm 1.3cm 2.2cm 1.8cm,clip]{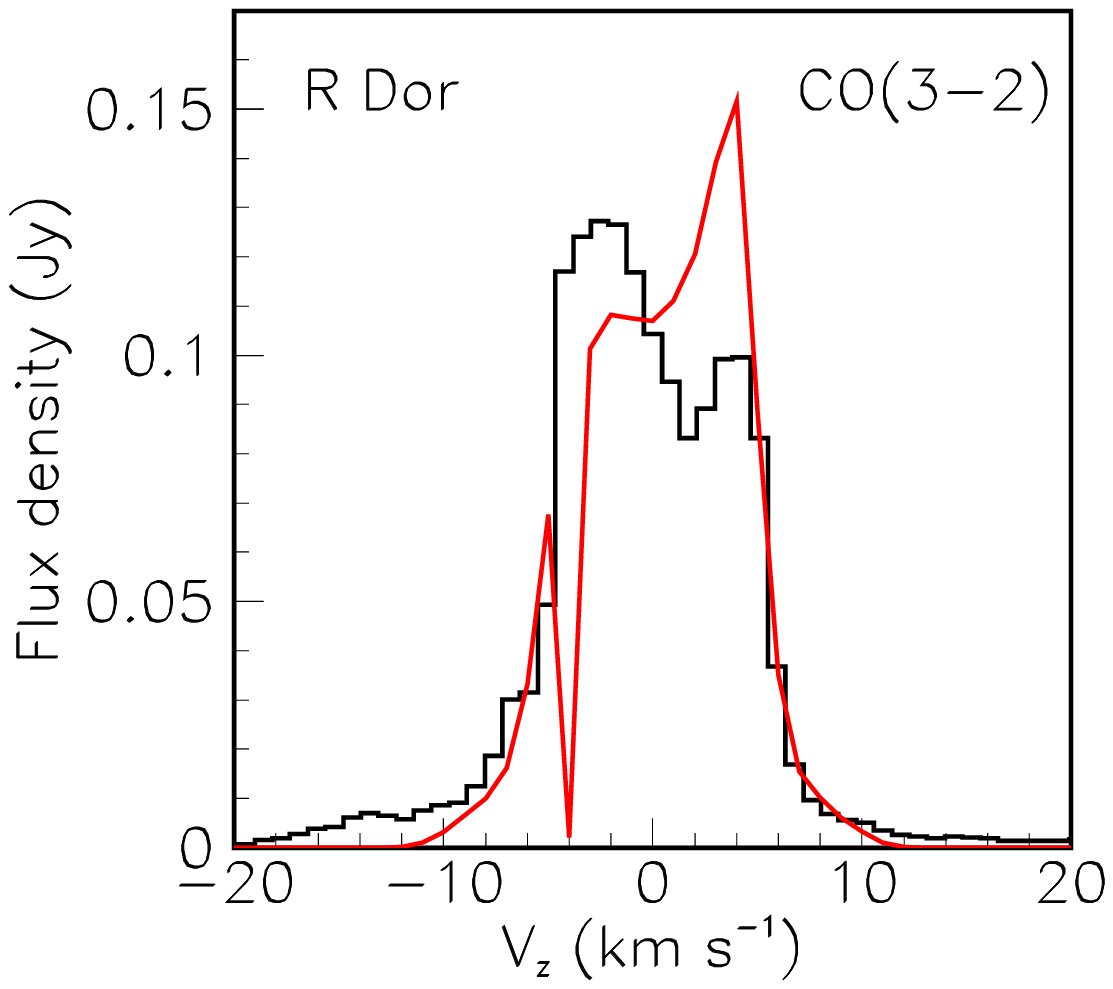}
  \caption{Comparison between CO and SiO spectra observed in quadrant 7 with best fit model predictions (see text). The pair of left panels is for W Hya and the pair of right panels for R Dor (properly scaled to quadrant 7, see text). In the three leftmost panels the red curve shows the result of the best fit. In the rightmost panel, the red curve is the model prediction for the CO spectrum using the parameters obtained for the best fit to the SiO spectrum (see text). }
  \label{fig10}
\end{figure*}

\begin{figure*}
  \centering
  \includegraphics[height=5cm,trim=.5cm 1.cm 1.8cm 1.8cm,clip]{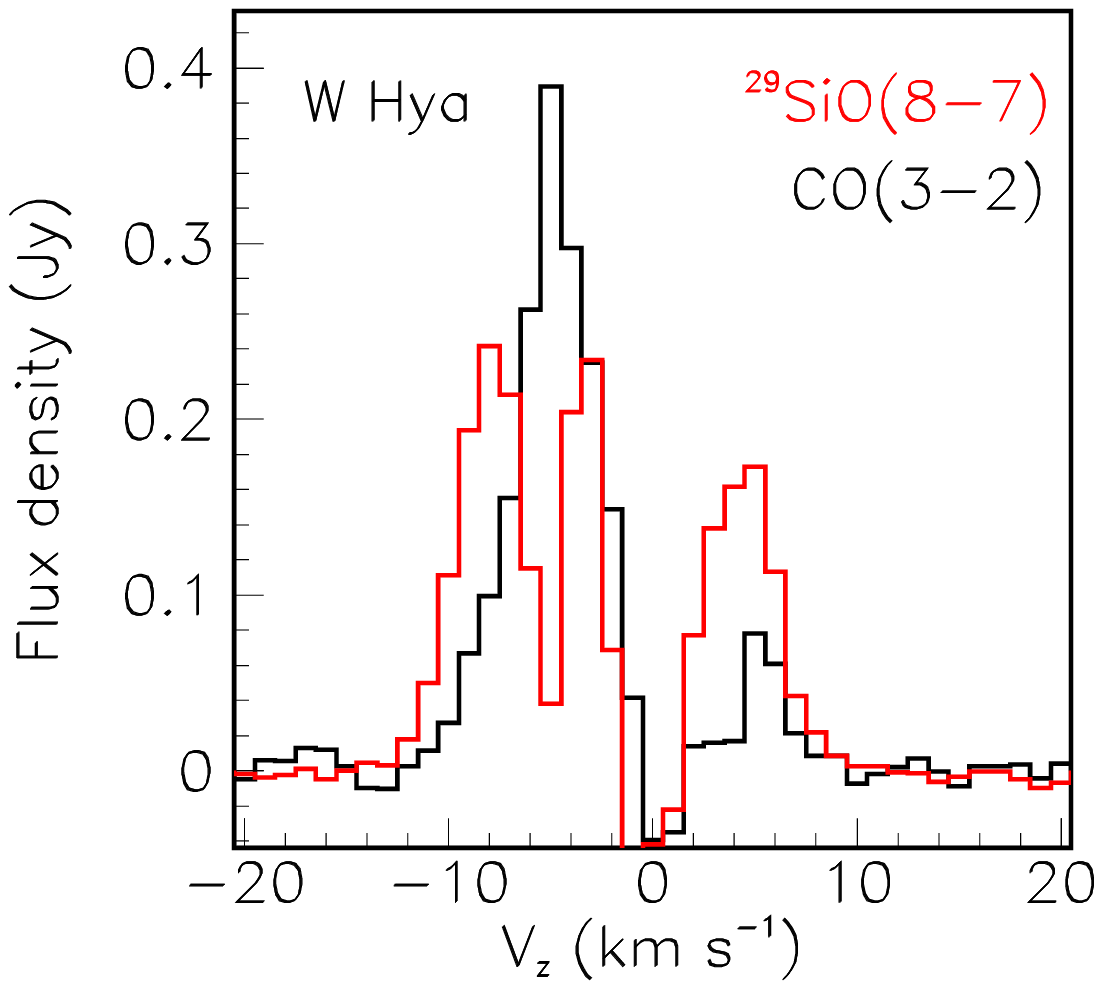}
  \includegraphics[height=5cm,trim=.5cm 1.cm 1.8cm 1.8cm,clip]{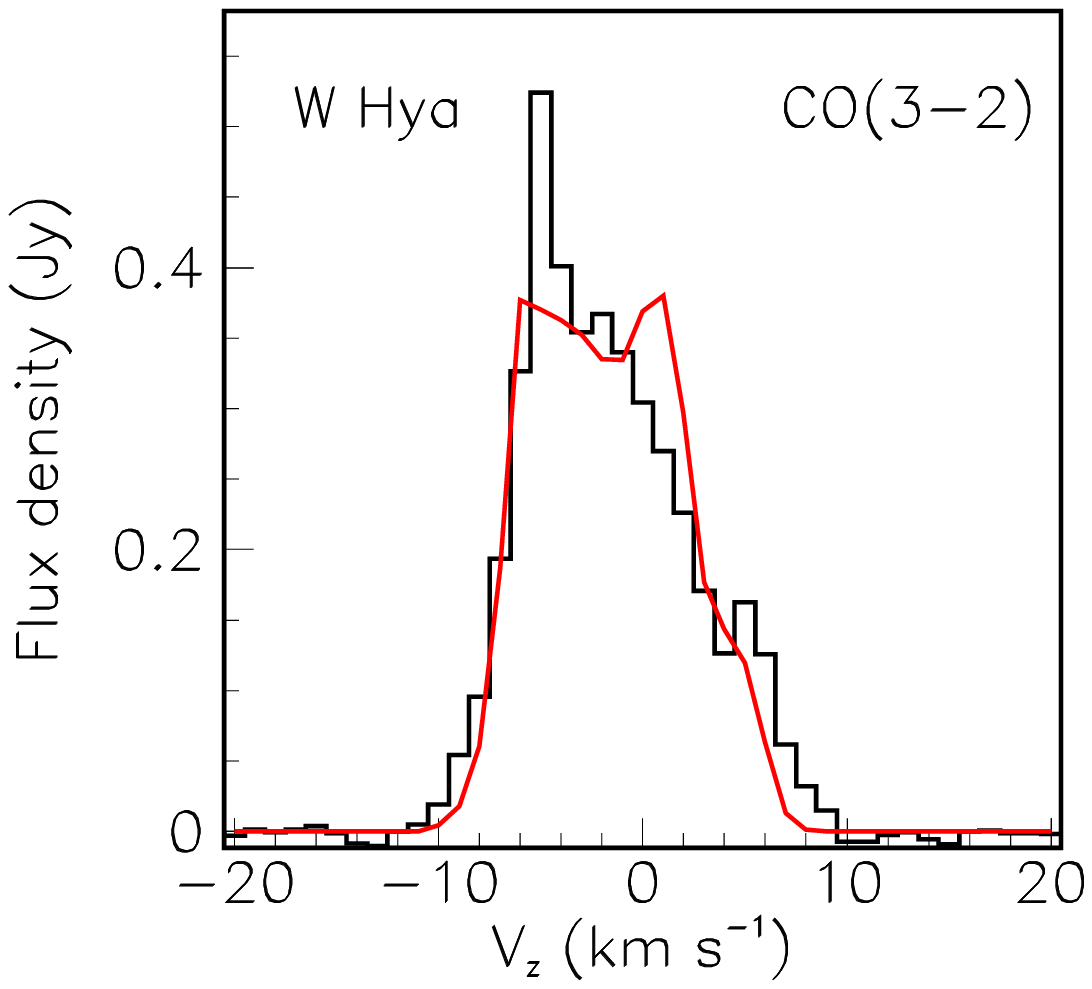}
  \includegraphics[height=5cm,trim=.5cm 1.cm 1.8cm 1.8cm,clip]{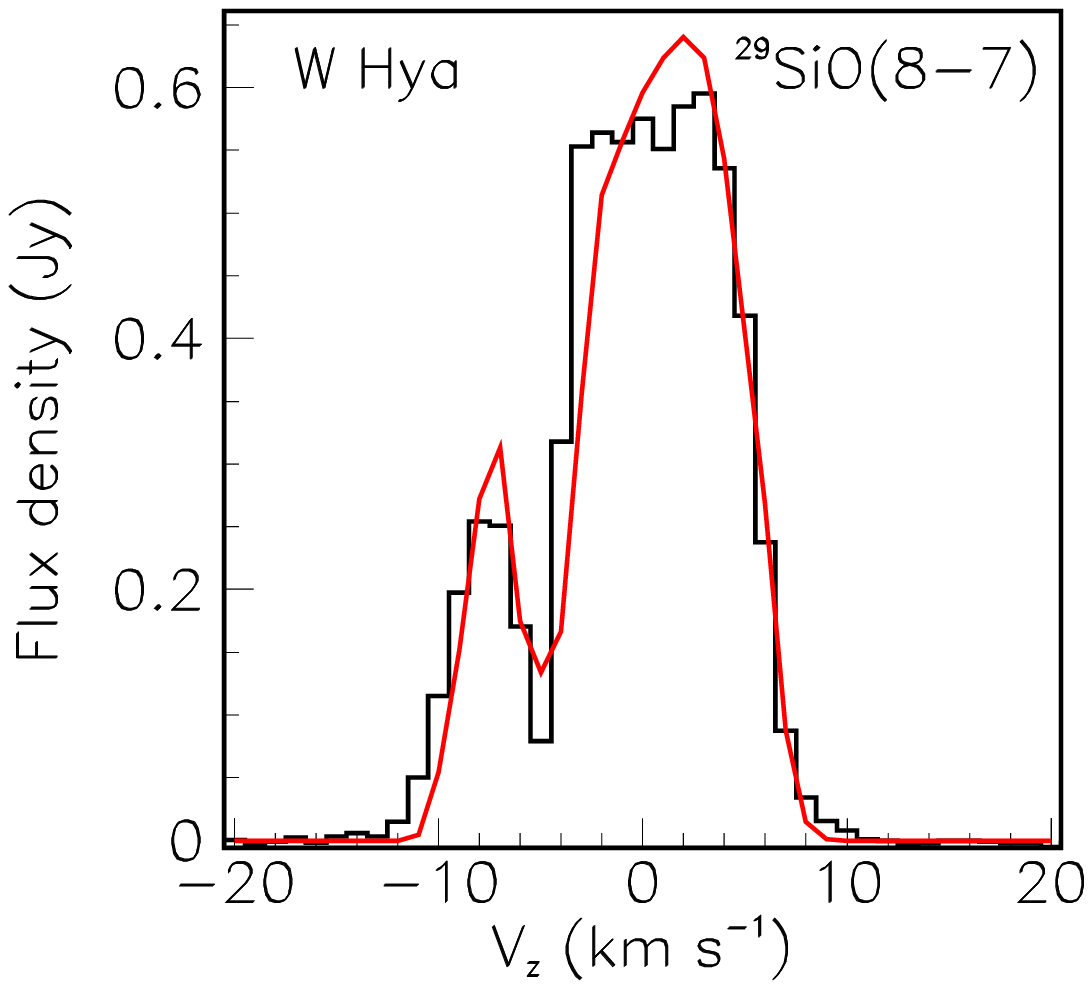}
  \caption{ Left: difference between the Doppler velocity spectra observed in quadrants 7 and 5 of W Hya for $^{12}$CO(3-2) (black) and $^{29}$SiO(8-7) (red) emissions. Middle and right: Quadrant 5 observed spectra (black histograms, CO middle and SiO right) are compared with the results of the fit described in the text (red lines).}
  \label{fig11}
\end{figure*}

In order to shed some light on the morpho-kinematics at stake in region C of W Hya, we show in the left panel of Figure \ref{fig11}, for both CO and SiO emissions, the difference between the Doppler velocity spectra observed in quadrants 7 and 5. Both are in the outer ring, the former outside of region C and the latter inside. The result is very similar to that obtained in Figure \ref{fig7} for the blob region, defined as $|\omega|$$<$30\dego\ instead of |$\omega-350$\dego|$<$45\dego. It gives evidence for intense blue-shifted emission accompanied by strong line broadening. With the idea that this is the effect of a source confined to short distances from the star in the blue-shifted hemisphere, we modify the model used to produce the spectra illustrated in Figure \ref{fig10} by keeping all parameters unchanged except for a narrow layer defined as 0$>$$z$$>$$-$0.25 arcsec (0$>$$z$$>$$-$25 au) where we adjust density, temperature and velocity in order to reproduce the CO and SiO spectra observed in quadrant 5. The results are displayed in the right panels of Figure \ref{fig11} and Table \ref{tab4b}. The quality of the fits is again satisfactory given the extreme crudeness of the model: the rms differences $\Delta$ between the model flux densities and those observed are 39 mJy for the CO spectrum and 47 mJy for the SiO spectrum. The main outstanding parameter value is the CO density, $d_0$=930$\pm$160 mol cm$^{-3}$, an order of magnitude larger than outside region C. Also remarkably different is the line width, $V_{turb1}$=3.75$\pm$0.50 \kms\ instead of 1.9$\pm$0.3 \kms\ outside region C. The velocity of the layer is 6.5$\pm$0.5 \kms, and the other parameters take similar values as outside region C. Because of the high value of the density, the SiO layer is extremely opaque and the value of $\rho$ is largely undefined: its best value is 37$\times$10$^{-3}$ but any value larger than 20$\times$10$^{-3}$ gives an acceptable fit.  
 
\section{Continuum emission}

Two sources contribute to continuum emission: the star itself and hot dust surrounding it. Observations by \citet{Ohnaka2016,Ohnaka2017} using SPHERE/ZIMPOL on the VLT and bracketing the ALMA observations analysed in the present article have given evidence for anisotropic dust formation, with an enhancement in the northern hemisphere and a depression in the south-western quadrant. Observation of H$\alpha$ emission has given evidence for shocks anti-correlated with dust formation. Moreover, the polarisation pattern observed using SPHERE/ZIMPOL displays the circular symmetry characteristic of light scattered from a spherical shell and the polarised intensity gives further evidence for a south-western depression of the dust density. 

\begin{deluxetable*}{ccccccccccccc}
\tablenum{4}
\tablecaption{Parameters of the radiative transfer fits in quadrant 7 for W Hya and R Dor, the latter being scaled as described in the text. The values quoted after the $\pm$ sign measure the sensitivity of the fit to the particular parameter: they correspond to a 10\% increase of the $\chi^2$ per degree of freedom, the other parameters being kept at their best fit values. They should not be interpreted as uncertainties.  \label{tab4}}
\tablehead{\colhead{}& \colhead{$T_0$}&
 \colhead{$p$}&
 \colhead{$V_{term}$}&
 \colhead{$r_{term}$}&
 \colhead{$d_0$}&
 \colhead{$r_{turb}$}&
 \colhead{$V_{turb1}$}&
 \colhead{$q_V$}&
 \colhead{$q_r$}&
 \colhead{$V_{turb2}$}&
 \colhead{$\rho$} &
 \colhead{$\Delta$}\\
\colhead{}&\colhead{[K]}&\colhead{}&\colhead{[\kms]}&
\colhead{[arcsec]}&
\colhead{[cm$^{-3}$]}&
\colhead{[arcsec]}&
\colhead{[\kms]}&
\colhead{}&
\colhead{}&
\colhead{[\kms]}&
\colhead{[10$^{-3}$]}&
\colhead{[mJy]}}
\startdata
W Hya&
1970&
0.74&
5.0&
0.32&
79&
0.31&
1.9&
2.2&
3.15&
0.92&
10&
CO:12\\
&$\pm$120&
$\pm$0.05&
$\pm$0.3&
$\pm$0.02&
$\pm$4&
$\pm$0.07&
$\pm$0.3&
$\pm$0.7&
$\pm$0.30&
$\pm$0.20&
$\pm$3&
SiO:33 \\
R Dor&
1730&
1.44&
4.15&
0.19&
47&
0.84&
1.83&
2.47&
6.3&
$<$0.2&
8.5&
9\\
&
$\pm$75&
$\pm$0.04&
$\pm$0.10&
$\pm$0.02&
$\pm$6&
$\pm$0.03&
$\pm$0.09&
$\pm$0.15&
$\pm$0.4&
&
$\pm$1.1&\\
  \enddata
  \end{deluxetable*}

\begin{deluxetable*}{cccccccc}
\tablenum{5}
\tablecaption{Parameters of the radiative transfer fits in quadrants 5 and 7 of W Hya (same format as Table \ref{tab4}) \label{tab4b}}
\tablehead{\colhead{}&
 \colhead{$V (layer)$}&
 \colhead{$d_0$}&
 \colhead{$r_{turb}$}&
 \colhead{$V_{turb1}$}&
 \colhead{$T_0$}&
 \colhead{$p$}&
 \colhead{$\rho$}\\
 \colhead{}&\colhead{[\kms]}&
 \colhead{[cm$^{-3}$]}&
 \colhead{[arcsec]}&
 \colhead{[\kms]}&
 \colhead{[K]}&
 \colhead{}&\colhead{[10$^{-3}$]}}
\startdata
Quarant 5&6.5&930&3.75&0.38&1970&0.71&>20\\
(inside C)&
$\pm$0.5&
$\pm$60&
$\pm$0.5&
$\pm$0.05&
$\pm$200&
$\pm$0.09&\\
Quadrant 7&nolayer&79&1.9&0.31&1970&0.74&10\\
(outside C)&&
$\pm$4&
$\pm$0.3&
$\pm$0.07&
$\pm$120&
$\pm$0.05&
$\pm$3\\
  \enddata
  \end{deluxetable*}

\begin{figure*}
  \centering
  \includegraphics[height=6cm,trim=.0cm 1.cm .0cm 1.cm,clip]{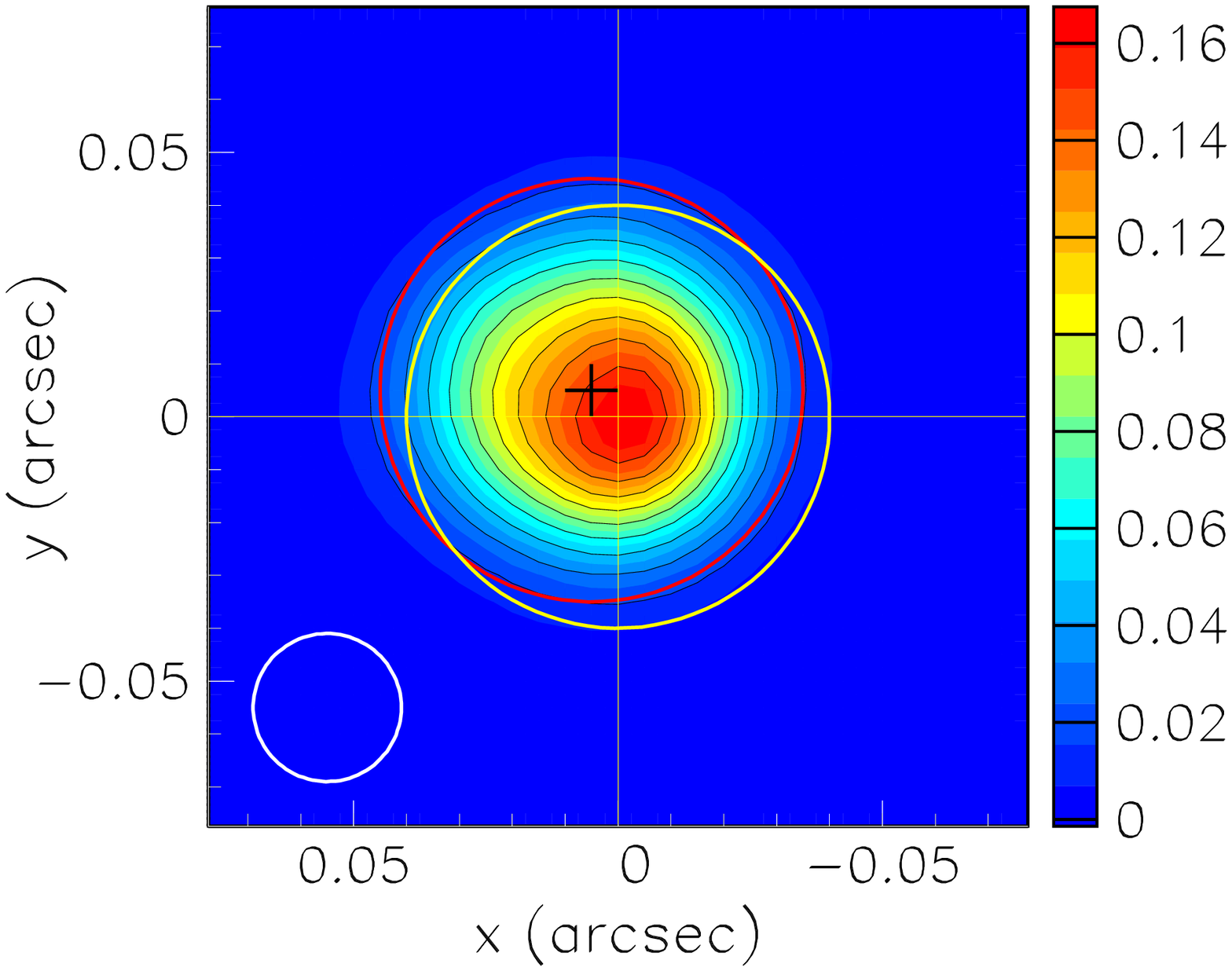}
    \includegraphics[height=5.42cm,trim=-1cm -2.1cm .0cm 0.5cm,clip]{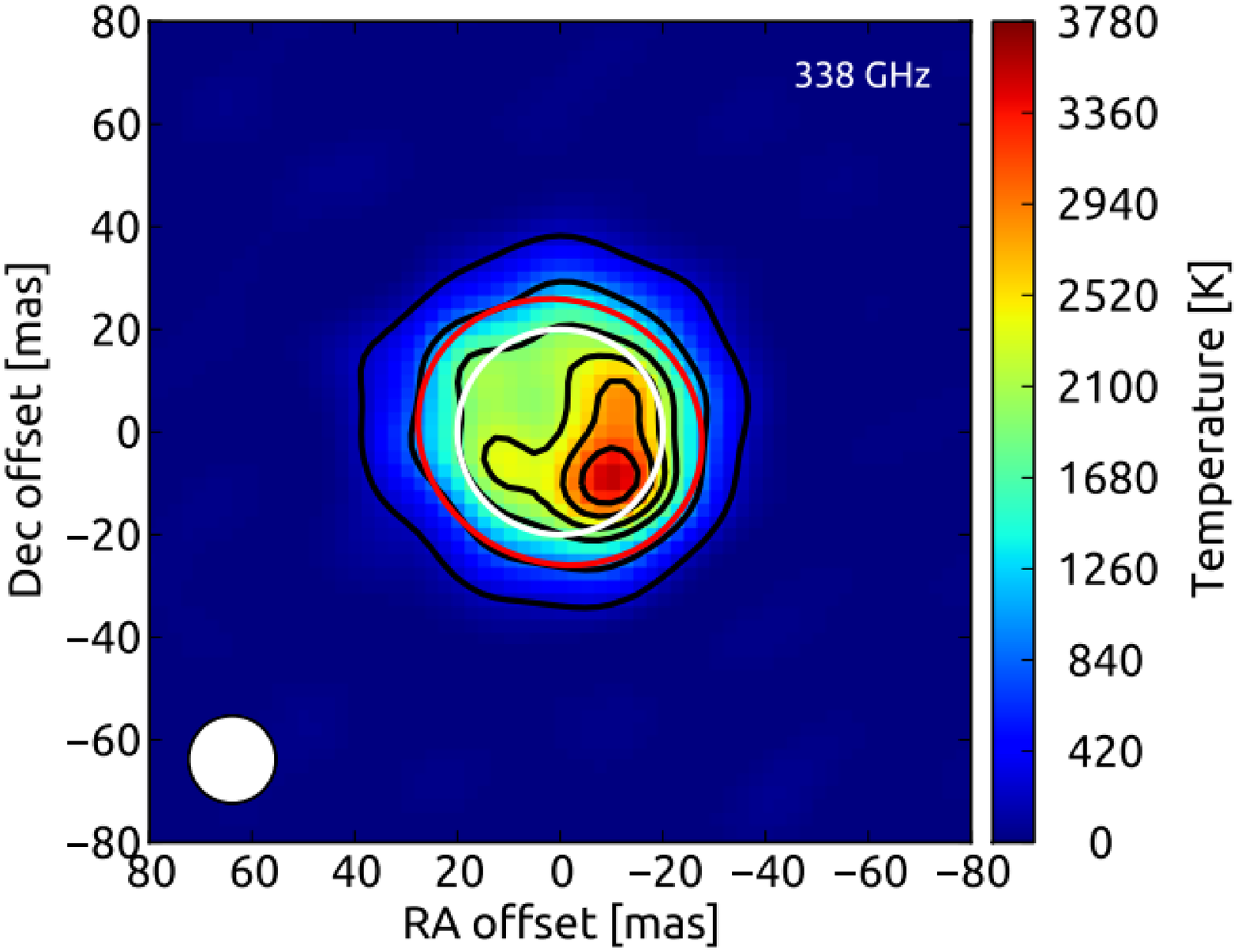}
  \caption{Continuum emission of W Hya. Left: image obtained from data set 3. The yellow circle, centred on the origin of coordinates at RA=13h49m01.9386s and Dec=$-$28d22m04.489s, is centred at the maximum of the emission. The red circle is instead centred (black cross) on the boundary of the emission, as is the red circle in the right panel. It is shifted $\sim$5 mas north and $\sim$5 mas east. The right panel shows the image obtained by \citet{Vlemmings2017}. In both panels the beams are shown in the lower left corners. The colour scales are in Jy beam$^{-1}$ for the left panel and in Kelvin for the right panel, the conversion factor between these being 13,600 K per Jy beam$^{-1}$.}
  \label{fig12}
\end{figure*}

The level of continuum emission obtained from the analyses of SiO and CO line emissions presented in the present article, confined within $\sim$50 mas from the centre of the star, is similar to that of R Dor (after correction for the difference in distances) in agreement with earlier evaluations of the star temperatures, $\sim$2700 K for W Hya \citep{Vlemmings2019} and $\sim$3000 K \citep{Dumm1998} for R Dor. As the continuum emission of W Hya has been studied in detail by \citet{Vlemmings2017}, we did not attempt to add new contributions to their results. However, our results seem to be partly inconsistent with theirs and we find it useful to explain this apparent inconsistency. As illustrated in the right panel of Figure \ref{fig12}, these authors show the presence of a very hot and compact spot on the south-western quadrant of the stellar disc, within 20 mas from the centre of the star, and interpret it as evidence for a shock produced by a convective cell ejection; the spot temperature is at the level of 50000 K, compared with 2500 K for the temperature of the photosphere and 2900 K for that of the heated molecular gas layer surrounding it. As illustrated in Figure \ref{fig12}, we find no evidence for such a hot spot; on the contrary, as mentioned in Section 3.4, we observe a small depression of continuum emission in the south-western quadrant.   

We recall that W Hya was observed on three days December 05th, December 03rd and November 30th, 2015. The data are stored in three sets 1, 2, 3 respectively. They are calibrated using the script provided by the ALMA staff. The maximal baseline is 7.7 km for set 1, 6.3 km for set 2 and 10.8 km for set 3. There are three spectral windows, two of 937.5 MHz and 960 channels were centred on 330.85 (spw0) and 342.85 GHz (spw1), and a window of 1.875 GHz and 3840 channels was centred on 344.95 GHz (spw2). The continuum emission displayed in the left panel of Figure \ref{fig12} uses the most extended array and is averaged over spw2, excluding  lines identified by using task uv$_-$prev in GILDAS/IMAGER. The image has been produced with CASA using uniform weighting. The beam is 28$\times$20 mas$^2$ (FWHM) and the map is produced with a circular restoring beam of 28 mas.

The right panel of Figure \ref{fig12} shows the image obtained by \citet{Vlemmings2017}. In addition to the flagged antennas indicated in the standard script, they flagged one additional antenna which had a high system temperature. They performed two rounds of phase self-calibration and one round of phase and amplitude self-calibration on the continuum. They obtained a beam with uniform weighting of 17$\times$24 mas$^2$ and the final image is restored with a circular beam of 17 mas, smaller than the major axis.

However, the apparent contradiction between the two results boils down to what is taken as precise star location. In our case, we assume that it is the maximum of the emission, which is slightly shifted south-west with respect to the centre of the boundary of the emission, approximately by 5 mas west and 5 mas south. In such a case, we conclude that the emission is slightly depressed in the south-western quadrant. \citet{Vlemmings2017}, who use a smaller beam than we do, obtain an image, which strongly suggests that the maximum of the emission corresponds to a hot spot, and that the star location is instead at the centre of the boundary of the emission. It is therefore essentially the use of a smaller beam, possibly together with two successive self-calibrations, that makes the difference of interpretation. We note that a same comment applies to the interpretation of the images obtained by \citet{Ohnaka2016, Ohnaka2017}.

\section{A global view}

\begin{figure*}
  \centering
  \includegraphics[height=8cm,trim=.0cm 0cm 0cm 0cm,clip]{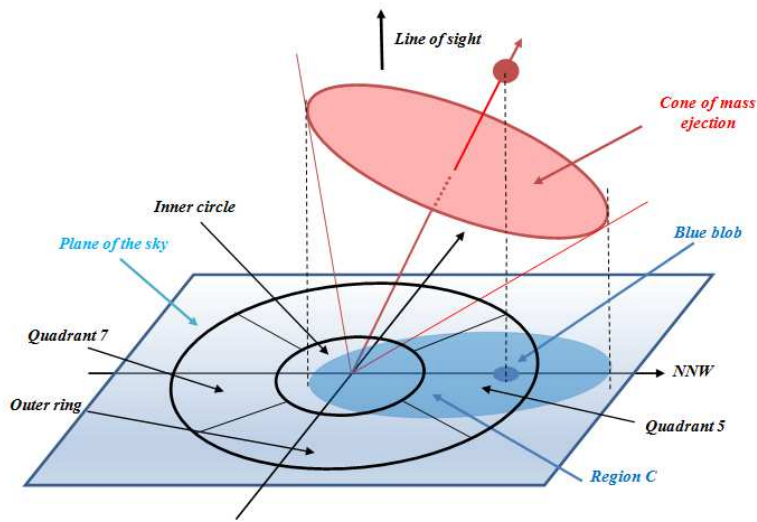}
  \caption{Simplified schematic of a possible geometry.}
  \label{fig13}
\end{figure*}

Accounting for the observations made in the previous sections, the present section is an attempt at drawing a consistent picture of the morpho-kinematics of the CSE of W Hya.

a) The star is surrounded by an expanding layer of gas and dust, which, to first order, is optically thick for $^{29}$SiO(8-7) emission and optically thin for $^{12}$CO(3-2) emission.  In contradiction with earlier estimates \citep{Khouri2014a}, the terminal velocity is $\sim$5 \kms\ instead of 7.5 \kms. It has been reached early, at probably some $\sim$12 $R_*$ (0.3 arcsec) from the centre of the star, where the escape velocity is still 7.7 \kms: this implies that the acceleration of the gas by collisions with dust grains manages to keep the velocity approximately constant over a broad radial range. Such early acceleration to the terminal velocity is consistent with the conclusion reached by \citet{Takigawa2017}, who claim that aluminium-rich grains have formed within some 3 $R_*$ (75 mas) from the centre of the star and are efficient at boosting rapidly the gas to radial velocities at the 5 \kms\ scale. It is supported by the narrow width of the peaks observed in emission for CO and in absorption for SiO in Figures \ref{fig2} and \ref{fig5}: if acceleration were progressive these peaks would tend to trail in the red-shifted direction. Note that masers can be invoked to explain the narrow CO emission peak, but they would not explain the narrow SiO absorption peaks. Moreover, the observation of a peak at $+$5 \kms\ in some panels of Figures \ref{fig2} and \ref{fig5} supports the picture of a spherical shell. For such narrow peaks to be present, this layer must have been stable for many years, a century at least. Comparison with R Dor has shown remarkable similarities and has given evidence for the validity of a picture accounting for the main features usually displayed by oxygen-rich AGB stars to properly describe the morpho-kinematics of the CSE of W Hya outside region C.

b) The intense blue-shifted emission observed in region C displays features of a mass ejection that occurred recently, at the scale of a few years, and is associated with the blue blob. Figure \ref{fig13} shows a simplified schematic of a possible geometry: based on the confinement of region C on the sky plane and on the crude radiative transfer considerations made in Section 4.3, it assumes that the impact of the mass ejection covers a broad cone with an axis making an angle of some 30\dego\ with the line of sight, implying that the source associated with the blue blob is located some 0.2 arcsec (20 au) above the plane of the sky. At a velocity of $\sim$10 \kms, such distance is covered in $\sim$10 years. Within the cone, and within radial distances confined to less than 25 au from the centre of the star, the number density is increased by an order of magnitude, with a SiO/CO ratio larger by at least a factor 2, and the effective line width is also about twice as large. The mean radial velocity inside the cone is 6-7 \kms. The cone geometry matches the dust distribution observed by \citet{Ohnaka2017}. However, it evolves at a time scale of a few years while the variability observed by these authors is at the scale of a few months. Such a picture impacts significantly the evidence for CO maser emission claimed by \citet{Vlemmings2021} but does not exclude some possible contribution.

c) Within the spectral interval of terminal velocity, $-$6 to $-$4 \kms, both over region C and outside, the expanding spherical layer discussed under a) is essentially optically thick for the SiO line and the observed emission probes exclusively its external shell: it makes therefore no difference whether one looks over region C or out of it. What makes the difference in the depth of the absorption peak illustrated in the central panel of Figure \ref{fig3} is instead the emission outside the spectral interval of terminal velocity. There, the opacity of the SiO line is much less and the emission over region C is enhanced by the additional contribution of the emission of region C. It is this enhancement, rather than the intensity of the absorption, that explains the correlation illustrated in the right panel of Figure \ref{fig3}.

d) Outside the spectral interval of terminal velocity, we observe the emissions of the CO and SiO lines from gas having not yet reached the terminal velocity and/or from gas expanding in a direction significantly different from the line of sight. Such gas is confined within at most 30 au from the centre of the star (at such a distance, gas having deviated from the line of sight by a projected distance $R$=150 mas makes an angle of 30\dego\ with the line of sight and its Doppler velocity is 87\% of its radial velocity). It is seen in emission for both the CO and SiO lines with the exception of some red-shifted absorption of SiO line emission over the inner quadrants, which we interpret as evidence for in-falling gas. Important line broadening is observed in this component of the CSE, witness of the turbulent regime induced by shocks from pulsations and cell ejections. It reaches typically $\pm$10 \kms, however covering a smaller radial distance in W Hya than it does in R Dor as had been already noted by \citet{Hoai2021} and has been confirmed by the analysis presented in Section 4.3. The spectra are significantly narrower outside region C, where a depression of dust formation has been reported by \citet{Ohnaka2017}.

\section{Conclusion}

In summary, the picture that emerges from all above observations is made of two components, one stable over time, at the scale of at least a century, the other variable, at the scale of months and years. It differs significantly from the picture that was suggested by earlier analyses. In particular, it gives evidence for a recent mass ejection, which had been overlooked by all former studies and it underscores relations between the dust emission observed at the VLT and the gas emission observed at ALMA. 

The stable component consists of an approximately spherical envelope of gas and dust expanding radially and reaching rapidly, at a radius of $\sim$0.3 arcsec =30 au =12$R_*$, a terminal velocity of $\sim$5 \kms. The close environment of the star hosts a complex chemistry of gas and dust, displaying both stable and variable features. In particular, \citet{Vlemmings2017} have given evidence for gas having to stay close to the star photosphere for at least 10$^3$ years and \citet{Ohnaka2017} have demonstrated the variability of dust formation.

The variable component covers a broad cone that projects on the plane of the sky over region C, elongated and shifted with respect to the star in the NNW direction ($\sim$10\dego\ west of north). It is characterized by a high density and displays features of a mass ejection that started a few years ago. It is somehow related to the variable features observed earlier: dust formation enhanced in the same region \citep{Ohnaka2017,Takigawa2017} and evidence for shocks in the south-western quadrant \citep{Vlemmings2017,Ohnaka2017}. The details of this relation are still in need of being clarified.

While consistent with, and inspired by many of the conclusions reached by earlier authors from studies of the close vicinity of the star, in particular \citet{Vlemmings2017}, \citet{Ohnaka2017} and \citet{Takigawa2017}, this picture contradicts the model proposed by \citet{Khouri2014a} that implies a progressive acceleration to a terminal velocity of 7.5 \kms. It also impacts significantly the evidence for CO maser obtained by \citet{Vlemmings2021} and provides useful clarifications on the presence of a hot spot south-west of the stellar disc discovered by \citet{Vlemmings2017}.

As first argued by \citet{Vlemmings2017} this picture favours convective cell ejections as playing a major role in the generation of the nascent wind, the stable component of the CSE being seen as the result of many successive such events occurring in different directions at short time intervals. Major progress in modelling the relevant dynamics has been recently reported \citep{Hofner2019}.  If confirmed by future studies, such a picture raises several questions; in particular, the detailed relation between the features observed by \citet{Vlemmings2017}
and \citet{Ohnaka2017}, including the observed anti-correlation between shocks and dust formation, needs to be clarified. Moreover, the nearly exact compensation of the star gravity by the acceleration produced by the stellar light on dust grains, required to make the gas velocity stay constant well before having escaped the star gravity, needs to be understood.

While making use of many observations and analyses reported earlier, the present article offers a new and significantly improved description of the morpho-kinematics of the wind of W Hya. We failed to think of another sensible interpretation than presented here; in particular, invoking the presence of an evaporating planet in relation with the blue blob and the high Doppler velocity wings, as described in a recent publication \citep{Gottlieb2022}, cannot be seriously pursued in the present case. Most of our knowledge of the morpho-kinematics of the CSE of W Hya is based on observations confined to the very close neighbourhood of the star, less than 0.2 arcsec projected distance. To confirm the validity of the picture inferred from these observations, it is essential to observe the molecular line emission of several lines, such as CO, SiO, SO, SO$_2$, HCN and some of their isotopologues, over angular distances at the scale of a few arcsec. Each individual star has its own specific identity and progress in understanding the dynamics at stake in the generation of the nascent wind requires a detailed exploration of its properties. Moreover, new observations of the close neighbourhood of the star, of a same quality as those made at or near the end of 2015, and possibly including line emissions probing higher temperatures, such as SO$_2$($34_{3,31}-34_{2,32}$), would secure and refine our understanding of the variable dynamics.

\begin{acknowledgments} 
We thank Pr. Wouter Vlemmings and Pr. Aki Takigawa for useful discussions. We are grateful to Drs Jan Martin Winters and Pierre Leasaffre, for very careful reading of the manuscript and for pertinent comments that helped improving the presentation of the results. This paper uses ALMA data ADS/JAO.ALMA\#2015.1.01446.S. ALMA is a partnership of ESO (representing its member states), NSF (USA) and NINS (Japan), together with NRC (Canada), MOST and ASIAA (Taiwan), and KASI (Republic of Korea), in cooperation with the Republic of Chile. The Joint ALMA Observatory is operated by ESO, AUI/NRAO and NAOJ. The data are retrieved from the JVO/NAOJ portal. We are deeply indebted to the ALMA partnership, whose open access policy means invaluable support and encouragement for Vietnamese astrophysics. Financial support from the World Laboratory, the Odon Vallet Foundation and the Vietnam National Space Center is gratefully acknowledged. This research is funded by the Vietnam National Foundation for Science and Technology Development (NAFOSTED) under grant number 103.99-2019.368.
\end{acknowledgments}

\bibliography{whya2_vjste_review1_clean}{}

\begin{thebibliography}{}
\expandafter\ifx\csname natexlab\endcsname\relax\def\natexlab#1{#1}\fi
\providecommand{\url}[1]{\href{#1}{#1}}
\providecommand{\dodoi}[1]{doi:~\href{http://doi.org/#1}{\nolinkurl{#1}}}
\providecommand{\doeprint}[1]{\href{http://ascl.net/#1}{\nolinkurl{http://ascl.net/#1}}}
\providecommand{\doarXiv}[1]{\href{https://arxiv.org/abs/#1}{\nolinkurl{https://arxiv.org/abs/#1}}}

\bibitem[{{Danilovich} {et~al.}(2017){Danilovich}, {Lombaert}, {Decin},
  {Karakas}, {Maercker}, \& {Olofsson}}]{Danilovich2017}
{Danilovich}, T., {Lombaert}, R., {Decin}, L., {et~al.} 2017, \aap, 602, A14,
  \dodoi{10.1051/0004-6361/201630055}

\bibitem[{{Danilovich} {et~al.}(2019){Danilovich}, {Richards}, {Karakas}, {Van
  de Sande}, {Decin}, \& {De Ceuster}}]{Danilovich2019}
{Danilovich}, T., {Richards}, A.~M.~S., {Karakas}, A.~I., {et~al.} 2019,
  \mnras, 484, 494, \dodoi{10.1093/mnras/stz002}

\bibitem[{{Dumm} \& {Schild}(1998)}]{Dumm1998}
{Dumm}, T., \& {Schild}, H. 1998, \na, 3, 137,
  \dodoi{10.1016/S1384-1076(98)00003-7}

\bibitem[{{Etoka} {et~al.}(2001){Etoka}, {B{\l}aszkiewicz}, {Szymczak}, \& {Le
  Squeren}}]{Etoka2001}
{Etoka}, S., {B{\l}aszkiewicz}, L., {Szymczak}, M., \& {Le Squeren}, A.~M.
  2001, \aap, 378, 522, \dodoi{10.1051/0004-6361:20011184}

\bibitem[{{Gottlieb} {et~al.}(2022){Gottlieb}, {Decin}, {Richards}, {De
  Ceuster}, {Homan}, {Wallstr{\"o}m}, {Danilovich}, {Millar}, {Montarg{\`e}s},
  {Wong}, {McDonald}, {Baudry}, {Bolte}, {Cannon}, {De Beck}, {de Koter}, {El
  Mellah}, {Etoka}, {Gobrecht}, {Gray}, {Herpin}, {Jeste}, {Kervella},
  {Khouri}, {Lagadec}, {Maes}, {Malfait}, {Menten}, {M{\"u}ller}, {Pimpanuwat},
  {Plane}, {Sahai}, {Van de Sande}, {Waters}, {Yates}, \&
  {Zijlstra}}]{Gottlieb2022}
{Gottlieb}, C.~A., {Decin}, L., {Richards}, A.~M.~S., {et~al.} 2022, \aap, 660,
  A94, \dodoi{10.1051/0004-6361/202140431}

\bibitem[{{Hadjara} {et~al.}(2019){Hadjara}, {Cruzal{\`e}bes}, {Nitschelm},
  {Chen}, {Michael}, \& {Moreno}}]{Hadjara2019}
{Hadjara}, M., {Cruzal{\`e}bes}, P., {Nitschelm}, C., {et~al.} 2019, \mnras,
  489, 2595, \dodoi{10.1093/mnras/stz2240}

\bibitem[{{Hoai} {et~al.}(2021){Hoai}, {Anh}, {Tuyet Nhung}, {Darriulat},
  {Diep}, {Bich Ngoc}, \& {Thai}}]{Hoai2021}
{Hoai}, D.~T., {Anh}, P.~T., {Tuyet Nhung}, P., {et~al.} 2021, Journal of
  Korean Astronomical Society, 54, 171.
\newblock \doarXiv{2109.10624}

\bibitem[{{H{\"o}fner} \& {Freytag}(2019)}]{Hofner2019}
{H{\"o}fner}, S., \& {Freytag}, B. 2019, \aap, 623, A158,
  \dodoi{10.1051/0004-6361/201834799}

\bibitem[{{Khouri} {et~al.}(2015){Khouri}, {Waters}, {de Koter}, {Decin},
  {Min}, {de Vries}, {Lombaert}, \& {Cox}}]{Khouri2015}
{Khouri}, T., {Waters}, L.~B.~F.~M., {de Koter}, A., {et~al.} 2015, \aap, 577,
  A114, \dodoi{10.1051/0004-6361/201425092}

\bibitem[{{Khouri} {et~al.}(2014{\natexlab{a}}){Khouri}, {de Koter}, {Decin},
  {Waters}, {Lombaert}, {Royer}, {Swinyard}, {Barlow}, {Alcolea}, {Blommaert},
  {Bujarrabal}, {Cernicharo}, {Groenewegen}, {Justtanont}, {Kerschbaum},
  {Maercker}, {Marston}, {Matsuura}, {Melnick}, {Menten}, {Olofsson},
  {Planesas}, {Polehampton}, {Posch}, {Schmidt}, {Szczerba}, {Vandenbussche},
  \& {Yates}}]{Khouri2014a}
{Khouri}, T., {de Koter}, A., {Decin}, L., {et~al.} 2014{\natexlab{a}}, \aap,
  561, A5, \dodoi{10.1051/0004-6361/201322578}

\bibitem[{{Khouri} {et~al.}(2014{\natexlab{b}}){Khouri}, {de Koter}, {Decin},
  {Waters}, {Maercker}, {Lombaert}, {Alcolea}, {Blommaert}, {Bujarrabal},
  {Groenewegen}, {Justtanont}, {Kerschbaum}, {Matsuura}, {Menten}, {Olofsson},
  {Planesas}, {Royer}, {Schmidt}, {Szczerba}, {Teyssier}, \&
  {Yates}}]{Khouri2014b}
---. 2014{\natexlab{b}}, \aap, 570, A67, \dodoi{10.1051/0004-6361/201424298}

\bibitem[{{Khouri} {et~al.}(2020){Khouri}, {Vlemmings}, {Paladini}, {Ginski},
  {Lagadec}, {Maercker}, {Kervella}, {De Beck}, {Decin}, {de Koter}, \&
  {Waters}}]{Khouri2020}
{Khouri}, T., {Vlemmings}, W.~H.~T., {Paladini}, C., {et~al.} 2020, \aap, 635,
  A200, \dodoi{10.1051/0004-6361/201834618}

\bibitem[{{Knapp} {et~al.}(2003){Knapp}, {Pourbaix}, {Platais}, \&
  {Jorissen}}]{Knapp2003}
{Knapp}, G.~R., {Pourbaix}, D., {Platais}, I., \& {Jorissen}, A. 2003, \aap,
  403, 993, \dodoi{10.1051/0004-6361:20030429}

\bibitem[{{Lebzelter} {et~al.}(2005){Lebzelter}, {Hinkle}, {Wood}, {Joyce}, \&
  {Fekel}}]{Lebzelter2005}
{Lebzelter}, T., {Hinkle}, K.~H., {Wood}, P.~R., {Joyce}, R.~R., \& {Fekel},
  F.~C. 2005, \aap, 431, 623, \dodoi{10.1051/0004-6361:20041575}

\bibitem[{{Maercker} {et~al.}(2008){Maercker}, {Sch{\"o}ier}, {Olofsson},
  {Bergman}, \& {Ramstedt}}]{Maercker2008}
{Maercker}, M., {Sch{\"o}ier}, F.~L., {Olofsson}, H., {Bergman}, P., \&
  {Ramstedt}, S. 2008, \aap, 479, 779, \dodoi{10.1051/0004-6361:20078680}

\bibitem[{{Mamon} {et~al.}(1988){Mamon}, {Glassgold}, \& {Huggins}}]{Mamon1988}
{Mamon}, G.~A., {Glassgold}, A.~E., \& {Huggins}, P.~J. 1988, \apj, 328, 797,
  \dodoi{10.1086/166338}

\bibitem[{{M{\"u}ller} {et~al.}(2005){M{\"u}ller}, {Schl{\"o}der}, {Stutzki},
  \& {Winnewisser}}]{Muller2005}
{M{\"u}ller}, H. S.~P., {Schl{\"o}der}, F., {Stutzki}, J., \& {Winnewisser}, G.
  2005, Journal of Molecular Structure, 742, 215,
  \dodoi{10.1016/j.molstruc.2005.01.027}

\bibitem[{{Nhung} {et~al.}(2021){Nhung}, {Hoai}, {Tuan-Anh}, {Darriulat},
  {Diep}, {Ngoc}, {Phuong}, \& {Thai}}]{Nhung2021}
{Nhung}, P.~T., {Hoai}, D.~T., {Tuan-Anh}, P., {et~al.} 2021, \mnras, 504,
  2687, \dodoi{10.1093/mnras/stab954}

\bibitem[{{Norris} {et~al.}(2012){Norris}, {Tuthill}, {Ireland}, {Lacour},
  {Zijlstra}, {Lykou}, {Evans}, {Stewart}, \& {Bedding}}]{Norris2012}
{Norris}, B. R.~M., {Tuthill}, P.~G., {Ireland}, M.~J., {et~al.} 2012, \nat,
  484, 220, \dodoi{10.1038/nature10935}

\bibitem[{{Ohnaka} {et~al.}(2016){Ohnaka}, {Weigelt}, \&
  {Hofmann}}]{Ohnaka2016}
{Ohnaka}, K., {Weigelt}, G., \& {Hofmann}, K.~H. 2016, \aap, 589, A91,
  \dodoi{10.1051/0004-6361/201628229}

\bibitem[{{Ohnaka} {et~al.}(2017){Ohnaka}, {Weigelt}, \&
  {Hofmann}}]{Ohnaka2017}
---. 2017, \aap, 597, A20, \dodoi{10.1051/0004-6361/201629761}

\bibitem[{{Samus'} {et~al.}(2017){Samus'}, {Kazarovets}, {Durlevich},
  {Kireeva}, \& {Pastukhova}}]{Samus2017}
{Samus'}, N.~N., {Kazarovets}, E.~V., {Durlevich}, O.~V., {Kireeva}, N.~N., \&
  {Pastukhova}, E.~N. 2017, Astronomy Reports, 61, 80,
  \dodoi{10.1134/S1063772917010085}

\bibitem[{{Takigawa} {et~al.}(2017){Takigawa}, {Kamizuka}, {Tachibana}, \&
  {Yamamura}}]{Takigawa2017}
{Takigawa}, A., {Kamizuka}, T., {Tachibana}, S., \& {Yamamura}, I. 2017,
  Science Advances, 3, eaao2149, \dodoi{10.1126/sciadv.aao2149}

\bibitem[{{Takigawa} {et~al.}(2019){Takigawa}, {Kim}, {Igami}, {Umemoto},
  {Tsuchiyama}, {Koike}, {Matsuno}, \& {Watanabe}}]{Takigawa2019}
{Takigawa}, A., {Kim}, T.-H., {Igami}, Y., {et~al.} 2019, \apjl, 878, L7,
  \dodoi{10.3847/2041-8213/ab1f80}

\bibitem[{{Van de Sande} {et~al.}(2018){Van de Sande}, {Decin}, {Lombaert},
  {Khouri}, {de Koter}, {Wyrowski}, {De Nutte}, \& {Homan}}]{VandeSande2018}
{Van de Sande}, M., {Decin}, L., {Lombaert}, R., {et~al.} 2018, \aap, 609, A63,
  \dodoi{10.1051/0004-6361/201731298}

\bibitem[{{van Leeuwen}(2007)}]{vanLeeuwen2007}
{van Leeuwen}, F. 2007, \aap, 474, 653, \dodoi{10.1051/0004-6361:20078357}

\bibitem[{{Vlemmings} {et~al.}(2017){Vlemmings}, {Khouri}, {O'Gorman}, {De
  Beck}, {Humphreys}, {Lankhaar}, {Maercker}, {Olofsson}, {Ramstedt}, {Tafoya},
  \& {Takigawa}}]{Vlemmings2017}
{Vlemmings}, W., {Khouri}, T., {O'Gorman}, E., {et~al.} 2017, Nature Astronomy,
  1, 848, \dodoi{10.1038/s41550-017-0288-9}

\bibitem[{{Vlemmings} {et~al.}(2019){Vlemmings}, {Khouri}, \&
  {Olofsson}}]{Vlemmings2019}
{Vlemmings}, W.~H.~T., {Khouri}, T., \& {Olofsson}, H. 2019, \aap, 626, A81,
  \dodoi{10.1051/0004-6361/201935329}

\bibitem[{{Vlemmings} {et~al.}(2021){Vlemmings}, {Khouri}, \&
  {Tafoya}}]{Vlemmings2021}
{Vlemmings}, W.~H.~T., {Khouri}, T., \& {Tafoya}, D. 2021, \aap, 654, A18,
  \dodoi{10.1051/0004-6361/202141656}

\bibitem[{{Zhao-Geisler} {et~al.}(2015){Zhao-Geisler}, {K{\"o}hler}, {Kemper},
  {Kerschbaum}, {Mayer}, {Quirrenbach}, \& {Lopez}}]{ZhaoGeisler2015}
{Zhao-Geisler}, R., {K{\"o}hler}, R., {Kemper}, F., {et~al.} 2015, \pasp, 127,
  732, \dodoi{10.1086/682261}

\end{thebibliography}
\bibliographystyle{aasjournal}

\appendix

\begin{figure*}
  \centering
 
  \includegraphics[height=8.7cm]{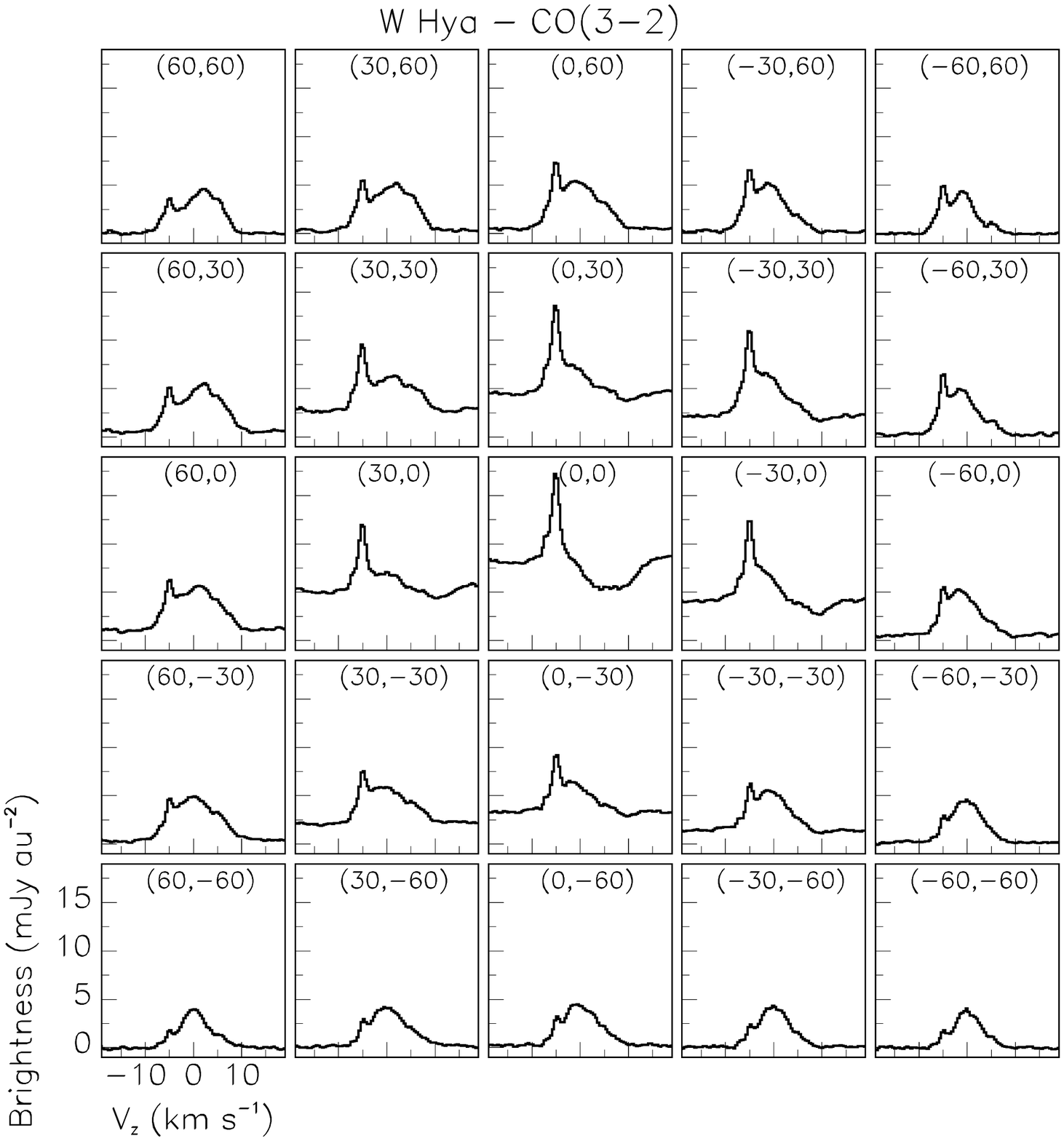}
  \includegraphics[height=8.7cm]{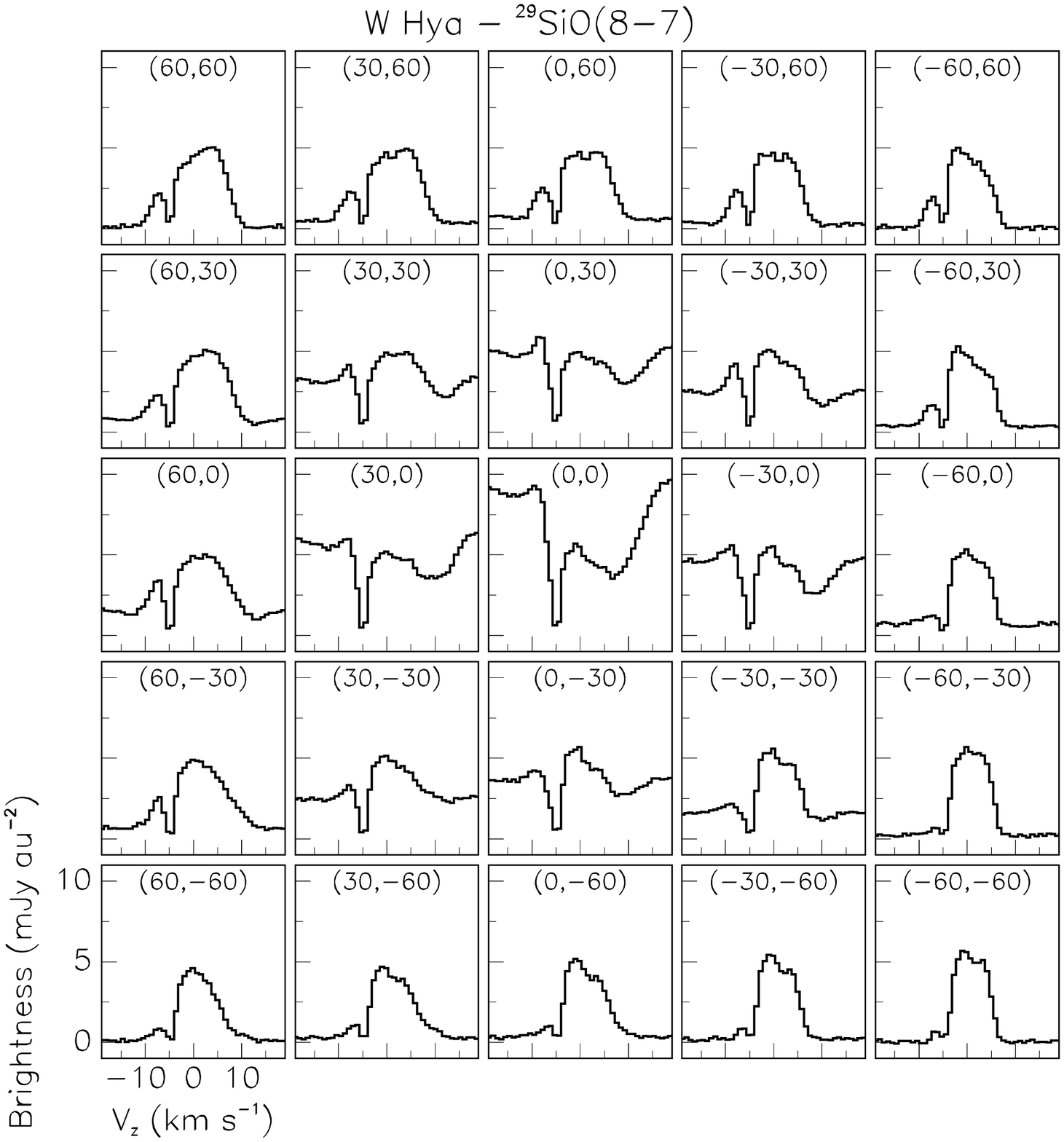}
  \includegraphics[height=8.7cm]{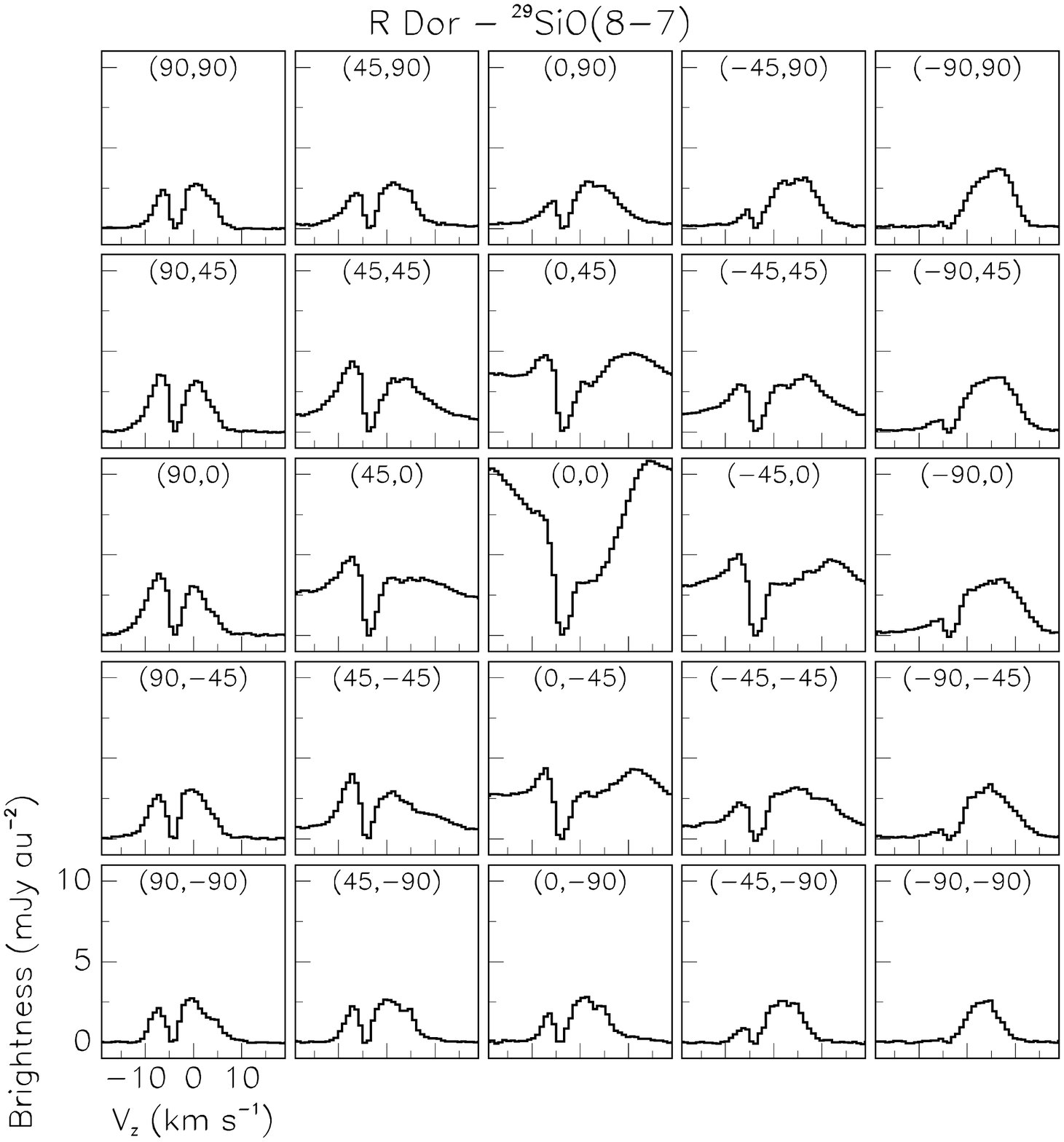}
  \caption{ Spectral maps of W Hya and R Dor molecular line emissions. The first set of 5$\times$5 maps is for the $^{12}$CO(3-2) emission of W Hya. The second and third sets are for the $^{29}$SiO(8-7) emissions of W Hya and R Dor, respectively. Each spectrum is averged over a square aperture of 30$\times$30 mas$^2$  for W Hya and of 45$\times$45 mas$^2$ for R Dor centred at ($x,y$) position indicated in the inserts (in mas) of each panel. The beam (FWHM) is 52$\times$40 mas$^2$ for W Hya and 41$\times$35 mas$^2$ for R Dor. The R Dor spectra have been divided by a factor 3.1 to account for the different distances (104 pc for W Hya and 59 pc for R Dor).}
  \label{figa1}
\end{figure*}

\begin{figure*}
  \centering
  \includegraphics[width=14cm,trim=.0cm 1.8cm 0cm 0cm,clip]{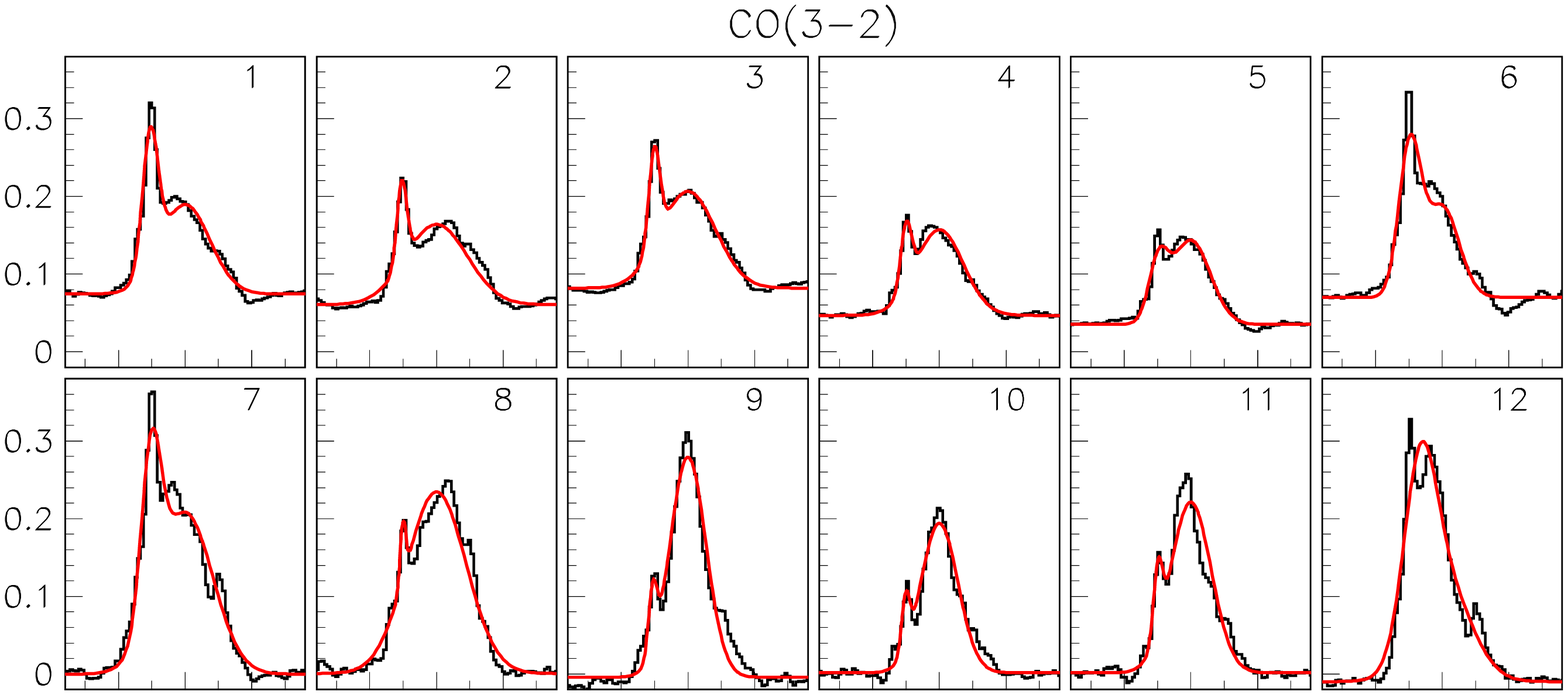}
  \includegraphics[width=14cm]{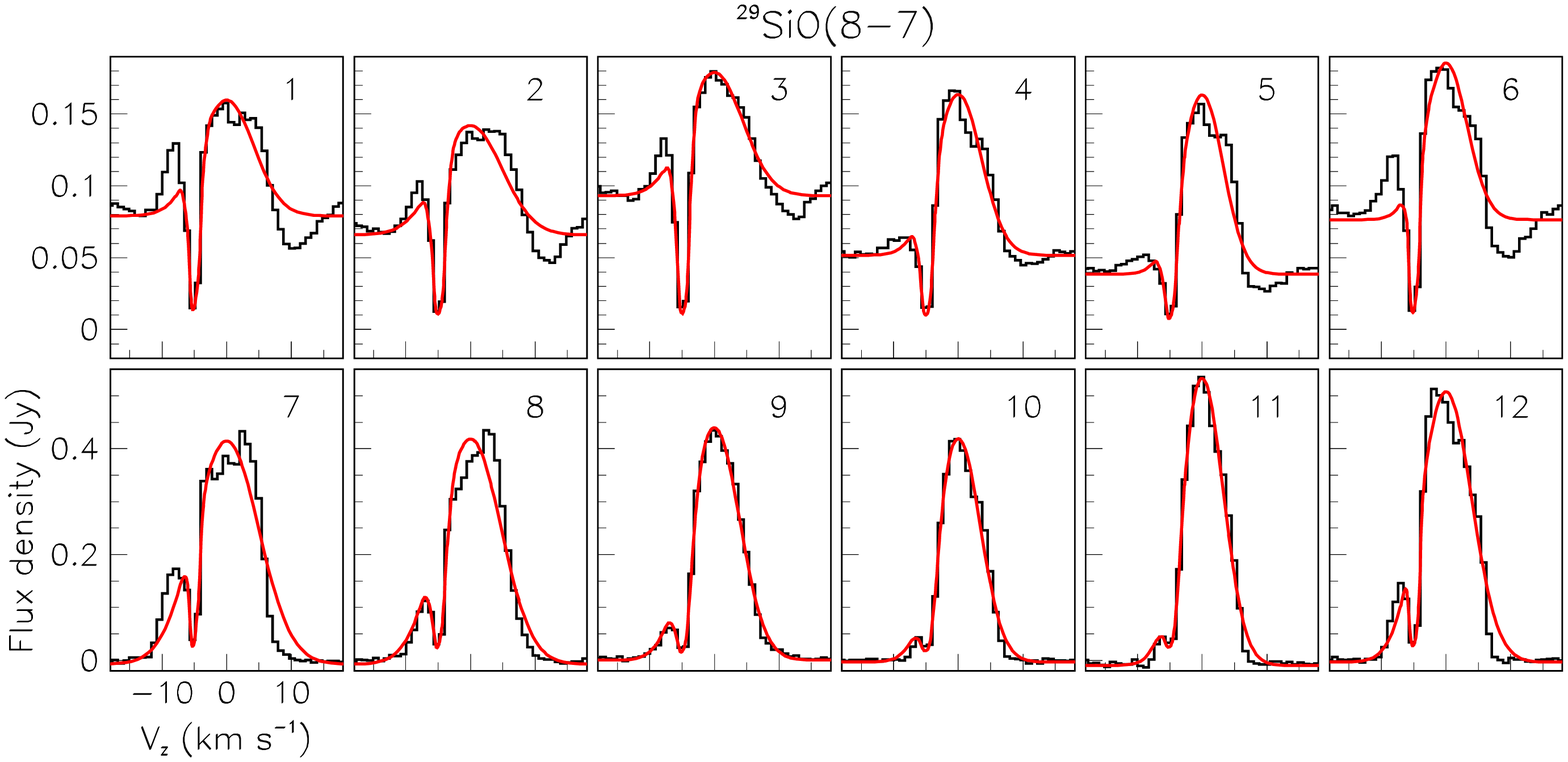}
  \caption{Doppler velocity spectra observed in 12 sextants for the $^{12}$CO(3-2) emission (upper pair of rows) and $^{29}$SiO(8-7) emission (lower pair of rows) of W Hya. The sextants cover the projected radial ranges 0$<$$R$$<$75 mas (upper row of each pair) and 75$<$$R$$<$150 mas (lower row of each pair), respectively. In each row the sextants are numbered in increasing order starting from north counter-clockwise, from 1 to 6 for the inner sextants and from 7 to 12 for the outer sextants. Sextants 1 and 7 cover position angles $|\omega|$$<$30\dego. The red lines show the results of fits of the form: $f$=$a+b\exp(-\frac{1}{2}V_z^2/c^2)+d\exp[-\frac{1}{2}(V_z-e)^2/f^2]$. The best fit values of the parameters are listed in Tables \ref{taba2} and \ref{taba3}.}
   \label{figa2}
\end{figure*}

Table \ref{taba1} lists some of the most relevant observations and analyses of the close neighbourhood of the star available in the published literature and Figure \ref{figa1} displays spectral maps for $^{12}$CO(3-2) and $^{29}$SiO(8-7) emissions in both W Hya and R Dor. Figure \ref{figa2} displays Doppler velocity spectra of the $^{12}$CO(3-2) and $^{29}$SiO(8-7) emissions within 150 mas projected distance from the centre of W Hya, separated in twelve sextants.  Tables \ref{taba2} and \ref{taba3} list the parameters describing these spectra, as defined in the caption of Figure \ref{figa2}.

\begin{deluxetable*}{c|c|c|c|c}
\tablenum{6}
\tablecaption{Some of the most relevant observations and analyses of the close neighbourhood of the star available in the published literature.\label{taba1}}
\tablehead{\colhead{$R [R_*]$}&\colhead{$\sim$1}&\colhead{1-2}&\colhead{2-3}&\colhead{3-6}}
\startdata
\multicolumn{5}{c}{\citet{Hoai2021}}\\
\hline
$^{29}$SiO(8-7)&\multicolumn{3}{c|}{Inner quadrants (sextants)}&Outer quadrants (sextants)\\
$^{12}$CO(3-2)&\multicolumn{3}{c|}{Broad linewidth}&Blue blob\\
\hline
\multicolumn{5}{c}{\citet{Vlemmings2017}}\\
\hline
Continuum ALMA&Photosphere 2500 K&&&\\
&Hot spot SW&&&\\
CO $v$=1 ALMA& Warm molecular gas layer&Cool gas &&\\
&2900 K&in-fall and outflow 900 K&&\\
\hline
\multicolumn{5}{c}{\citet{Ohnaka2017}}\\
\hline
VLT/&&Clumpy dust clouds&\multicolumn{2}{c}{Polarization on dust}\\
SPHERE/&&Depressed SW&\multicolumn{2}{c}{Shell structure}\\
ZIMPOL&&H$\alpha$: shocks anti-correlated& \multicolumn{2}{c}{Weaker H$\alpha$}\\
&&with dust clouds& \multicolumn{2}{c}{}\\
&&\multicolumn{3}{c}{TiO emission}\\
VLTI/AMBER & CO lines&\multicolumn{3}{c}{}\\
\hline
\multicolumn{5}{c}{\citet{Takigawa2017}}\\
\hline
AlO ALMA&&\multicolumn{2}{c|}{Depletion AlO gas}&\\
  &&\multicolumn{2}{c|}{Formation AlO dust}&\\
  &&\multicolumn{2}{c|}{SW depressed}&\\
$^{29}$SiO ALMA&&\multicolumn{2}{c|}{Blob revealing acceleration}&Progressive depletion\\
  \enddata
\end{deluxetable*}

\begin{deluxetable*}{ccccccccccccc}
  \tablenum{7}
  \tablecaption{Parameters describing the Doppler velocity spectra observed in the CO sextants, as described in the caption of Figure \ref{figa2}.\label{taba2}}
\tablewidth{0pt}
  \tablehead{ 
    \colhead{Sextant}&    \colhead{1}&    \colhead{2}&    \colhead{3}&    \colhead{4}&    \colhead{5}&    \colhead{6}&    \colhead{7}&    \colhead{8}&    \colhead{9}&    \colhead{10}&    \colhead{11}&    \colhead{12}}
  \startdata
  $a$ (mJy)&  74&  60&  82&  47&  35&  70&  0&  0&  $-$4&  2&  1&  1\\
  $b$ (mJy)&  110&  100&  120&  110&  110&  120&  210&  230&  280&  190&  220&  130\\
  $c$&  3.71&  4.63&  4.15&  3.61&  2.93&  2.52&  4.13&  4.44&  2.69&  2.80&  3.08&  4.13\\
  $d$ (mJy)&  170&  110&  120&  79&  69&  190&  220&  77&  70&  71&  89&  210\\
  $e$&  $-$5.23&  $-$5.20&  $-$5.05&  $-$5.00&  $-$4.89&  $-$4.88&  $-$5.10&  $-$5.03&  $-$5.43&  $-$5.01&  $-$4.90&  $-$3.38\\
  $f$&  1.18&  0.74&  0.84&  0.79&  1.43&  1.65&  1.48&  0.39&  0.78&  0.52&  0.60&  2.40\\
  \enddata
\end{deluxetable*}

\begin{deluxetable*}{ccccccccccccc}
  \tablenum{8}
  \tablecaption{Parameters describing the Doppler velocity spectra observed in the SiO sextants, as described in the caption of Figure \ref{figa2}.\label{taba3}}
  \tablehead{\colhead{Sextant}&\colhead{1}&\colhead{2}&\colhead{3}&\colhead{4}&\colhead{5}&\colhead{6}&\colhead{7}&\colhead{8}&    \colhead{9}&    \colhead{10}&    \colhead{11}&    \colhead{12}}
  \startdata
  $a$ (mJy)&79&66&93&52&39&76&$-$7&$-$7&1&$-$3&$-$9& $-$3\\
  $b$ (mJy)&  81&  76&  86&  112&  124&  109&  420&  430&  440&  420&  540&  510\\
  $c$&  4.17&  4.85&  4.38&  3.58&  3.25&  3.26&  4.96&  4.74&  3.86&  3.43&  3.27&  3.96\\
  $d$ (mJy)&  $-$110&  $-$110&  $-$140&  $-$90&  $-$74&  $-$112&  $-$247&  $-$220&  $-$200&  $-$150&  $-$160&  $-$250\\
  $e$&  $-$4.93&  $-$4.84&  $-$4.82&  $-$4.75&  $-$4.76&  $-$4.88&  $-$4.91&  $-$4.82&  $-$4.59&  $-$4.33&  $-$4.40&  $-$4.80\\
  $f$&  0.76&  0.90&  0.88&  0.91&  0.92&  0.66&  0.60&  0.97&  1.03&  1.11&  1.00&  0.58\\
  \enddata
\end{deluxetable*}


\end{document}